\documentclass{article}
\usepackage{graphicx}
\usepackage{authblk}
\usepackage{verbatim}
\usepackage{array, tabularx}

\usepackage[a4paper, total={6in, 8in}]{geometry}

\usepackage{xr}
\usepackage[utf8]{inputenc}
\usepackage[T1]{fontenc}
\usepackage{amssymb}

\usepackage{calc}
\usepackage{amsmath}

\usepackage{xparse}

\usepackage[export]{adjustbox}
\usepackage{pgffor}
\usepackage{nomencl}

\usepackage{multicol}
\makenomenclature

\usepackage{amsfonts}
\usepackage{calrsfs}

\usepackage{graphicx}
\usepackage{grffile}
\usepackage{svg}
\usepackage{subfig}
\usepackage{cleveref}

\usepackage{xspace}
\usepackage{bm}
\usepackage{placeins}
\usepackage{pdfpages}
\usepackage[font={small,it}]{caption}
\usepackage{cite}
\usepackage{array, tabularx}
\usepackage{makecell}
\usepackage{multirow}
\usepackage{booktabs}
\usepackage{url}
\usepackage{xcolor}
\usepackage[export]{adjustbox}
\usepackage{framed}
\usepackage{pdflscape}
\usepackage{rotating}
\usepackage{afterpage}

\usepackage{array}
\newcolumntype{C}[1]{>{\centering\arraybackslash}p{#1}}
\newcolumntype{R}[1]{>{\raggedleft\arraybackslash}p{#1}}
\newcolumntype{L}[1]{>{\raggedright\arraybackslash}p{#1}}

\usepackage{balance}
\usepackage{float}

\newenvironment{conditions*}
  {\par\vspace{\abovedisplayskip}\noindent
   \tabularx{\columnwidth}{>{$}l<{$} @{${}:{}$}
   >{\raggedright\arraybackslash}X}} {\endtabularx\par\vspace{\belowdisplayskip}}

\usepackage[linesnumbered,ruled,vlined]{algorithm2e}

\SetCommentSty{mycommfont}

\SetAlFnt{\sffamily}

\makeatletter
\renewcommand{\algocf@Vline}[1]{
  \strut\par\nointerlineskip
  \algocf@push{\skiprule}
  \hbox{\bgroup\color{cyan}\vrule\egroup%
    \vtop{\algocf@push{\skiptext}
      \vtop{\algocf@addskiptotal #1}\bgroup\color{cyan}\Hlne\egroup}}\vskip\skiphlne
  \algocf@pop{\skiprule}
  \nointerlineskip}
\renewcommand{\algocf@Vsline}[1]{
  \strut\par\nointerlineskip
  \algocf@bblockcode%
  \algocf@push{\skiprule}
  \hbox{\bgroup\color{cyan}\vrule\egroup
    \vtop{\algocf@push{\skiptext}
      \vtop{\algocf@addskiptotal #1}}}
  \algocf@pop{\skiprule}
  \algocf@eblockcode%
}
\makeatother

\SetKwProg{Fn}{Function}{}{}
\SetKwInOut{Parameter}{Parameters}
\SetKw{KwVar}{Var}
\SetKw{KwSet}{Set}
\SetKw{KwTuple}{Tuple}
\SetKw{KwBreak}{break}
\usepackage{amsthm}

\usepackage{authblk}

\usepackage{nameref}

\usepackage{tikz}
\usetikzlibrary{arrows,shapes,automata,backgrounds, arrows.meta}
\usetikzlibrary{positioning}
\usetikzlibrary{decorations.pathreplacing}

\usepackage[official]{eurosym}
\usepackage{footnote}

\usepackage{thmtools}
\declaretheoremstyle[
spaceabove=6pt, spacebelow=6pt,
headfont=\normalfont\bfseries,
notefont=\mdseries, notebraces={(}{)},
bodyfont=\normalfont,
postheadspace=0.6em,
headpunct=:
]{mystyle}

\crefname{hyp}{hypothesis}{hypotheses}
\Crefname{hyp}{Hypothesis}{Hypotheses}

\DeclareSymbolFontAlphabet{\amsmathbb}{AMSb}%

\newcommand\Tau{\mathcal{T}}

\newcommand\ignore[1]{}

\newcommand\vmath[1]{\ensuremath{#1}\xspace}

\newcommand\commentOut[1]{}

\newcommand\toSet[1]{\vmath{\mathbb{\MakeUppercase{#1}}}}
\newcommand\cardinal[1]{\vmath{\left|\,#1\,\right|}}

\newcommand\vuser[0]{
	\vmath{u}
}
\nomenclature[01]{\vuser}{A user, where \vmath{\vuser \in \toSet{\vuser}}}
\nomenclature[01]{\toSet{\vuser}}{A set of users.}

\nomenclature[01]{\Tau}{A binary variable \vmath{\Tau \in \{0,1\}} indicating the participation of a user in a
treatment.}

\nomenclature[01]{R}{A binary variable \vmath{R \in \{0,1\}} indicating the (possible) response of a user receiving a
treatment.}

\nomenclature[01]{\toSet{\vuser}_{\Tau=1, R=1}}{A set of users that received and responded to a treatment.
Often referred to as treatment group.}
\nomenclature[01]{\toSet{\vuser}_{\Tau=0, R=1}}{A set of users that did not receive but would probably respond to a
treatment. Often referred to as control group.}
\nomenclature[01]{X}{A vector/matrix/tensor of covariates that are used to estimate a treatment effect on a
behavioral metric \vmath{Y}.}
\nomenclature[01]{Y}{A behavior metric that is used to measure the effect of a treatment.}

\newcommand\vproduct{
	\vmath{p}
}
\nomenclature[02]{\vproduct}{A product, where \vmath{\vproduct \in \toSet{\vproduct}}.}

\newcommand\vproducttag[0]{
	\vmath{z}
}
\nomenclature[03]{\vproducttag}{A product tag belonging to a product tag set \vmath{\toSet{\vproducttag}}.}
\nomenclature[04]{\vmath{\toSet{\vproducttag}_{\vproduct}}}{A product tag set for a specific product \vmath{\vproduct}.}

\newcommand\vpreference{
	\vmath{c}
}
\nomenclature[05]{\vpreference}{A preference statement, where
\vmath{\vpreference} $\in$ \toSet{\vpreference}.}

\newcommand\vpreferencetag{
	\vmath{\omega}
}
\nomenclature[06]{\vpreferencetag}{A preference tag belonging to a preference tag set \vmath{\toSet{\vpreferencetag}}}
\nomenclature[07]{\vmath{\toSet{\Omega}_{\vpreference}}}{A preference tag set related to a specific preference
\vmath{\vpreference}.}

\newcommand\vpreferencescore[0]{
	\vmath{s_{u,c}}
}

\nomenclature[08]{\vpreferencescore}{A preference score given by a user \vmath{\vuser}
to a preference statement \vmath{\vpreference}.}

\newcommand\vscoreoffset[1]{
	\vmath{o(#1)}
}
\nomenclature[09]{\vscoreoffset{\vpreferencescore}}{The subtraction of the preference score
median value \vmath{\vMedian{s}} from a preference score \vmath{\vpreferencescore}.}
\nomenclature[10]{\vmath{\cardinal{\cdot}}}{The absolute value of a number $\cdot$ or the cardinality of a set.}

\nomenclature[10]{\vmath{\toSet{Q}}}{A set of concepts.}
\nomenclature[10]{\vmath{\toSet{Q}^{+}_{\vpreferencetag}}}{A set of concepts that support a preference tag
\vmath{\vpreferencetag}.}
\nomenclature[10]{\vmath{\toSet{Q}^{-}_{\vpreferencetag}}}{A set of concepts that oppose a preference tag
\vmath{\vpreferencetag}.}
\nomenclature[10]{\vmath{\toSet{Q}_{\vpreferencetag}}}{A set of all concepts that define a preference tag
\vmath{\vpreferencetag}.}
\nomenclature[10]{\vmath{\toSet{Q}_{\vproducttag}}}{A set of concepts that define a product tag \vmath{\vproducttag}.}

\newcommand\vcorr[1]{
	\vmath{r(#1)}
}

\newcommand\vcorrelationindexed[2]{
	\vmath{\vcorr{#1, #2}}
}
\newcommand\vcorrelation[0]{
	\vmath{\vcorrelationindexed{\vproducttag}{\vpreferencetag}}
}
\nomenclature[11]{r^{+}(\vproducttag, \vpreferencetag)}{The positive association score between a product tag
\vproducttag and a preference tag \vmath{\vpreferencetag}, which is maximized when the concepts of a product tag
\vmath{\vproducttag}
are
enough to support and define the existence of a preference tag \vmath{\vpreferencetag}.}
\nomenclature[11]{r^{-}(\vproducttag, \vpreferencetag)}{The negative association score between a product tag
\vmath{\vproducttag} and a preference tag \vmath{\vpreferencetag}, which is maximized when the concepts of a product tag
\vmath{\vproducttag} are enough to support and define the absence of a preference tag \vmath{\vpreferencetag}.}
\nomenclature[11]{\vcorrelation}{The association score between a product tag
\vproducttag and a preference tag \vmath{\vpreferencetag}.}

\nomenclature[11]{\vcorrelation}{The association score between a product tag
\vmath{\vproducttag} and a preference tag \vmath{\vpreferencetag}.}

\nomenclature[11]{\epsilon}{The error introduced when aggregating tag associations \vmath{\vcorrelation} in the
approximation of
\vmath{\vrelevance}.
The overlaps may be or on positive and negative concepts, causing the corresponding \vmath{\epsilon^{+},
\epsilon^{-}}
errors.}

\newcommand\vrel[2]{
	\vmath{\eta({#1}, {#2})}
}
\newcommand\vrelevance[0]{
	\vrel{\vproduct}{\vpreferencetag}
}
\newcommand\vrelsupp[2]{
	\vmath{\eta^{#1}{(#2)}}
}

\newcommand\vrelpos[0]{
	\vrelsupp{+}{\vpreferencetag}
}
\newcommand\vrelneg[0]{
	\vrelsupp{-}{\vpreferencetag}
}
\newcommand\vrelnorm[0]{
	\vrelsupp{*}{\vproduct, \vpreferencetag}
}

\nomenclature[11]{\vrelevance}{The aggregated association score between a
product \vmath{\vproduct} and a preference tag
	\vmath{\vpreferencetag}.}
\nomenclature[12]{\vrelpos}{The positive reference association score
that can be achieved for a preference tag \vmath{\vpreference} and all products
\vmath{\forall\,\vproduct\,\in\,\toSet{\vproduct}}.
The negative reference correlation is referred as \vrelneg}
\nomenclature[13]{\vrelnorm}{The normalized aggregated association score for a
preference \vmath{\vpreferencetag} and a product \vmath{\vproduct}.}

\newcommand\vconn[2]{\vmath{v(#1, #2)}}
\newcommand\vconnection{\vmath{\vconn{\vproduct}{\vpreference}}}
\nomenclature[14]{\vconnection}{The average association value between a
preference \vmath{\vproduct} and a product \vmath{\vpreference}.}

\newcommand\vrating[2]{
	\vmath{\varrho(#1, #2)}
}
\nomenclature[14]{\varrho^{*}(\vproduct, \vuser)}{The raw product rating assigned to a
product \vmath{\vproduct} based on the preferences of a user \vmath{\vuser} before scaling}
\nomenclature[14]{\vrating{\vproduct}{\vuser}}{The product rating assigned to a
product \vmath{\vproduct} based on the preferences of a user \vmath{\vuser}.}

\makeatletter
	\newcommand{\removelatexerror}{\let\@latex@error\@gobble}
\makeatother

\newcommand\vMedian[1]{\vmath{\widetilde{#1}}}

\ExplSyntaxOn
\NewDocumentCommand{\replace}{mmm}
 {
  \marian_replace:nnn {#1} {#2} {#3}
 }

\tl_new:N \l_marian_input_text_tl

\cs_new_protected:Npn \marian_replace:nnn #1 #2 #3
 {
  \tl_set:Nn \l_marian_input_text_tl { #1 }
  \tl_replace_all:Nnn \l_marian_input_text_tl { #2 } { #3 }
  \tl_use:N \l_marian_input_text_tl
 }
\ExplSyntaxOff

\definecolor{dpurple}{HTML}{5e3c99}
\definecolor{lpurple}{HTML}{b2abd2}
\definecolor{dorange}{HTML}{e66101}
\definecolor{lorange}{HTML}{fdb863}
\definecolor{llpurple}{HTML}{DFD8FF}

\tikzset{entity/.style={
    draw=#1,
    thick,
    ellipse,
    minimum width=0.75cm,
    minimum height=0.75cm,
    font=\small,
    outer sep=3pt,
  },
  text style/.style={
    sloped,
    text=black,
    font=\footnotesize,
    above
  }
}

\tikzset{entity node/.style={entity,
                            draw=lorange!75,
                            fill=lorange!20,
                            align=center,
                            font=\bfseries,
                            dotted,
                            line width = 1.2mm
                            }
}
 \tikzset{literal node/.style={entity,
                                draw=dorange,
                                fill=dorange!60,
                                align=center,
                                inner sep=2pt,
                                font=\bfseries,
                                text=white
                                }
}

\tikzset{scalar node/.style={entity,
                             draw=lpurple!75,
                             fill=lpurple!60,
                             align=center,
                             font=\bfseries,
                             }
}

\tikzset{calculated node/.style={entity,
                                 thick,
                                 draw=dpurple,
                                 fill=dpurple!60,
                                 align=center,
                                 font=\bfseries,
                                 text=white
                            }
}

\tikzset{edgestyle/.style={ fill=white,
                            anchor=center,
                            pos=0.5,
                            font=\bfseries
                            }
}
\tikzset{isa/.style={       -{Latex[open]},
                            dashed,
                            shorten >=1pt,
                            auto
                            }
}
\tikzset{isalabel/.style={
                            fill=white,
                            anchor=center,
                            pos=0.4
                            }
}

\newcolumntype{a}{>{\hsize=.70\hsize}X}
\newcolumntype{b}{>{\hsize=.90\hsize}X}
\newcolumntype{d}{>{\hsize=.20\hsize}X}
\newcolumntype{e}{>{\hsize=.40\hsize}X}
\newcolumntype{z}{>{\hsize=.10\hsize}X}
\newcolumntype{q}{>{\raggedright\arraybackslash}X}
\newcolumntype{f}{>{\hsize=.50\hsize}X}
\newcolumntype{N}{>{\centering\arraybackslash}X}
\newcolumntype{Y}{>{\centering\arraybackslash}X}
\newcolumntype{y}{>{\raggedleft\arraybackslash}X}
\newcolumntype{g}{>{\centering\arraybackslash}X}
\newcolumntype{G}{>{\hsize=.5\hsize}r}

\let\origref\ref
\def\ref#1{\textnormal{\origref{#1}}}

\begin{document}

\title{How Value-Sensitive Design Can Empower Sustainable Consumption}
\author[1]{Thomas Asikis}
\author[2]{Johannes Klinglmayr}
\author[1]{Dirk Helbing}
\author[3]{Evangelos Pournaras}
\affil[1]{Professorship of Computational Social Science, ETH Zurich, Zurich, Switzerland, E-mail: \{asikist,d.helbing\}@ethz.ch}
\affil[2]{Linz Center of Mechatronics GmbH, Linz, Austria, E-mail: Johannes.Klinglmayr@lcm.at}
\affil[3]{School of Computing, University of Leeds, Leeds, UK, E-mail: e.pournaras@leeds.ac.uk}

\renewcommand\Authands{ and }

\maketitle

\begin{abstract}
In a so-called overpopulated world, sustainable consumption is of existential importance.However, the expanding spectrum of product choices and their production complexity challenge consumers to make informed and value-sensitive decisions. Recent approaches based on (personalized) psychological manipulation are often intransparent, potentially privacy-invasive and inconsistent with (informational) self-determination. In contrast, responsible consumption based on informed choices currently requires reasoning to an extent that tends to overwhelm human cognitive capacity. As a result, a collective shift towards sustainable consumption remains a grand challenge. Here we demonstrate a novel personal shopping assistant implemented as a smart phone app that supports a value-sensitive design and leverages sustainability awareness, using experts' knowledge and “wisdom of the crowd” for transparent product information and explainable product ratings. Real-world field experiments in two supermarkets confirm higher sustainability awareness and a bottom-up behavioral shift towards more sustainable consumption. These results encourage novel business models for retailers and producers, ethically aligned with consumer preferences and with higher sustainability.
\end{abstract}

Creating more sustainable consumption patterns turns out to be imperative for mitigating climate change and supporting a more viable future of our society~\cite{parodi_potential_2018,sachs_six_2019}. Food systems, in particular, play a key role, influencing 12 out of the 17 Sustainable Development Goals of the United Nations~\cite{Desa2016}. Food supply is already leading total greenhouse gas emissions~\cite{Food2016,Bajzelj2013}, while food demand is expected to grow by 50\% along with an increase of the global population from 7 to 9.8 billion people by 2050~\cite{Searchinger2019}. Thus, introducing policies that promote more sustainable consumption patterns is critical. Yet, it has proven to be insufficient to overcome the `attitude-behaviour gap'~\cite{Tomkins2018}. For instance, chosing grocery products according to a broad spectrum of (often opposing~\cite{Hasegawa2019,Nerini2019,Spaiser2017}) sustainability criteria requires the processing of an overwhelming amount of information and, as a result, the price often remains the dominant choice criterion. Such information is often summarized and its processing is automated, i.e. by product labels or smart phone shopping assistants that scan product barcodes~\cite{Grunert2014,Horne2009,Winkler2012}. However,  ensuring a value-sensitive design in terms of accountability, transparency, privacy, self-determination, and an overall practicality by a seamless integration into the shopping process remains a grand challenge. However, designing digital shopping assistants, e.g. smart phone apps, for such values turns out to be a requirement for consumer trust (see IEEE Global Initiative on Ethics of Autonomous and Intelligent Systems~\cite{Chatila2019}). Given the broad scope and complexity of the subject of sustainability and the co-determination of consumer decisions by price~\cite{Choi2011},  product choices based on sustainability criteria can be particularly prone/vulnerable to manipulative nudging when personalized decision-support systems use personal data. 

In this contribution we show how value-sensitive design can empower a bottom-up shift to more sustainable consumption. To make this possible, we built a novel and general-purpose shopping assistant~\cite{ASSET} and tested it at two real-world supermarkets in Estonia (Retailer A~\cite{Coop}) and Austria (Retailer B~\cite{Winkelmarkt}). The shopping assistant was implemented as an Android smart phone application (app) for decision-support: It rates products (as well as product categories) that are in front of the consumer in a shop according to \emph{personal sustainability criteria} (preferences) and in a privacy-preserving way. These preferences cover the categories of \emph{environment}, \emph{social}, \emph{health}, and \emph{(product) quality} so that consumers can improve their health while supporting socially favorable production conditions and making choices with better environmental impact. Products that better match consumers' preferences are rated close to $10$ and those that oppose the preferences receive values close to $0$. The consumer can interact with the app to change sustainability preferences using a continuous Likert-scale slider in the range $[0,10]$. Ratings are explainable to consumers by making transparent why certain products receive higher ratings than others.

\begin{figure}[!htb]
\centering
\includegraphics[width=1.0\textwidth]{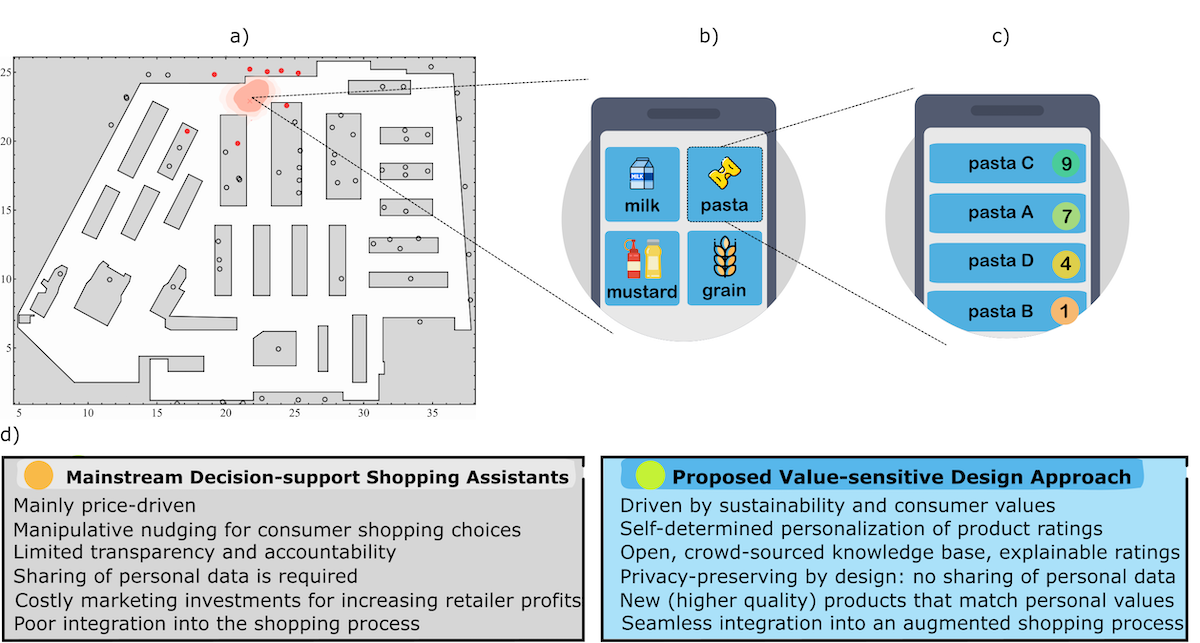}
\caption{\textbf{An outline and comparison of the proposed value-sensitive, preference-based decision-support process}. (a) A consumer's localization at Retailer A. The red dots on the map denote the product categories in close proximity. (b) Presenting the product categories in close proximity to the consumer. (c) Product ratings close to $10$ denote better matching of the product to the consumer's sustainability preferences. (d) Comparison of the proposed value-oriented, preference-based design approach with mainstream decision-support shopping assistants.}\label{fig:ASSET}
\end{figure}

The value-sensitive design of this privacy-preserving digital shopping assistant is distinguished from other mainstream approaches (see Figure~\ref{fig:ASSET}d) by integrating the following values in decision-support: (i) Adding environmental, social, health and quality criteria in consumers' decisions. (ii) Limiting manipulative nudging. (iii) Self-determination in consumers' preference choices. (iv) Privacy preservation. (v) Explainability. (vi) Practicality and compatibility with the existing shopping process. 

These values are realized by a design process that tackles the challenges to provide (i) \emph{transparent and explainable personalized product ratings}, (ii) \emph{a seamless integration into the consumers' shopping process}, and (iii) \emph{scalable, high quality product information on sustainability}. The scope and combination of inter-disciplinary methods required for this is quite exceptional, covering a new value-sensitive approach, its implementation within a sophisticated system and the testing of this approach in real-world settings using rigorous social science methods.  This sets this work apart from most other state-of-the-art studies (see Table~\ref{tbl:online:ratings} in the Supplementary Information).

\textbf{Transparent and explainable personalized product rating.} The first challenge of value-sensitive design is addressed by designing a novel content-based recommender algorithm for the personalized rating of products that is decentralized and privacy-preserving. In contrast to user-based collaborative filtering algorithms that rely on the collection of sensitive historical consumer data to compare purchase profiles~\cite{Friedman2015}, the novelty of our algorithm is that personalization is taking place on the smart phone and as a result, the sustainability preferences of a consumer remain localized on the consumer's device, i.e. private. This is possible via a scalable, distributed communication protocol that handles a consumer's request for public summarized product information on sustainability (sustainability indices). When this information is localized on the smart phone, efficient calculations turn the sustainability index of products into personalized product ratings. The implications of our value-oriented and preference-based design approach are significant: without sharing sensitive personal data with third parties, manipulative nudging that serves corporate interests and may oppose personal preferences or sustainable consumption patterns is limited~\cite{Helbing2019}. As a result, consumers can trust that the rating of the products is a result of their own intrinsic values expressed via their sustainability preferences. Moreover, the calculation of a product rating is accountable to the consumer, who can visually explore which product information and preferences are mainly responsible for the overall rating (see Figure~\ref{fig:contr} in the Supplementary Information). The explainability is also localized on the smart phone so that no third parties can manipulate the perception of a consumer on why certain products receive higher or lower ratings.  
 
\textbf{Seamless integration into the consumer shopping process.} For the second challenge, a seamless integration in the shopping process is crucial for the practicality and adoption of the solution in retailer shops. Barcode scanning makes the comparison of different products cumbersome~\cite{vonReischach2010}. Instead, in our system, consumers automatically view all nearby products of a category, e.g. all different pasta products, on their smart phone and can therefore make efficiently an augmented comparison while moving along the shelves (see Figure~\ref{fig:ASSET} and Figure~\ref{fig:app:main} in the Supplementary Information). This novel augmented reality experience is made possible via a low-power bluetooth beacon localization technology that has been deployed and extensively tested on retailers' site~\cite{Hormann2019}. Consumer localization in the shop does not require a centralized collection of GPS traces which tends to be unreliable, privacy-intrusive and not suitable for indoor environments~\cite{Tippenhauer2011}. 

\textbf{Scale and quality of product information on sustainability.} This third challenge is addressed by designing and populating an open knowledge-base. It is based on a new sustainability ontology with which a transparent and accountable reasoning for the personalized rating of products is performed. Ontologies have been influential and have often accelerated scientific progress in biology and beyond~\cite{Smith2008,Ashburner2000,Blake2004}. They systematically dissect and structure complex concepts to reason and develop a shared understanding that motivates their use in this work. The designed knowledge-base consists of information from retailers
, online data sources with domain and crowd-sourced knowledge
, domain experts, who supported this project during its lifetime
, as well as the wisdom of crowds~\cite{Fritz2019,Aminpour2020}, for example, by crowd-sourcing the analysis of Wikipedia data (see Table~\ref{tbl:impl:data-sources} in the Supplementary Information). The knowledge-base is actually a set of 795 associations between 15472 products and 25 consumer preferences belonging to the four sustainability categories of environment, health, social and quality.  See Table~\ref{tbl:preferences:env}-\ref{tbl:preferences:health} of the Supplementary Information for a complete list of all preferences and how they are obtained. Formally, the association between a product keyword (e.g. meat) and a preference keyword (e.g. vegetarian) is quantified in the range $[-1,1]$, with $-1$ representing opposition (meat does not fit vegetarian diet) and $1$ representing support (lettuce fits vegetarian diet). In-between values represent, for instance, the relative positive effect of different vitamins or the relative negative effect of different additives/preservatives in several health indicators. Therefore, the design of the knowledge-base is the selection of such keywords (tags) according to formal semantics as well as the reasoning about their in-between association score (196 product and 61 preference tags as shown in Figure~\ref{fig:cloud} of the Supplementary Information). This semantic information is used to calculate a non-personalized \emph{sustainability index} of a product for certain sustainability preferences by aggregating the association scores among the assigned product and preference tags. Moreover, rebound effects that model the incompatibility of Sustainable Development Goals~\cite{Hasegawa2019,Nerini2019,Spaiser2017} at the level of consumer choices can be  measured at the design phase by analyzing the semantic links that interconnect products, preferences and their tags.

\textbf{Field tests.} We conducted novel and ambitious field tests at two supermarkets from May to November 2018. These go beyond earlier survey questionnaires~\cite{Stockigt2019,Kolodinsky2018,Boswell2018}, lab experiments~\cite{Caniglia2017} or gamification~\cite{shan_what_2020}. We implemented the actual value-sensitive and preference-based decision-support system to empower consumers to shift their shopping choices to sustainable consumption. As a result of the actual system implementation, a more realistic assessment of the shopping shift can be made. A total number of 323 (Retailer A) and 69 (Retailer B) participants with a loyalty card were recruited by the retailers and project staff members on site by offering coupon discounts of up to 100 euros. The loyalty card provides access to baseline historical purchases allowing us to compare consumers using the app (treatment condition) with similar consumers not using the app (no treatment). The rewards are designed to incentivize a regular shopping behavior, limit dropouts and biases. Consumers using the app fill in an entry survey, choose their (baseline) preferences that can be changed later and start shopping with the app. For the purpose of this study, the app usage and purchases are traced but remain anonymized. After the end of the test period consumers collect their coupons by answering an exit survey.

\textbf{Outline of analysis.} The data collected during the field tests allows us to analyze whether consumers shift their shopping choices to other products that better respect their own values, self-determined via their sustainability preferences. It turns out that consumers are usually ready to pay more for higher quality products, which are healthier and more sustainable~\cite{Choi2011}. We dissect the complex link of sustainability preferences with price by analyzing whether such a link originates from consumer preferences or from biases in the knowledge-base we developed.  We further identify the profiles of consumers' preferences and illustrate to what degree they are adjusted to meet their expectations of how products should be rated.

\section*{Results} 

\begin{figure}[!htb]
\centering
\includegraphics[width=0.49\textwidth]{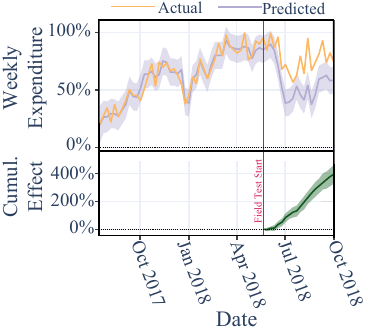}
\includegraphics[width=0.49\textwidth]{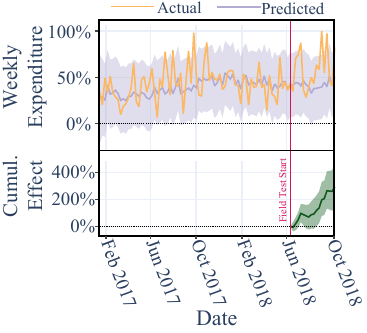}
\caption{\textbf{Shift to more sustainable consumption during the field test period}. Actual vs. predicted normalized weekly expenditures made for highly rated products ($>5$). (a) Retailer A and (b) Retailer B. We find a high statistical significance of $p<0.001$ for a significance level of $\alpha=0.01$.}\label{fig:behavior-change}
\end{figure}

We first demonstrate the shift of consumer behavior towards more sustainable consumption, meaning products that are rated higher than $5$. The sustainable shopping behavior is measured using the weekly expenditures made for such highly ranked products, normalized to $[0,1]$, using the observed maximum value. The expenditure values are deflated using the Harmonized Index of Consumer Prices~\cite{diewert2002harmonized}. The consumers that did not use the app (control group) were matched with consumers that did use the app. The following matching criteria (behavior covariates) were applied: (i) The deflated monthly total budget spent, (ii) the distribution of deflated monthly budget spent per product category and (iii) the sustainability index of the average purchased products per month. Matching reduces the control group from 3438 to 532 for Retailer A and from 1843 to 59 for Retailer B. The total consumers who participated in the field tests are filtered out to 148 (Retailer A) and 30 (Retailer B) consumers (treatment) by keeping the consumers who have seen the rating of the products before they have purchased it.  The weekly expenditures of the consumers using the app are predicted based on ``historic'' data in order to estimate the shopping behavior of the treatment group in case they did not use the shopping app. This is done via causal impact analysis~\cite{brodersen2015inferring} using spike-and-slab causal inference via structured Bayesian times series~\cite{senin2008dynamic}. The predicted values are compared to the actual weekly expenditures made when using the app (see Figure~\ref{fig:behavior-change}, where the prediction horizon covers the entire field test period). 

Our results confirm a statistically significant increase of expenditures for highly rated products: (i) The absolute effect is about 20\% for each retailer. (ii) The relative effect is 36.7\% for Retailer A and 41\% for Retailer B. The cumulative effect is depicted in order to show how this shift towards expenditures for more sustainable products unfolds over the period of the field test. Table~\ref{table:causal-impact} in the Supplementary Information summarizes the results of the causal impact analysis. 

A few additional observations can strengthen the conclusions of Figure~\ref{fig:behavior-change}: 68\% and 75\% of the total purchased products in Retailer A and Retailer B, respectively, are highly rated ($>5$) products that are seen in the app. Novel purchases of (i) products purchased for the first time and (ii) products whose rating is seen in the app also increase, namely, by 22\% for Retailer A and and 16\% for Retailer B. 

The survey results of Figure~\ref{fig:feedback:pos},~\ref{fig:feedback:awareness} and~\ref{fig:feedback:final} of the Supplementary Information furthermore confirm the following: (i) High product ratings reflect consumers' preferences. (ii) New products are discovered via the products rating. (iii) The rating of products appears justified to consumers. (iv) Consumers are more aware of sustainability aspects after using the app. 

Note that the two retailer shops have different weekly expenditure patterns. Retailer A exhibits stronger seasonality effects than Retailer B, which considerably relies on customers with a student profile. Customers of Retailer B seem to have a more diverse demographic profile and a stronger interest in sustainability.

\subsubsection*{Relation between sustainability and price} 

During the field test, 16\% of the field test consumers at both retailers pay at least 10 euro-cents more on average for each novel product by supporting at least one of the three environmental preferences (see Table~\ref{tbl:preferences:env} in Supplementary Information).  This number is significant given that the price level of most products is low.  Moreover, the percentage of consumers who regularly interact with the smart phone app (the ones in the center of our causal impact analysis increases over time to 39\% for Retailer A and 30\% for Retailer B.

We next assess whether the purchased products rated high within each product category are also the more expensive ones. First, both product deflated prices and the product ratings are normalized (see Section~\ref{subsec:prices:eval} of Supplementary Information for details). Figure~\ref{fig:price-rating} shows the frequency of the rated products that were viewed and purchased against price and rating. Retailer A clearly has a higher number of products with high prices and ratings as reflected by the high values on the top right of the heatmap. Similarly in Retailer B, for each normalized price value per product unit, a higher number of products are viewed and purchased that have higher rating values. We further measure the non-linear relation between product price and product ratings with the Spearman's correlation coefficient across product categories. We observe a positive correlation coefficient of $\rho=0.14$ and $\rho=0.19$ for Retailer A ($p=0.01$) and Retailer B  ($p=0.02$) respectively. This suggests that highly rated products may come with slightly higher prices. 

\begin{figure}[!htb]
\centering
\includegraphics[width=0.49\textwidth]{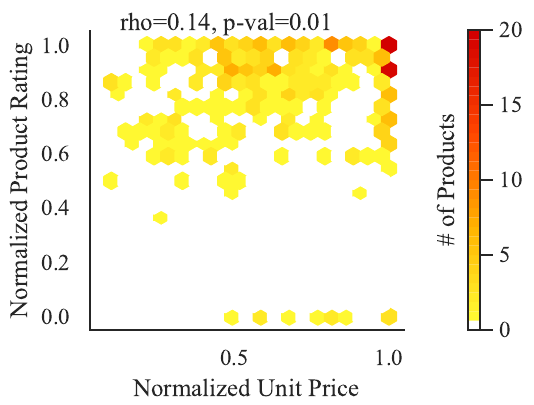}
\begin{picture}(0,0)
	\put(-213,155){\colorbox{white}{\small a)}}
\end{picture}
\includegraphics[width=0.49\textwidth]{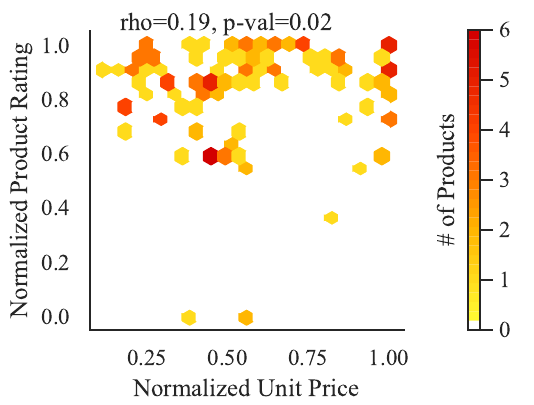}
\begin{picture}(0,0)
	\put(5,168){\colorbox{white}{\small b)}}
\end{picture}
\caption{\textbf{Highly rated products correlate to some extent with increased prices.} (a) Retailer A and (b) Retailer B. Normalized product unit price vs. normalized product rating for the frequency of purchased products, whose rating is seen in the app. Frequency values are derived at the level of product categories. Note the increased frequency of purchased products with higher price as the product rating increases. This is particularly visible for Retailer A with a higher number of data records.}\label{fig:price-rating}
\end{figure}

The question that arises here is whether this positive correlation stems from the consumers' personalization, i.e. their choices regarding sustainability preferences, or the actual knowledge-base and ontology design that is universal for all consumers. To answer this question we repeat these measurements, but instead of the normalized product ratings we used the min-max normalized sustainability index for each preference. The Spearman's correlation coefficient between the sustainability index of products and product prices per category is calculated for each preference across product categories. Out of 50 correlation values evaluated at each retailer, only one correlation value is $>0.5$ and statistically significant. This confirms that no price biases are observed in the designed ontology in favor of more expensive products. Nevertheless, it is possible that certain preferences result in a sustainability index that is positively or negatively correlated with price due to the nature of this preference. For instance vegetarian diet products may come with lower prices only because they do not contain meat that is usually expensive. See Tables~\ref{tbl:price:corr:coop:norm},~\ref{tbl:price:corr:wm:norm} and~Figure~\ref{fig:price:radial} in the Supplementary Information for an evaluation of all preferences.

\subsubsection*{Sustainability preference profiles} 

To understand how different consumers influence the product rating via personalization, we extract preference profiles by clustering consumers based on the preference choices they made during the field tests. We distinguish between \emph{initial preferences} and \emph{preference adjustments}. The initial preference choices are the very first ones made before consumers are exposed to the product rating functionality. They reflect the intrinsic intentions of consumers towards the sustainability aspects. The preference adjustments are made throughout the field test after the initial preferences were set. 

\begin{figure}[!htb]
\centering
\includegraphics[width=1.0\textwidth]{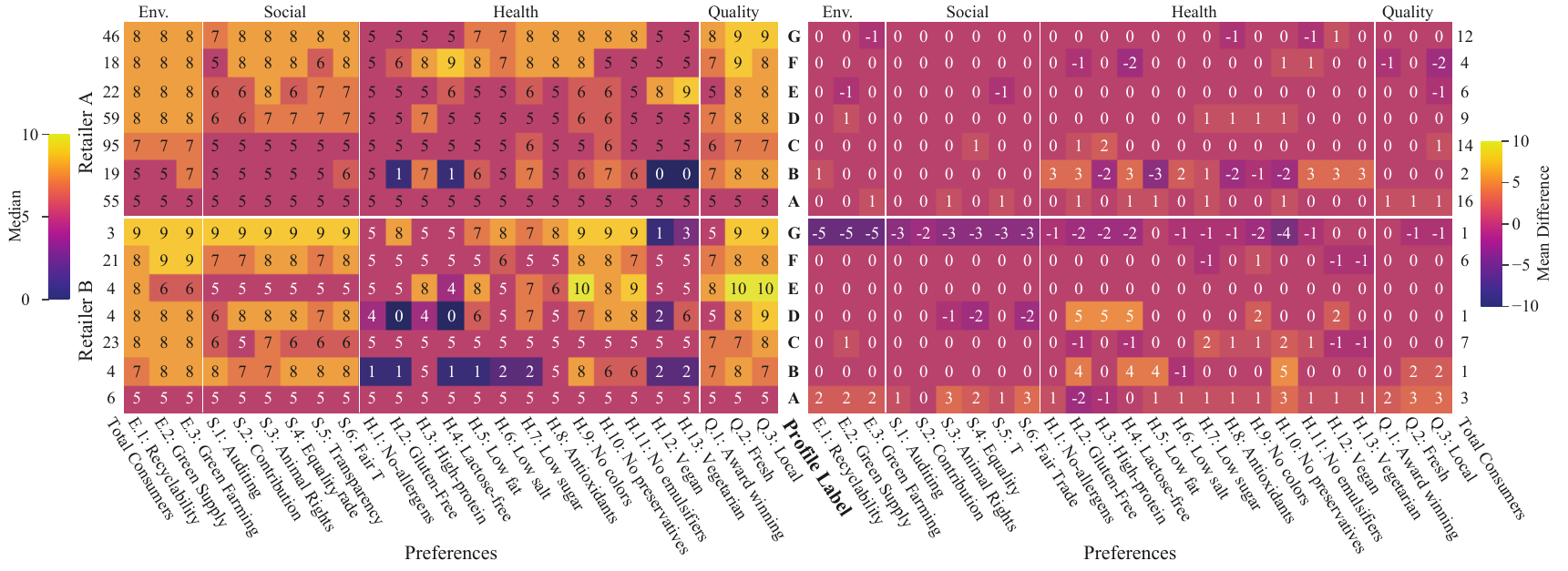}
\begin{picture}(0,0)
	\put(-210,170){\colorbox{white}{\footnotesize a)}}
	\put(-5,170){\colorbox{white}{\footnotesize b)}}
\end{picture}
\caption{\textbf{Preference score profiles over the four sustainability aspects for Retailer A (top) and Retailer B  (bottom)}. Capital letters are labels relating to the extracted preference profiles. They are the same in both figures. The numbers on the left and on the right of each figure refer to the number of consumers in each profile. (a) Initial preference profiles demonstrating the high preference scores in environment and quality aspects. (b) Minimal targeted adjustments made during the field test confirming that product ratings match the expectations expressed via the sustainability preferences. }\label{fig:consumer-profiles-coop}
\end{figure}

\textbf{Initial preferences.} The profiles of the initial preferences are extracted by clustering consumers using the Ward hierarchical clustering approach~\cite{johnson1967hierarchical}. Seven clusters are calculated for each retailer as shown in Figure~\ref{fig:consumer-profiles-coop}a. Note that most of the profiles consist of choices with a significantly high score in ``environment'' and ``(product) quality" as well as a high score in the category ``social''. The scores for ``health'' shows a significant variation among the different profiles that reflects the diversity of dietary needs: (i) Two neutral profiles with scores around 5 are observed for both retailers. (ii) There is a profile that penalizes strict preferences such as vegan, gluten/lactose-free, etc, with low score. (iii) The majority of the profiles retain a high score for health indicators such as no preservatives/colors/emulsifiers, etc. The majority of consumers avoids radical scores in health and expresses their sustainability profile via preferences regarding environment and quality aspects. 

\textbf{Preference adjustments.} We assume that consumers adjust their preferences if they are not satisfied with how products are rated or if they aspire to explore new products. Consumers may also try to match the rated products with the ones they use to buy. We measure the mean difference between the initial preferences and all adjusted preferences per profile to understand how different consumer groups adjust their preferences to meet their expectations. Figure~\ref{fig:consumer-profiles-coop}b shows the profile adjustments. We observe the following: (i) The majority of preference scores is on average unaffected, indicating that consumers usually do not need to adjust their preferences and they preserve their initial intrinsic sustainability preferences. This also validates the ontology from the consumers' perspective. (ii) Consumers with initially neutral preferences increase their preference scores, especially in Retailer B. (iii) Consumers who initially maximize the preference scores slightly roll back to more moderate preference values, especially at Retailer B. Apparently this rollback is related with a consumers' openness towards experiencing radically different highly rated products or with the ontology design and rebound effects that results in buying unexpected products. Similarly, the group of consumers that initially penalizes health preferences with low scores makes adjustments to higher scores.

\section*{Discussion and Outlook}

The findings of this work have several implications for consumers, retailers and producers. They also have an impact on science, policy and at an institutional level. We outline here the challenges and the new pathways that this work opens up. 

\textbf{Consumers.} We show that a value-sensitive, preference-based design empowers more informed, transparent and accountable shopping choices tailored to personal sustainability values. As a result, consumers explore and finally buy new products. Compared to related work limited to health or environmental indicators for sustainability, we study a broader, yet finite set of such indicators (preferences). In the future, consumers themselves may expand or even limit this set according to their intrinsic priorities. Augmented shopping processes with smart phones may still disrupt regular shopping routines for some consumers. It is unclear to what extent and scale consumers can adopt such technologies at this moment. However, this was also the case with online shopping a few years ago and nowadays efforts such as Amazon Go~\cite{Polacco2018} showcase the feasibility of such technologies in the market. The applicability of the value-sensitive shopping assistant in the context of online shopping is promising and may simplify its use by consumers. In terms of consumers' trust on products rating, new models are required to preserve a satisfactory level of explainability in the long run. This is especially the case when such models need to capture an expanding spectrum of product characteristics as well as the dependencies and incompatibility of sustainable development goals~\cite{Spaiser2017}. The transparency and accountability of the knowledge-base can be further enhanced with blockchain solutions as a means to securely verify sustainable production practices as well as the traceability of supply chains~\cite{Howson2019,Gammon2018}. Recent studies also confirm how indispensable the active consumers' involvement is to evolve a viable and accurate knowledge-base on sustainability via citizen science initiatives~\cite{sachs_six_2019,Fritz2019}. 

\textbf{Retailers.} Retailers can pioneer future consumption patterns by offering products that are ethically more aligned to consumers' sustainability preferences. They can focus on selling a higher number of highly rated products that might come with slightly higher prices (or profit margins), while they serve consumers' values on sustainability. 

\textbf{Producers.} Value-sensitive design provides new opportunities for producers to offer more attractive, higher quality products to retail markets, by capturing (i) consumers' sustainability preferences and (ii) reducing the sustainability gap of existing products. Moreover, producers have new opportunities to sell higher quality products at a higher price. This incentivizes the improvement of production practices, considering consumer preferences regarding environment, health or worker rights. In other words, thanks to welcomed recommendations of higher-quality products, more sustainable production can pay off.  

\textbf{Science.} By demonstrating the shift of individuals to sustainable consumption, we also open up new opportunities for social coordination: moving from individual sustainability preferences to collective actions, which meets sustainable development goals in a bottom-up way. Scientific methods from the areas of game theory, mechanism design, incentive design, socio-technical optimization and learning~\cite{Becker2017,Mason2012,Mann2017,Pournaras2018} are becoming applicable to further support sustainability movements in society. Methods such as the ones of this work and a transdisciplinary scientific approach are required to improve and accelerate the development of sustainability knowledge-bases, using responsible AI and the wisdom of crowds~\cite{Bergmair2018,Kittur2019}.

\textbf{Policy-making and institutions.} We foresee new opportunities for environmental non-profit organizations to educate and interact with the general public by publishing their own personalization templates for consumers, for instance, adopting the WWF or Greenpeace priorities for sustainability. Such customizable personalization is supported by the built system. Legislation could protect the open and public good character of a knowledge-base for sustainability. Such a knowledge-base should not be used as an unquestionable universal ground truth for sustainability. Instead, we envision the active, global involvement of various stakeholders such as public organizations, scientists and especially citizens to capture and develop the evolving nature of sustainability values. Promising related initiatives include the European Commission Environmental Footprint initiative or the Global Product Classification standardizing institution GS1. In other fields such as biology, initiatives for ontologies and knowledge-bases have had significant impact, for instance the Gene Ontology Consortium~\cite{Ashburner2000}. Overall, the  EU project ASSET~\cite{ASSET}, in the context of which this work has been carried out, is a promising blueprint showing how sustainable consumption might be scaled up in a participatory way.

\section*{Methods} 

Here we illustrate more details of the field tests and the causal impact analysis as well as the localization solution at retailer shops. We also outline the ontology design, the product rating methodology, the preference profiles and preference interactions. 

\subsubsection*{Field tests and causal impact analysis}

\textbf{Recruitment.} Each shopping visit at retailer shops is rewarded with 10 euros. By analyzing the history of purchases, 10 visits are expected on average during the field tests and therefore the total maximum amount of 100 euros incentivizes and rewards regular shopping behavior. The study has received ethical approval by the ethical committee of ETH Zurich. To minimize biases originated by different human explanations of how to use the smart phone app (see Figure~\ref{fig:app:usage} in Supplementary Information), an integrated tutorial requires completion before consumers start using the smart phone app. A FAQ based on questions that arose during living lab experiments before the field tests was also included. 

\textbf{Causal impact analysis} The Dynamic Time Warping (DTW)~\cite{senin2008dynamic} algorithm is used to compare the time series data of consumers that used the smart phone app (treatment consumer) against all consumers  that did not use the smart phone app (non-treatment consumers) with purchase records up to 13 months before the field tests (excluding the last month as a buffer zone). Each comparison results in the similarity ranking of consumers, who did not use the smart phone app, from which the top-$k$ nearest neighbors are selected to form the group for assessing sustainable consumption. Since different values of $k$ yield different groups for comparison, the $k$ is selected via a dynamic time warping of weekly expenditure before the treatment as shown in Table~\ref{tbl:distance:explanations} of the Supplementary Information.  

The temporal distance between consumers is measured over the three matching criteria (behavior covariates). More specifically, the Euclidean distance with dynamic time warping is used to determine the top-k nearest neighbors for the deflated monthly total budget spent and the sustainability index of the average purchased products per month. In contrast, the Jensen-Shannon distance (divergence)~\cite{serjantov2002towards} with dynamic time warping is used for the distribution of deflated monthly budget spent per product category as such distance metric is known for measuring the similarity of distributions. For each consumer that used the smart phone app (treatment consumer), we rank all other consumers that did not use the app (non-treatment consumers) according to the three distance metrics. Since there are three distance metrics and we cannot determine which one has higher influence, the average ranking of the consumers that did not use the app is calculated over all combinations of distance metrics.

\subsubsection*{Localization of products} 

A privacy-preserving indoor localization system is designed for the two retailer shops~\cite{Hormann2019}. It derives the list of products that are in close proximity to consumers. It relies on Bluetooth low energy beacons installed on the shop floor of both retailers. The beacons are aligned with the shop map to provide absolute anchors to the shop coordinates. By analyzing the signal strength of the beacons using triangulation and a map logic on feasible positions and consumer movements,  the position of the consumers in the shop is estimated, see Figure~\ref{fig:ASSET}a. This consumer position is related to the coordinates of product groups in close proximity. The rating of these products is calculated and presented to consumers' smart phone as shown in Figure~\ref{fig:ASSET}c. The localization system is privacy-preserving as all calculations are performed at the consumer's smart phone. Moreover, beacons do not obtain any information from consumers and therefore retailers cannot derive consumer movements from this localization system. The deployed technology uses off-the-shelf hardware which yields low hardware costs. The obtained precision has an inaccuracy of about 2 meters.  The granularity of product group positions ranged from several meters down to zero accounting the intrinsic ambiguity of product groups positions on top of each other.

\subsubsection*{Ontology design} 

\textbf{Primitive concepts, tags and rules.} The ontology is designed to quantify the association between products and preferences, e.g. to what extent a certain product is for a vegetarian diet, fair trade etc. To measure such associations, we introduce a common alphabet of characteristics for products and preferences. This alphabet is a set of \emph{keywords (tags)} that represent \emph{primitive concepts}. They form the semantic space and scope of sustainability. Subsets of these primitive concepts compose a vocabulary of product and preference tags, while a primitive concept is not further decomposed to keep the ontology practical and consistent~\cite{Alterovitz2010} within its scope, i.e. a primitive concept is regarded disjoint within a chosen scope of sustainability. For example, the preference tag "vegan" can be composed by the two primitive concepts ``production with no animals" and ``production with no animal products". Moreover, products and preference tags are assigned to products and preferences respectively based on logical rules. For instance, to assign the product tag ``low fat" to a product, a logical rule could determine the number of grams in fat, relative to the total product weight, contained in the product and/or whether the product has a low-fat label. Over 600 such rules are created for this purpose using the Drools framework~\cite{Drools}. We focused on food products. They are the ones that populate the knowledge-base and connect to product tags. In summary, the design of the ontology consists of (i) the choice of the primitive concepts, (ii) their assembly to product and preference tags and (iii) the creation of logical rules to connect the tags with products and preferences. These actions required domain knowledge from experts (WWF, Greenpeace, Ethical Consumer, VKI), reliable data sources (e.g. EU reports) as well as the wisdom of the crowd by running the Social Impact Data Hack~\cite{SIDH2017}  to mine and structure information from Wikipedia, for instance, branding information (see Section~\ref{sec:crowd_sourced_product_data} in the Supplementary Information). 

\textbf{Association scores.} The association between a product and a preference tag is measured by their shared primitive concepts that satisfy a preference tag. We distinguish between positive and negative associations by determining for each pair of product and preference tags subsets of primitive concepts that semantically support or oppose the preference tag,  i.e. the preference tag ``vegan" supports the primitive concept ``no animals involved in production" but opposes the concept "animal product". Therefore, the association score comes with positive and negative values in the range $[-1,1]$ by summing up the associations between supported and opposed primitive concepts (see Section~\ref{sec:rating} of the Supplementary Information for more details). 

\begin{figure}[!htb]
\centering
\includegraphics[width=1.0\textwidth]{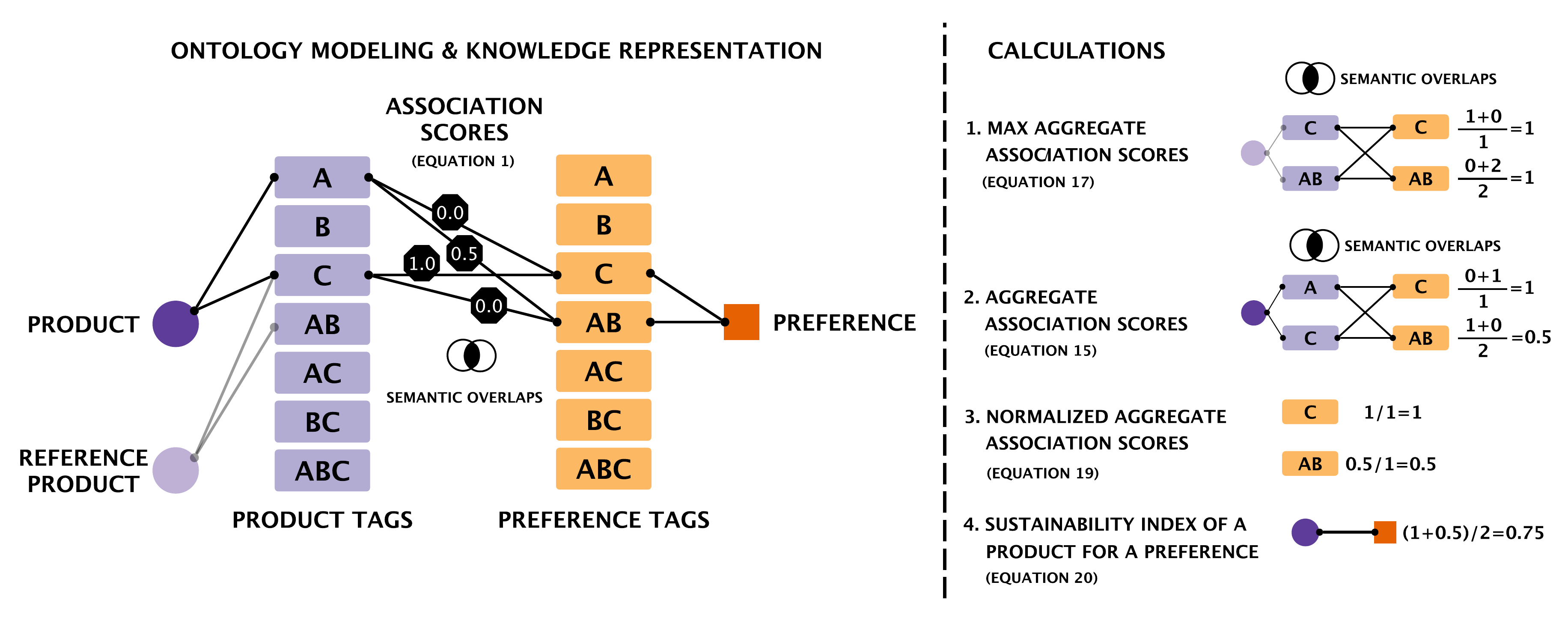}
\caption{\textbf{Calculation of the sustainability index using the ontology of products and preferences}. Assume an alphabet of primitive sustainability concepts represented in this simple example by $\{A,B,C\}$. Combining the primitive concepts results is the word vocabulary $\{A,B,C,AB,AC,BC,ABC\}$ of product and preference tags. Using rules based on experts' knowledge and verified data such as ingredients of products, the product tags $A$ and $C$ are assigned to a product. Similarly, a sustainability preference is designed by a composition of the two preference tags $C$ and $AB$. We can now calculate the association scores between the product and preference tags in an automated way (without expert knowledge) by taking the intersection $\cap$ of the tags sets as follows: $|\{A\} \cap \{C\}|/|\{C\}|=0$, $|\{C\}\cap \{C\}|/|\{C\}|=1$, $|\{A\} \cap \{AB\}|/|\{AB\}|=0.5$, $|\{C\} \cap \{AB\}|/|\{AB\}|=0$. The sustainability index of a product for a given preference is calculated by the average normalized aggregate association scores of the assigned preference tags as demonstrated in this numerical example. The equations refer to the Supplementary Information.}\label{fig:sustainability-index}
\end{figure}

\textbf{Reduction design principle.} The construction of product and preference tags should adhere to the \emph{reduction design principle}: (i) Between two tags with the same primitive concepts, one and only one should be assigned to a product or preference. (ii) When two tags assigned to a product or preference share primitive concepts, these primitive concepts should be removed and form a new tag. In the example of Figure~\ref{fig:sustainability-index}, the reduction design principle is violated if the product tag $AC$ is assigned to the product, or, the preference tag $BC$ is assigned to the preference. We prove in Section~\ref{sec:ontology} of the Supplementary Information how this principle minimizes the error of overlapping tags when the association scores are aggregated to calculate the rating of a product. Violations of the reduction design principle may result in excessive influence of certain preferences on the product rating. In practice, these artifacts may be captured by consumers, whose adjustments of preferences provide additional countermeasures against the error of semantically overlapping tags. 

\textbf{Experts' guideline.} We propose a high-level guideline to populate the sustainability knowledge-base according to the proposed ontology. This guideline can be used by domain experts to guide the construction process and is outlined as follows:

\begin{enumerate}
\item Identify relevant primitive concepts based on (i) the product categories (i) the available product data and (iii) the scope of sustainability preferences (goals).  
\item Create product and preference tags using the primitive concepts such that these tags represent how product/preference characteristics oppose or support a product/preference.
\item Create rules that connect the product tags with products, and the preference tags with preferences.
\item Apply the reduction design principle between all combinations of product tags and preference tags that have overlapping concepts. 
\item Calculate association scores between product-preference tag pairs in the range $[-1,1]$.
\item If necessary, go back to Step 1, add or remove primitive concepts and repeat the process. 

\end{enumerate}

In Step 1, experts determine the scope of the sustainability by defining primitive concepts that capture key characteristics of products and preferences. Experts need to be aware of the products for which they design the ontology, the available information they have about these products as well as the sustainability preferences that should be captured. In Step 2 they can start combining these concepts into tags with the purpose of representing support or opposition to product characteristics and preferences. In Step 3, experts can assign  these tags to product and preferences and can formalize rules under which these assignments are made. The processes of the first three steps are the most tedious ones and requires knowledge, experience and a good overview of the available information. As an example of facilitating such processes, we performed workshops with several stakeholders during the project lifetime and organized the Social Impact Data Hack~\cite{SIDH2017}. Step 4 applies the reduction design principle to improve the consistency of the ontology. Step 5 performs the calculations of the association scores based on the (automated) calculations illustrated in Figure~\ref{fig:sustainability-index}. In practice, these calculations are often calibrated by experts to reason about the association scores based on ground truth knowledge. For instance, consider a study that shows evidence about the effect of different preservatives on health. Obviously, the cause of such effects may be related to chemical or biological phenomena at a very low granularity level that is not captured within the scope of the designed sustainability ontology. In this case, association scores measuring can be calibrated to reflect the relative effect of preservatives according to the findings of such a study. Finally, the process can repeat by adding or removing primitive concepts. The motivation for this iterative process is to better capture the whole range of preferences, decompose further primitive concepts to make the ontology more granular, add/remove rules, expand product categories or enrich the knowledge-based with new datasets.  During the ontology design, we performed over 10 iterations for validation purposes and the quality criterion for convergence was how well the product rating could be justified to consumers during the preliminary living lab tests.

\subsubsection*{Product rating: Sustainability index and preferences} 

\textbf{Sustainability index.} It quantifies the support or opposition of a preference by a set of product characteristics found in a product. This support or opposition is compared to a product, existing or hypothetical (`reference product' in Figure~\ref{fig:sustainability-index}), that has all possible characteristics that can support or oppose respectively a preference. Figure~\ref{fig:sustainability-index} illustrates the involved calculations. The sustainability index between a product and a preference is measured using the normalized association scores aggregated over the connected product and preference tags. The normalized aggregated association score of a product-preference pair is the normalized aggregated association score of a product averaged over all preference tags assigned to the preference (Calculation 4 in Figure~\ref{fig:sustainability-index}). Each normalized aggregated association score between a product and a preference tag (Calculation 3 in Figure~\ref{fig:sustainability-index}) is calculated by the aggregated association scores of the product tags assigned to this product (Calculation 2 in Figure~\ref{fig:sustainability-index}) divided by the maximum association score between the reference product and the preference tags of the preference (Calculation 1 in Figure~\ref{fig:sustainability-index}). 

\textbf{Insights on sustainability of production.} Note that by calculating for each preference the density of the sustainability index over all products, new opportunities arise to reason about the following: (i) The sustainability profile of different retailers. (ii) New ways (preferences) with which producers can improve products with a more sustainable profile. (iii) Market gaps where new business ecosystems can evolve with stronger involvement of producers and consumers to accelerate sustainable consumption. For instance, the densities in Figure~\ref{fig:si:distro} of the Supplementary Information confirms the more sustainable profile of Retailer B products across the preferences, e.g. higher sustainability index for animal rights, fair trade, recyclability and green farming. Improvements can be made by either introducing new products with better sustainability footprint over these preferences or by improving the existing production practices of the available products. 

\textbf{Product rating.} Note that the sustainability index does not require any personal information for its calculation. It only relies on the information of the sustainability ontology, i.e. primitive concepts and tags, that we make available as public-good knowledge. As such, it can be calculated in public computational infrastructure, i.e. servers, public clouds, etc. In contrast, the calculation of the product rating requires personalization with consumers' preference choices that remain by design locally on their smart phones to protect privacy and limit manipulative nudging. As a result, the product rating is calculated on consumers' smart phones using the sustainability index values retrieved remotely using a distributed protocol of message passing between smart phones and a project server. The calculation is performed on-demand by consumers when they navigate in the retailer shop and request the rating of the products that are in their close proximity. For each product, the rating algorithm multiplies the sustainability index with the degree of opposition or support of each preference, measured by the distance (offset) from the median preference score ($5$, remaining neutral). These calculations are summed up and divided by the sum of all distances from the median preference scores. Section~\ref{sec:rating} in the Supplementary Information outlines in more detail the product rating calculation and its computational complexity. The (unscaled) product rating calculation is summarized as follows:

\begin{equation}
\text{product rating}=\frac{\text{sum of all (sustainability indices $*$ preference offsets)}}{\text{sum of absolute preference offsets}},
\end{equation}\label{eq:product-rating}

\noindent where the product rating values can be scaled to match different grading systems of different countries ($[0,10]$ in the field tests as supported in the Section~\ref{sec:rating} of the Supplementary Information).

\textbf{Explainability} Two levels of rating explainability are provided to consumers: (i) \emph{product tags} and (ii) \emph{preferences}. Consumers can learn about how each product characteristic influences the rating value by solving Equation~\ref{eq:contr:prftag} of Supplementary Information for a certain product tag, given that all other variables are known. Similarly, consumers can know how each offset of their preferences contributes to the product rating by solving Equation~\ref{eq:contr:preference} of the Supplementary Information for a certain preference offset. 

\textbf{Preferences selection.} The selection of preferences was made on the basis of providing a broad spectrum of different sustainability indicators with which consumers can express their preferences. However, this spectrum is not too broad to the extent of creating a cognitive overload for consumers and lack of comprehension about which preferences influence the rating of products and why. This is critical for the effectiveness of the rating explainability. Moreover, a lower number of preferences decreases the computational cost of the rating algorithm and improves the usability of the smart phone app (see Section~\ref{subsec:algorithm-complexity} of the Supplementary Information). This balance is a result of the following process: (i) Participation of several stakeholders in the ASSET project meetings and workshops providing insights about how grocery product choices influence different sustainability criteria. (ii) Preliminary living lab experiments and smaller-scale field tests for feedback acquisition on the preferences. (iii) Choice based on available data, i.e. product and preference tags. Preferences with very similar preference tags or very few preference tags are removed or merged. The set of the final preferences shows a balance between several individual criteria on health (13) versus collective criteria in environment, social and quality aspects (12). The consumers' feedback during the preliminary tests also determined the \emph{strict preferences} of gluten-free, lactose-free, vegan and vegetarian products. Fully supporting such preferences results in excluding products that do not fully satisfy them even though they may satisfy other (non-strict) consumer preferences. In other words, strict preferences cancel association scores with other preferences. 

\subsubsection*{Preference profiles and interactions} 

\textbf{Preference profiles.} The Ward hierarchical clustering approach~\cite{johnson1967hierarchical}, with which the preference profiles are extracted, is selected based on its interpretable clusters and high scoring in the following evaluation metrics~\cite{Rendon2011}: silhouette, Dunn index, Davies-Bouldin and Calinski-Harabaz. It uses Euclidean distance given that preference scores are continuous, not sparse and different preference scores are comparable, with the exception of strict preferences that may introduce biases. No normalization or standardization is performed given the same scale among preferences. 

\textbf{Interactions of preferences} The sustainability preferences show correlations that originate from the (i) anticipated association between different product/preference tags (ontology-level) and (ii) the actual linking of products to product tags (product-level). This information at the ontology and product-level respectively can be used to reason about potential rebound effects~\cite{Spaiser2017}: undermining a sustainability preference by satisfying another one. 

\begin{figure}[!htb]
\centering
\includegraphics[width=1.0\textwidth]{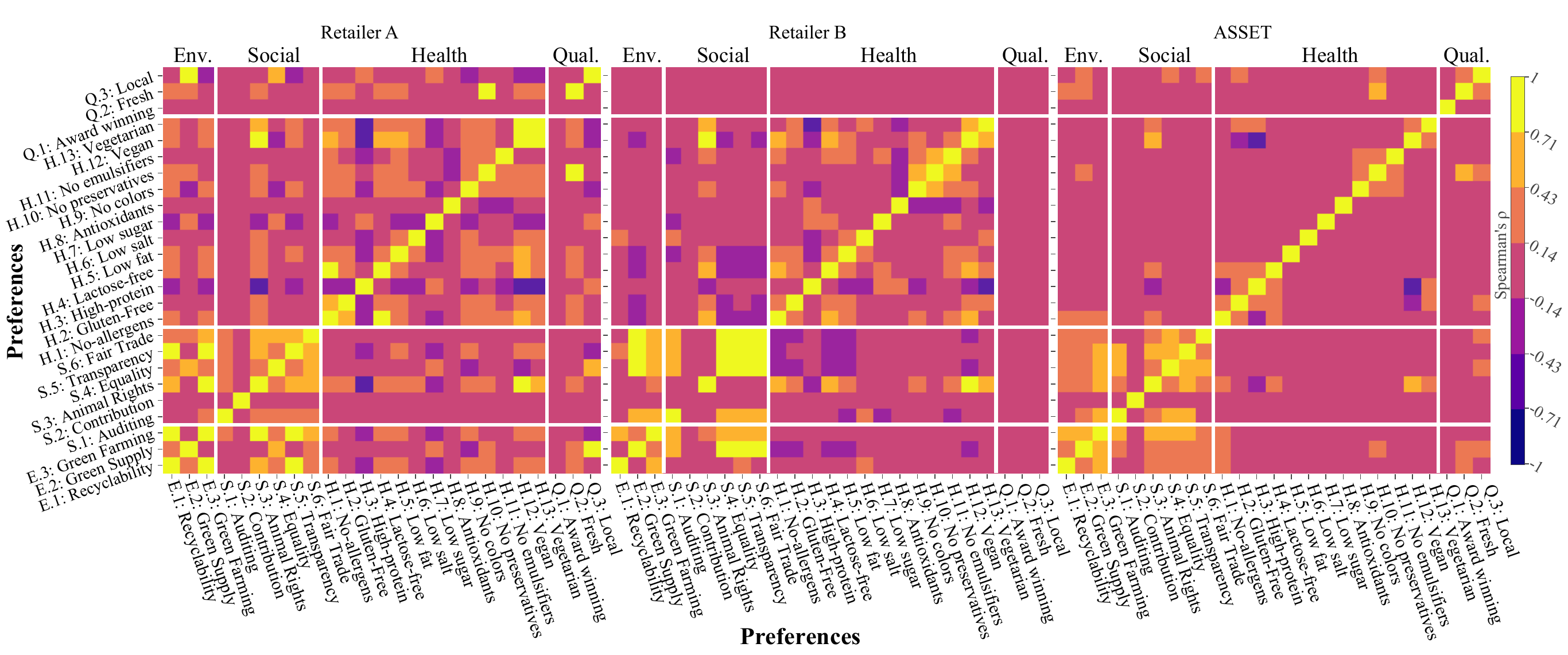}
\caption{\textbf{Interaction effects between different sustainability preferences}. Ontology-level: Correlation of association scores for (a) Retailer A and (b) Retailer B. Product-level: (c) Correlation of sustainability indices for a common set of products between Retailer A and Retailer B . Positively correlated preference pairs across the x and y axes indicate dependence whereas negatively correlated preference pairs indicate potential rebound effects.}\label{fig:rebound-effects}
\end{figure}

\begin{enumerate}
\item \textbf{Ontology-level}: The interaction of two preferences is measured as follows: (i) Find the shared product tags between the two preferences. (ii) For each of the two preferences and shared product tag, calculate the average association score among the preference tags of the preference and the shared product tag.  (iii) Calculate the Spearman correlation coefficient of the average association scores among the two preferences. 
\item \textbf{Product-level}: The interaction of two preference at the product-level is measured as follows: (i) Calculate the sustainability index for each product-preference pair. (ii) Calculate the Spearman correlation coefficient of the sustainability indices among the two preferences. 
\end{enumerate}

These Spearman correlations are shown in Figure~\ref{fig:rebound-effects}. Potential rebound effects are observed between health/quality preferences vs. environmental ones, while a highly rated product may be caused by multiple environmental preferences that share several common product tags.

\subsection*{Acknowledgments}

This work was supported by the European Commission H2020 Program under the scheme "ICT-10-2015 RIA" grant agreement \#688364 ASSET - ``Instant Gratification for Collective Awareness and Sustainable Consumerism'' (\url{https://www.asset-consumerism.eu}) and by the LCM – K2 Center within the framework of the Austrian COMET-K2 program. We would like to thank the two retailers Coop and Winkler Markt for supporting the project and field tests as well as all consortium partners for their contributions (alphabetically): AINIA, BIA, Fastline and VKI. The authors are grateful to the participants of the Social Impact Data Hack~\cite{SIDH2017} and in particular Maryna Pashchynska for her contribution to developing the sustainability ontology further. Several organizations supported this project in various ways such as WWF, Ethical Consumer, Greenpeace and Global2000. Finally the authors would like to heartily thank in alphabetical order George Lekakos, Marloes Maathuis, Heinrich Nax, Michael Siegrist and Stefan Wehrli for their invaluable suggestions and feedback. 

\subsection*{Open Data and Code}

The ontology data is available here: https://doi.org/10.6084/m9.figshare.8313257

\noindent The product rating algorithm is available here: https://github.com/asikist/value-sensitive-design-code

\bibliography{paper_no_url}
%
\bibliographystyle{naturemag}
%
%

\end{document}


\title{\textbf{Supplementary Information:} \\ How~Value-Sensitive~Design Can Empower Sustainable Consumption}
\author[1]{Thomas Asikis}
\author[2]{Johannes Klinglmayr}
\author[1]{Dirk Helbing}
\author[3]{Evangelos Pournaras}
\affil[1]{Professorship of Computational Social Science, ETH Zurich, Zurich, Switzerland, E-mail: \{asikist,d.helbing\}@ethz.ch}
\affil[2]{Linz Center of Mechatronics GmbH, Linz, Austria, E-mail: johannes.klinglmayr@lcm.at}
\affil[3]{School of Computing, University of Leeds, Leeds, UK, E-mail: e.pournaras@leeds.ac.uk}

\renewcommand\Authands{ and }

\maketitle

\renewcommand{\arraystretch}{1.2}

\vspace*{1cm}

\section*{\textbf{Supplementary Figures}}
	 \counterwithin{figure}{section}
   \renewcommand{\thefigure}{S.\arabic{figure}}
   \setcounter{figure}{0}

		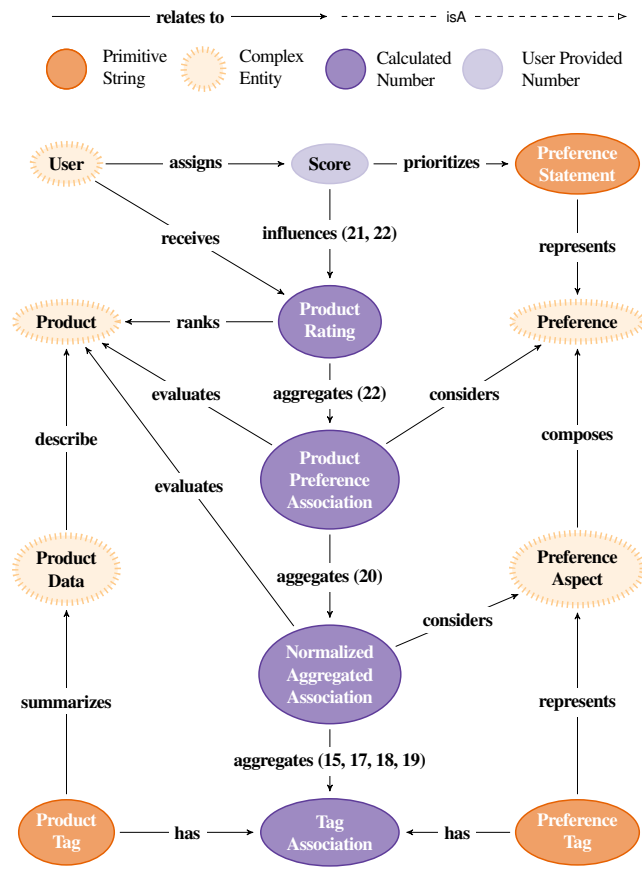
\begin{figure}[h!]
				\centering{
				\clearpage
\scalebox{0.7}{
\begin{tikzpicture}[->,>=stealth',shorten >=1pt,auto
]
    \path (0,14.5) node[literal node]      (qw)    {}
            +(1.3,0)  node[align=left]  (oolooo) {Primitive\\String}
            ++(2.6,0) node[entity node]       (qwq) {}
            ++(1.3,0) node[align=left] (ooloo) {Complex\\Entity}
            ++(1.4,0) node[calculated node]   (qwwq) {}
            ++(1.3,0) node[align=left] (olo) {Calculated\\Number}
            ++(1.3,0) node[scalar node]       (wqqw)  {}
            +(1.7,0) node[align=left] (ollo) {User Provided\\Number}
        ;
        \path (0,14.5) node[literal node]      (qw)    {}
           ++(-0.4,1)  node[]  (o) {}
            +(5.5,0)  node[]  (oo) {}
           ++(5.5,0)  node[]  (ooo) {}
            +(5.7,0)  node[]  (oooo) {}
       ;

        \path[every node/.style={font=\sffamily\small}]
        (o) edge node [edgestyle] {relates to} (oo);

        \path[every node/.style={font=\sffamily\footnotesize}]
        (ooo)      edge[isa]   node[isalabel]  {isA}   (oooo);

    \path (0,0) node[literal node] (PdT)   {Product \\ Tag}
        +(5,0) node[calculated node] (AcSc) {Tag \\ Association}
        ++(0,5) node[entity node]      (PdD) {Product \\ Data}
        ++(0,4.7) node[entity node]      (Pd)  {Product}
        +(5,0)  node[calculated node]  (PdR) {Product \\ Rating}
        +(5,-3)  node[calculated node]  (PdPfAc) {Product \\ Preference \\ Association}
        ++(0,3) node[entity node]      (U)   {User}
        ++(5,0) node[scalar node]      (Sc)  {Score};
    \path (9.7,0)     node[literal node]      (PfT)         {Preference \\ Tag}
          ++(0,5)     node[entity node]      (PfAs)        {Preference \\ Aspect}
          +(-4.7,-2)     node[calculated node]  (NAgAc) {Normalized \\ Aggregated \\ Association}
          ++(0,4.7)     node[entity node]      (Pf)          {Preference}
          ++(0,3)    node[literal node]       (PfSt)       {Preference \\ Statement};
  \path[every node/.style={font=\sffamily\small}]
    (PfT) edge node [edgestyle] {represents} (PfAs)
          edge node[edgestyle] {has}         (AcSc)
    (PfAs) edge node [edgestyle] {composes} (Pf)
    (PfSt) edge node [edgestyle] {represents} (Pf)
    (PdT)  edge node[edgestyle]  {summarizes} (PdD)
           edge node[edgestyle] {has}         (AcSc)
    (PdD)  edge node[edgestyle]  {describe}   (Pd)
    (U) edge node[edgestyle] {assigns} (Sc)
        edge node [edgestyle] {receives} (PdR)
    (PdR) edge node[edgestyle] {ranks}  (Pd)
          edge node[edgestyle] {aggregates (\ref{eq:rating:raw})} (PdPfAc)
    (Sc) edge node[edgestyle] {prioritizes} (PfSt)
    (PdPfAc) edge node[edgestyle] {evaluates} (Pd)
             edge node[edgestyle] {considers} (Pf)
             edge node[edgestyle] {aggegates (\ref{eq:connection})}  (NAgAc)
    (NAgAc) edge node[edgestyle] {evaluates}    (Pd)
            edge node[edgestyle] {considers}    (PfAs)
            edge node[edgestyle] {aggregates (\ref{eq:additive}, \ref{eq:relevance:max}, \ref{eq:relevance:min},
    \ref{eq:relevance:norm})}    (AcSc)
    (Sc)    edge node[edgestyle] {influences  (\ref{eq:offset}, \ref{eq:rating:raw})}   (PdR);

\end{tikzpicture}
}

				\caption{Product rating is calculated by utilizing the ontology.
				There are ontology relations that contain the relevant equation from the current article in parenthesis.
				}
				\label{fig:prod:rating:ontology}
				}
		\end{figure}

\begin{figure}[h!]
    \begin{center}\subfloat[Overlapping product tag and preference tag concepts.]{
    \includegraphics[width=0.21\textwidth]{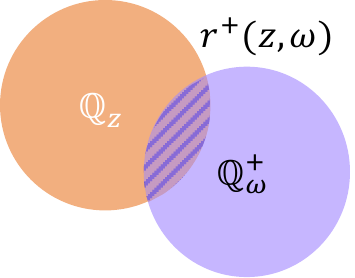}
    \label{fig:association}
    }\subfloat[The semantic space that defines a preference tag is the union of the concepts, which
    support or oppose its existence.]{
    \includegraphics[width=0.21\textwidth]{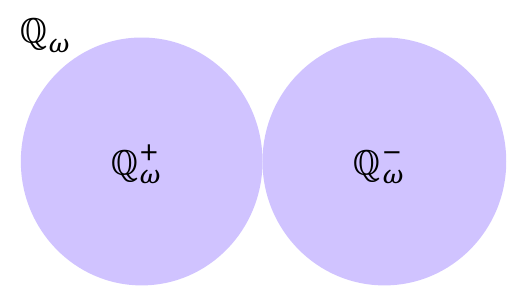}
    \label{fig:preference:concepts}
    }
    \\
    \subfloat[A product concept as a union of its product tag concepts.]{\includegraphics[width=0.155\textwidth]{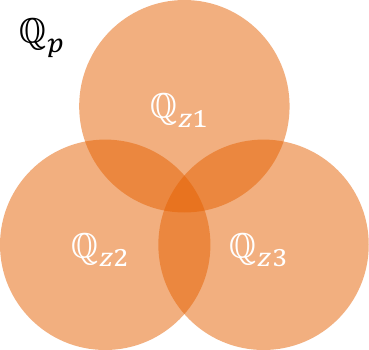}
    \label{fig:product:concept}}\subfloat[Semantic overlap between a product and a preference tag.]{\includegraphics[width=0.155\textwidth]{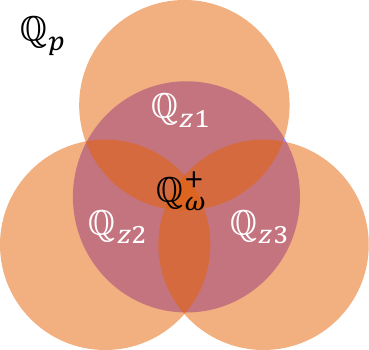}
    \label{fig:overlap}}\subfloat[Introduction of new product tag according to the reduction design principle.]{\includegraphics[width=0.155\textwidth]{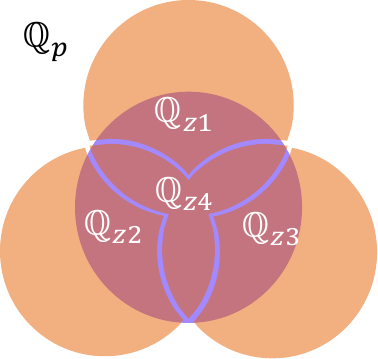}
    \label{fig:reduction:design}}\caption{Venn diagrams that illustrate the operations between sets of preference and product tags primitive
    concepts.}
		\end{center}
\end{figure}

\begin{figure}[h!]
	\centering
	\includegraphics[width=\textwidth]{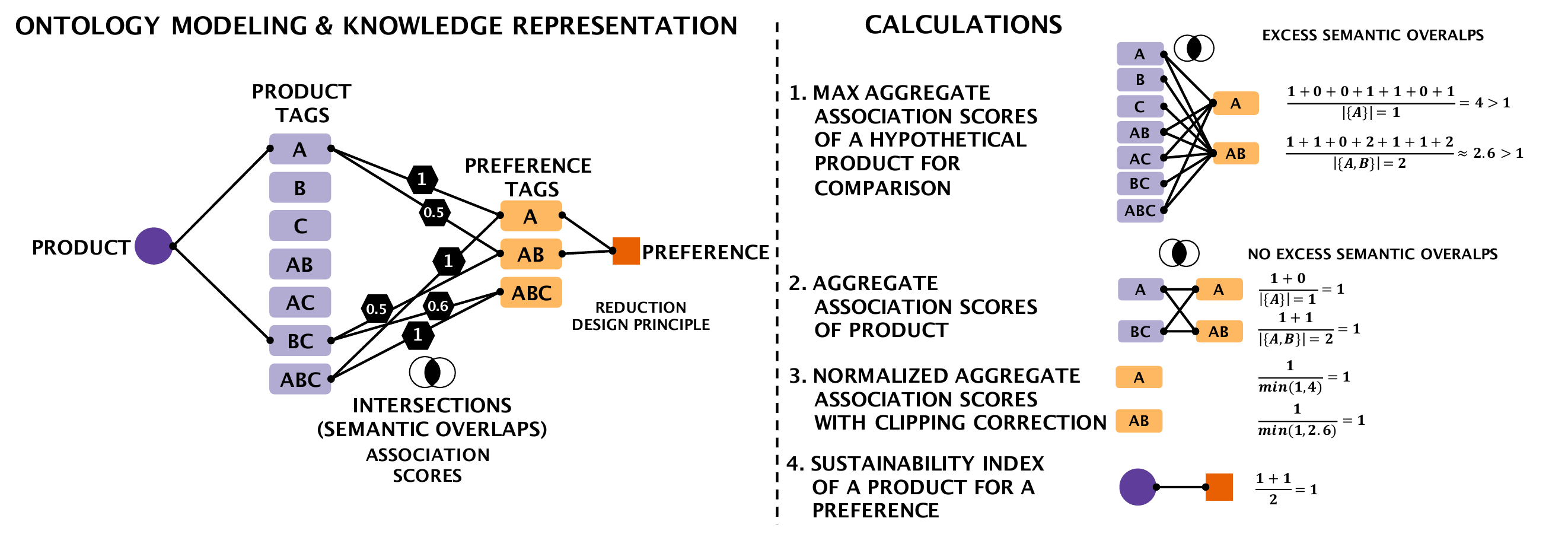}
	\caption{An alternative calculation of sustainability index illustrated in
             Figure~5. This calculation uses a theoretical normalized
			 aggregate association score that does not respect the reduction design principle of
			 \refsec{subsec:comparable-aggregated-associations}.
			 To avoid violation of the bounding properties of \refeq{eq:relevance:norm}, the
	         clipping methodology introduced in
			 \refsec{subsec:overlaps-and-possible-error-overflows} is used.
	}
	\label{fig:clip:aggregate}
\end{figure}

\begin{figure*}[ht!]
    \subfloat[Preference tags.]{\includegraphics[width=\linewidth]{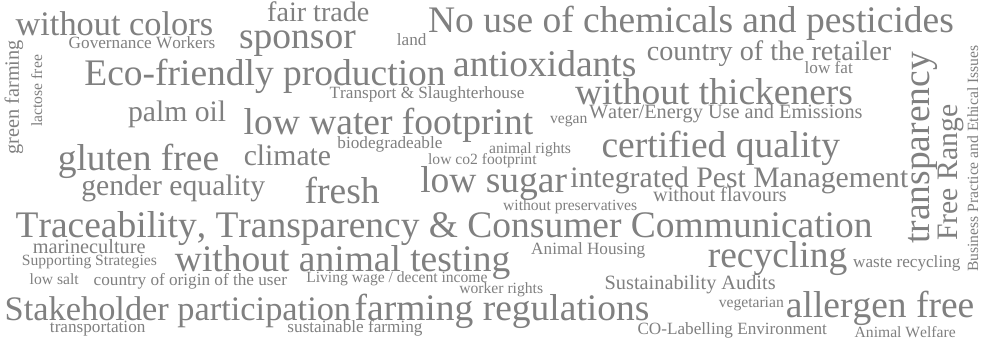}
    \label{fig:cloud:preferences}}\hfill\\
    \subfloat[Product tags.]{\includegraphics[width=\linewidth]{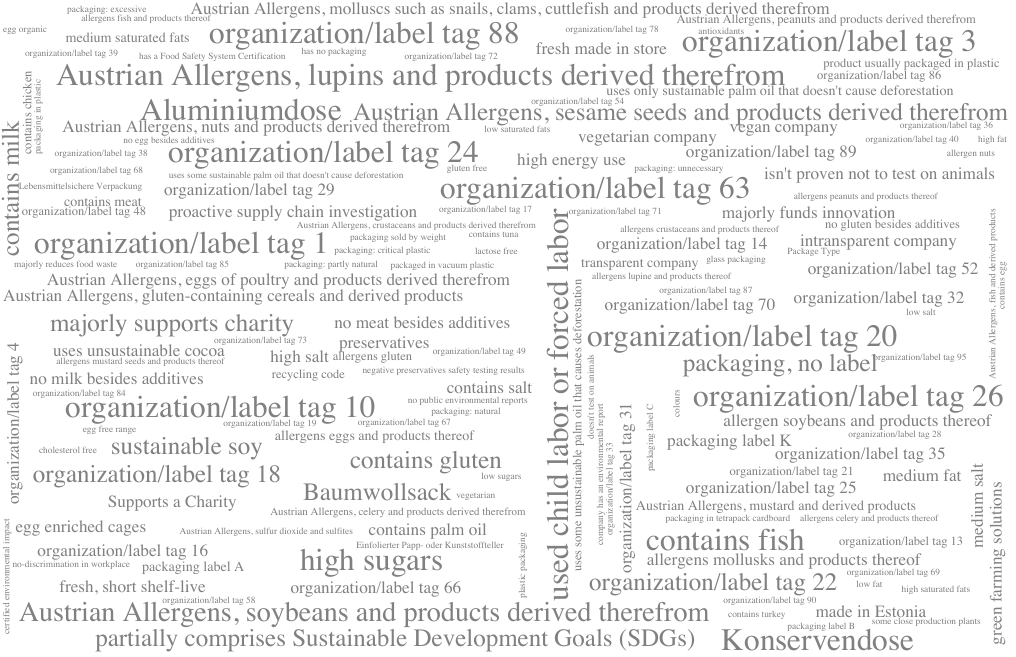}
    \label{fig:cloud:products}}\hfill\\
    \caption{Wordclouds of preference and product tags.
    The size is proportional to the number of products assigned to each tag.\label{fig:cloud}}
\end{figure*}

\begin{figure}[!htb]
    \subfloat[B.I.7: Finding additional product information in the app was easy.]{
        \includegraphics[width=0.28\linewidth]{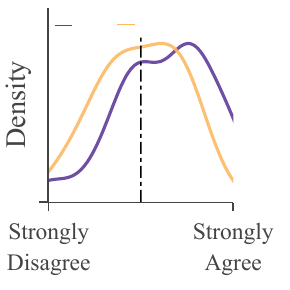}
        \label{fig:feedback:pos:extrainfo}
				\begin{picture}(0,0)
					\put(-115,140){\colorbox{white}{\footnotesize Retailer}}
					\put(-115,131){\colorbox{white}{\makebox(5,2){ }}}
					\put(-110,131){\colorbox{white}{\makebox(5,2){\hfill }}}
					\put(-102,131){\colorbox{white}{\makebox(5,2){\hfill }}}
					\put(-102,125){\colorbox{white}{\makebox(5,4){\hfill }}}
					\put(-115,129){\colorbox{white}{\footnotesize A}}

					\put(-82,140){\colorbox{white}{\footnotesize Retailer }}
					\put(-82,129){\colorbox{white}{\makebox(5,2){\hfill }}}
					\put(-85,131){\colorbox{white}{\makebox(5,2){\hfill }}}
					\put(-65,131){\colorbox{white}{\makebox(5,2){\hfill }}}
					\put(-55,131){\colorbox{white}{\makebox(5,2){\hfill }}}
					\put(-45,131){\colorbox{white}{\makebox(5,2){\hfill }}}
					\put(-35,131){\colorbox{white}{\makebox(5,2){\hfill }}}
					\put(-75,131){\colorbox{white}{\makebox(5,2){\hfill }}}
					\put(-82,129){\colorbox{white}{\footnotesize B}}
				\end{picture}
    }\hfill
    \subfloat[B.I.8: The product info justify the rating of the products.]{
        \includegraphics[width=0.28\linewidth]{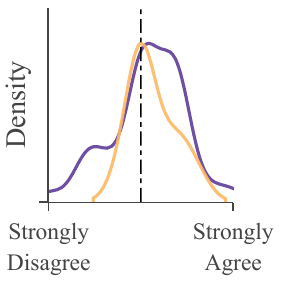}
        \label{fig:feedback:pos:justify}
    }\hfill
    \subfloat[B.II.6: The products with high rating match my preferences.]{
       \includegraphics[width=0.28\linewidth]{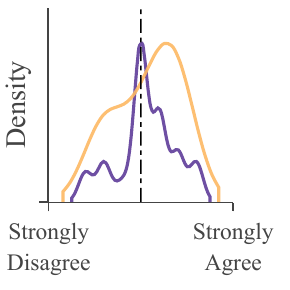}
        \label{fig:feedback:pos:prefmatch}
    }\hfill
    \subfloat[B.II.13: I discovered new products  while using the rating functionality.]{
        \includegraphics[width=0.28\linewidth]{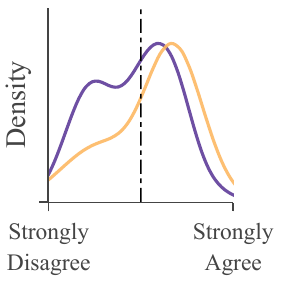}
        \label{fig:feedback:pos:novelty}
    }\hfill
    \subfloat[B.II.14: The app does not take into account my preferences when calculating the ratings.]{
        \includegraphics[width=0.28\linewidth]{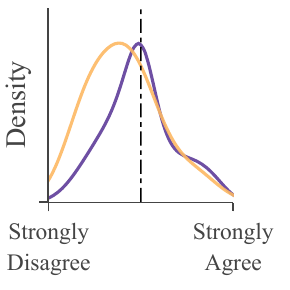}
        \label{fig:feedback:pos:preferences:account}
    }\hfill
    \subfloat[B.III.1: The app could capture my actual preferences.]{
        \includegraphics[width=0.28\linewidth]{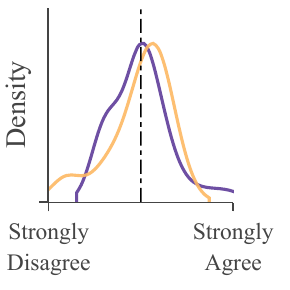}
        \label{fig:feedback:pos:capture:preferences}
    }\hfill
    \caption{Questions regarding the application that indicate acceptance from users.}
    \label{fig:feedback:pos}
\end{figure}

\begin{figure}[!htb]
 \subfloat[B.IV.3: The products I used to buy before using the app  achieve high rating in the app.]{
    \includegraphics[width=0.28\linewidth]{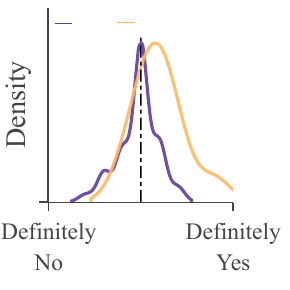}
    \label{fig:feedback:awareness:high:before}
		  		\begin{picture}(0,0)
					\put(-115,140){\colorbox{white}{\footnotesize Retailer}}
					\put(-115,131){\colorbox{white}{\makebox(5,2){ }}}
					\put(-110,131){\colorbox{white}{\makebox(5,2){\hfill }}}
					\put(-102,131){\colorbox{white}{\makebox(5,2){\hfill }}}
					\put(-102,125){\colorbox{white}{\makebox(5,4){\hfill }}}
					\put(-115,129){\colorbox{white}{\footnotesize A}}
					
					\put(-82,140){\colorbox{white}{\footnotesize Retailer }}
					\put(-82,129){\colorbox{white}{\makebox(5,2){\hfill }}}
					\put(-85,131){\colorbox{white}{\makebox(5,2){\hfill }}}
					\put(-65,131){\colorbox{white}{\makebox(5,2){\hfill }}}
					\put(-55,131){\colorbox{white}{\makebox(5,2){\hfill }}}
					\put(-45,131){\colorbox{white}{\makebox(5,2){\hfill }}}
					\put(-35,131){\colorbox{white}{\makebox(5,2){\hfill }}}
					\put(-75,131){\colorbox{white}{\makebox(5,2){\hfill }}}
					\put(-82,129){\colorbox{white}{\footnotesize B}}
				\end{picture}
		
	}\hfill
\subfloat[B.IV.4: The products I used to buy before using the app  achieve low rating.]{
    \includegraphics[width=0.28\linewidth]{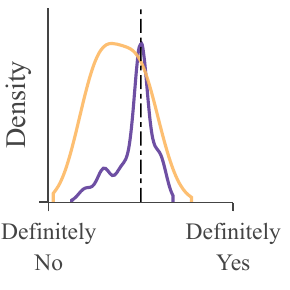}
				        \label{fig:feedback:awareness:lowe:before}
}\hfill
\subfloat[B.IV.5: The products I will buy in the future are the ones with high rating in the app.]{
    \includegraphics[width=0.28\linewidth]{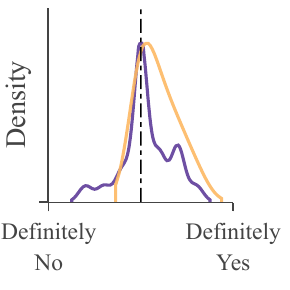}
    \label{fig:feedback:awareness:high:rating}
}\hfill
\\
\centering{
\null\hfill
\subfloat[B.IV.6: The products I will buy in the future are the ones with low rating in the app.]{
    \includegraphics[width=0.28\linewidth]{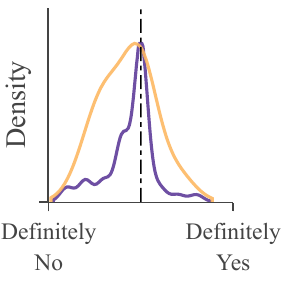}
    \label{fig:feedback:awareness:low:rating}
}\hfill
\subfloat[B.IV.9: I am more aware about sustainability aspects after using the app.]{
    \includegraphics[width=0.28\linewidth]{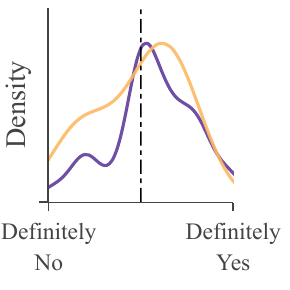}
    \label{fig:feedback:awareness:awareness}
}\hfill\null

}
\caption{Questions regarding sustainability awareness and future purchases.}
\label{fig:feedback:awareness}
\end{figure}

  \begin{figure}[!htb]\subfloat[The rating of the app.]{
    \includegraphics[width=0.28\linewidth]{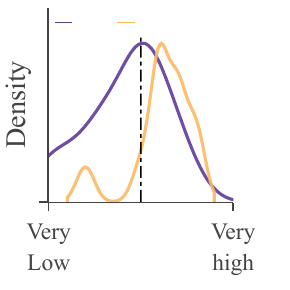}
    \label{fig:feedback:final:rating}
		  		\begin{picture}(0,0)
					\put(-115,140){\colorbox{white}{\footnotesize Retailer}}
					\put(-115,131){\colorbox{white}{\makebox(5,2){ }}}
					\put(-110,131){\colorbox{white}{\makebox(5,2){\hfill }}}
					\put(-102,131){\colorbox{white}{\makebox(5,2){\hfill }}}
					\put(-102,125){\colorbox{white}{\makebox(5,4){\hfill }}}
					\put(-115,129){\colorbox{white}{\footnotesize A}}
					
					\put(-82,140){\colorbox{white}{\footnotesize Retailer }}
					\put(-82,129){\colorbox{white}{\makebox(5,2){\hfill }}}
					\put(-85,131){\colorbox{white}{\makebox(5,2){\hfill }}}
					\put(-65,131){\colorbox{white}{\makebox(5,2){\hfill }}}
					\put(-55,131){\colorbox{white}{\makebox(5,2){\hfill }}}
					\put(-45,131){\colorbox{white}{\makebox(5,2){\hfill }}}
					\put(-35,131){\colorbox{white}{\makebox(5,2){\hfill }}}
					\put(-75,131){\colorbox{white}{\makebox(5,2){\hfill }}}
					\put(-82,129){\colorbox{white}{\footnotesize B}}
				\end{picture}
    }\hfill
    \subfloat[The product information I read in the app.]{
    \includegraphics[width=0.28\linewidth]{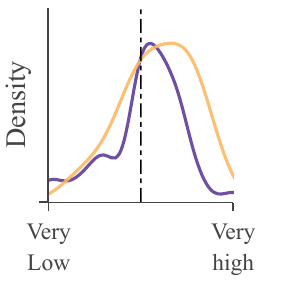}
    \label{fig:feedback:final:product:info}
    }\hfill
    \subfloat[Their price. ]{
    \includegraphics[width=0.28\linewidth]{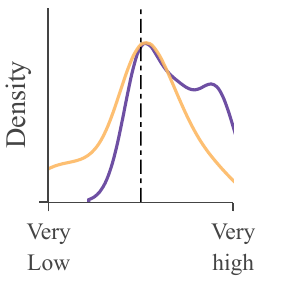}
    \label{fig:feedback:final:price}
    }
    \caption{The products I finally chose during this study are because of:}
    \label{fig:feedback:final}
\end{figure}

\begin{figure}[!htb]
    \centering
    \includegraphics[width=0.35\linewidth]{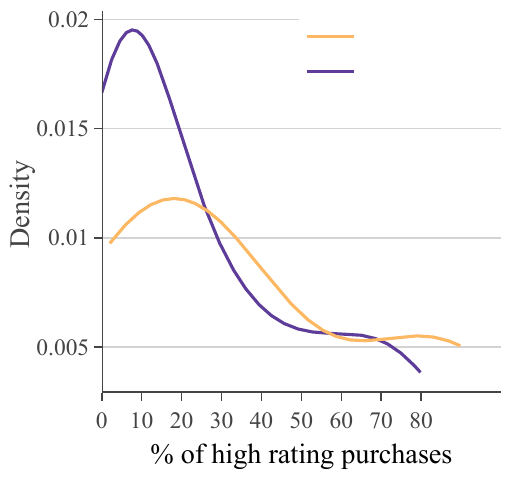}
    \caption{Estimate the \% of highly rated products you finally bought:}
    \label{fig:feedback:percentage:bought}
					\begin{picture}(0,0)
					\put(30,163){\colorbox{white}{\footnotesize Retailer A}}
					
					\put(60,175){\colorbox{white}{\makebox(5,2){\hfill }}}
					\put(70,175){\colorbox{white}{\makebox(5,2){\hfill }}}
					\put(80,175){\colorbox{white}{\makebox(5,2){\hfill }}}
					\put(30,175){\colorbox{white}{\footnotesize Retailer B }}
					
					\end{picture}
		
\end{figure}

\begin{figure}[!htb]
    \centering
    \subfloat[Retailer A.]{
    \includegraphics[width=0.429\columnwidth,valign=m]{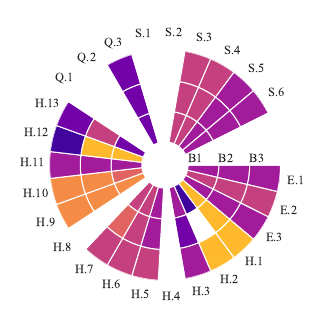}
    \label{fig:price:radial:norm:coop}
    }
    \hspace{-1.35em}
    \subfloat[Retailer B.]{
    \includegraphics[width=0.429\columnwidth,valign=m]{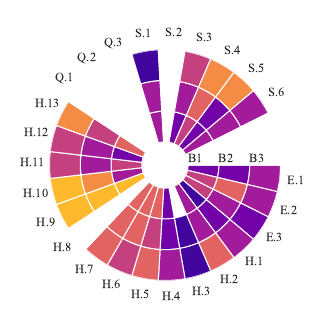}
    \label{fig:price:radial:norm:wm}
    }
    \hspace{-1.4em}
    \subfloat{
    \includegraphics[width=0.10\columnwidth,valign=m]{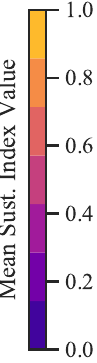}
    }\hfill
    \caption{Evaluation of mean normalized sustainability index (colorscale) per category across the different price
    bins per category as denoted in \reftab{tbl:price:bins} for all products of each retailer.
    }
    \label{fig:price:radial}
\end{figure}

\begin{figure}
  \centering
	\includegraphics[width=0.8\linewidth]{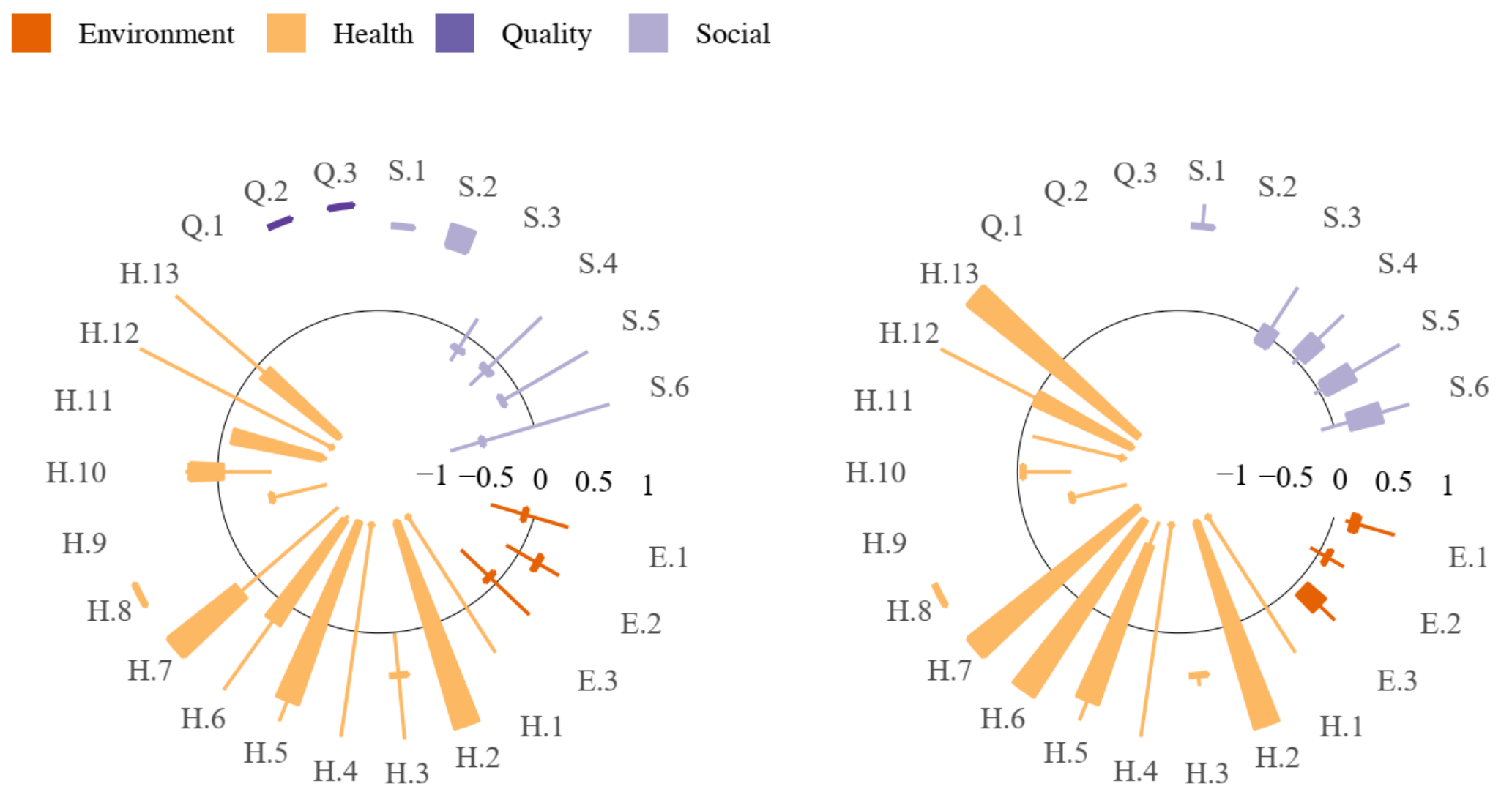}   
  \caption{The distribution of sustainability index per preference for all products in both retailers.\label{fig:si:distro}}
	\label{fig:si:distro}
\end{figure}

\begin{figure*}[ht!]
    \centering
    \subfloat[Answering the demographics questions in the entry survey.]{\includegraphics[width=0.31\columnwidth]{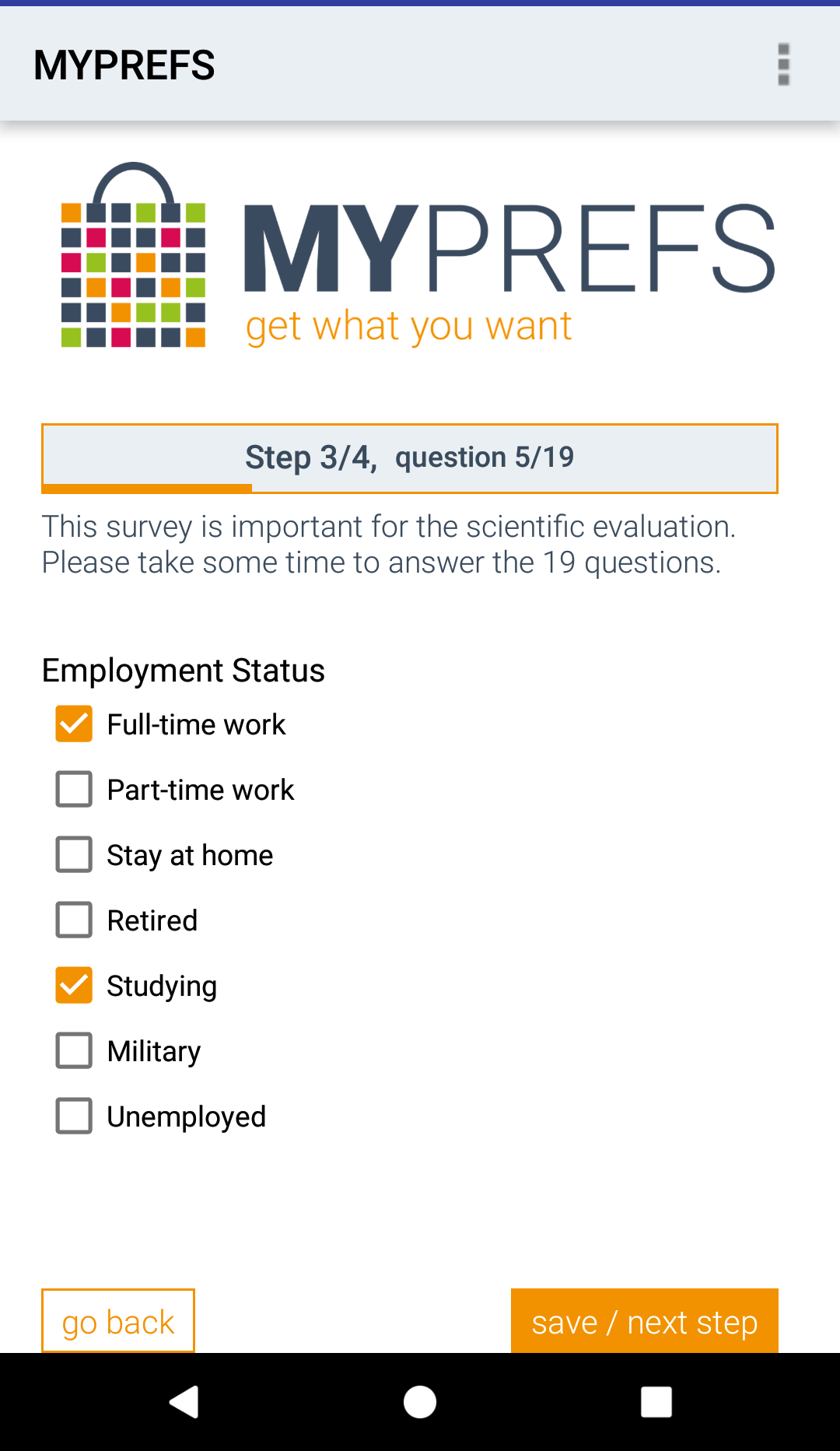}
        \label{fig:app:demographics}}\hfill
        \subfloat[Assigning preference scores.]{\includegraphics[width=0.31\columnwidth]{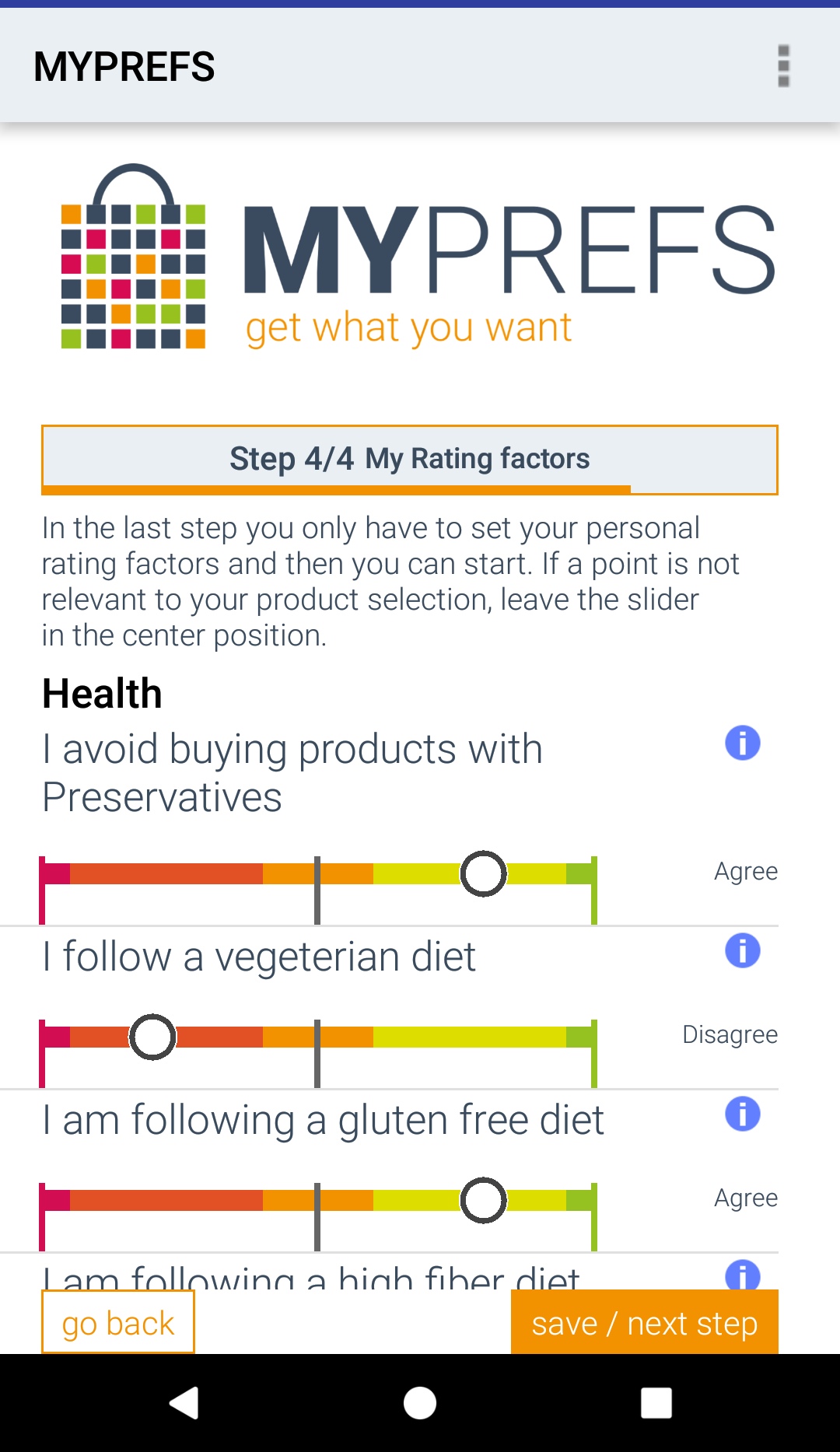}
        \label{fig:app:preferences}}\subfloat[Notification for extreme preference scores, which are also explained to the
        user in the app tutorial.]{\includegraphics[width=0.31\columnwidth]{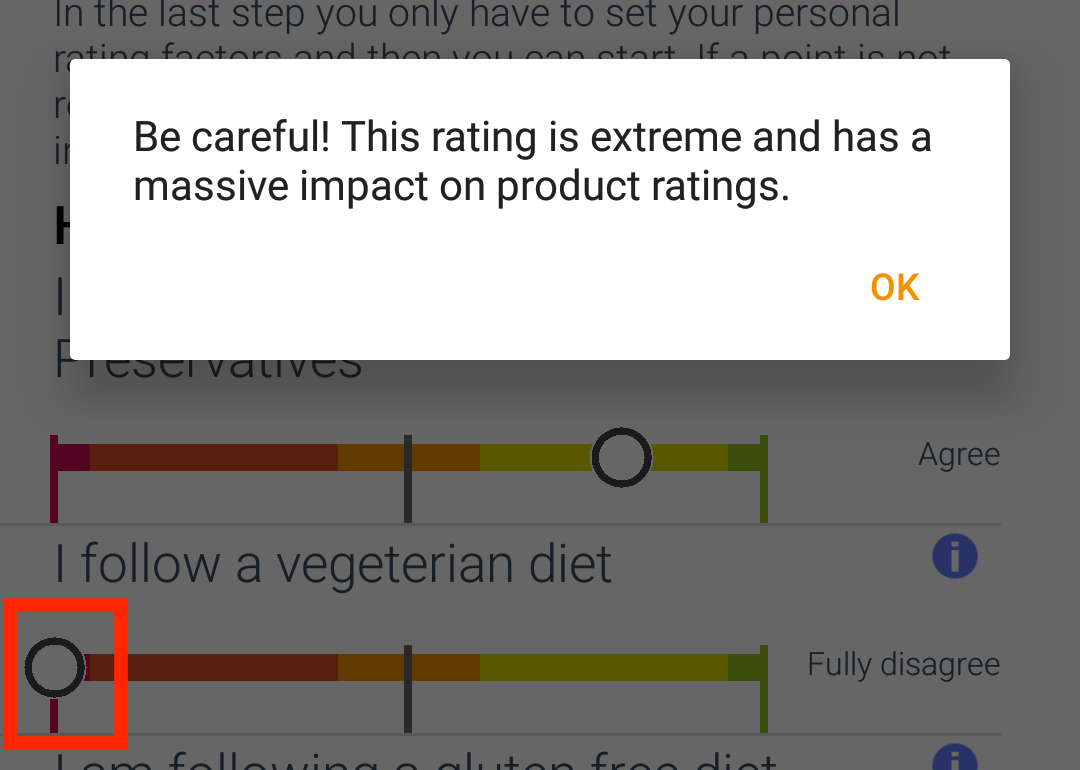}
        \label{fig:app:preferences:extreme}
        }
        \caption{Survey and the preference user interfaces of ASSET.}
        \label{fig:app:usage}
\end{figure*}
    \begin{figure*}[ht!]
        \subfloat[The main screen of the application.]{\includegraphics[width=0.31\columnwidth]{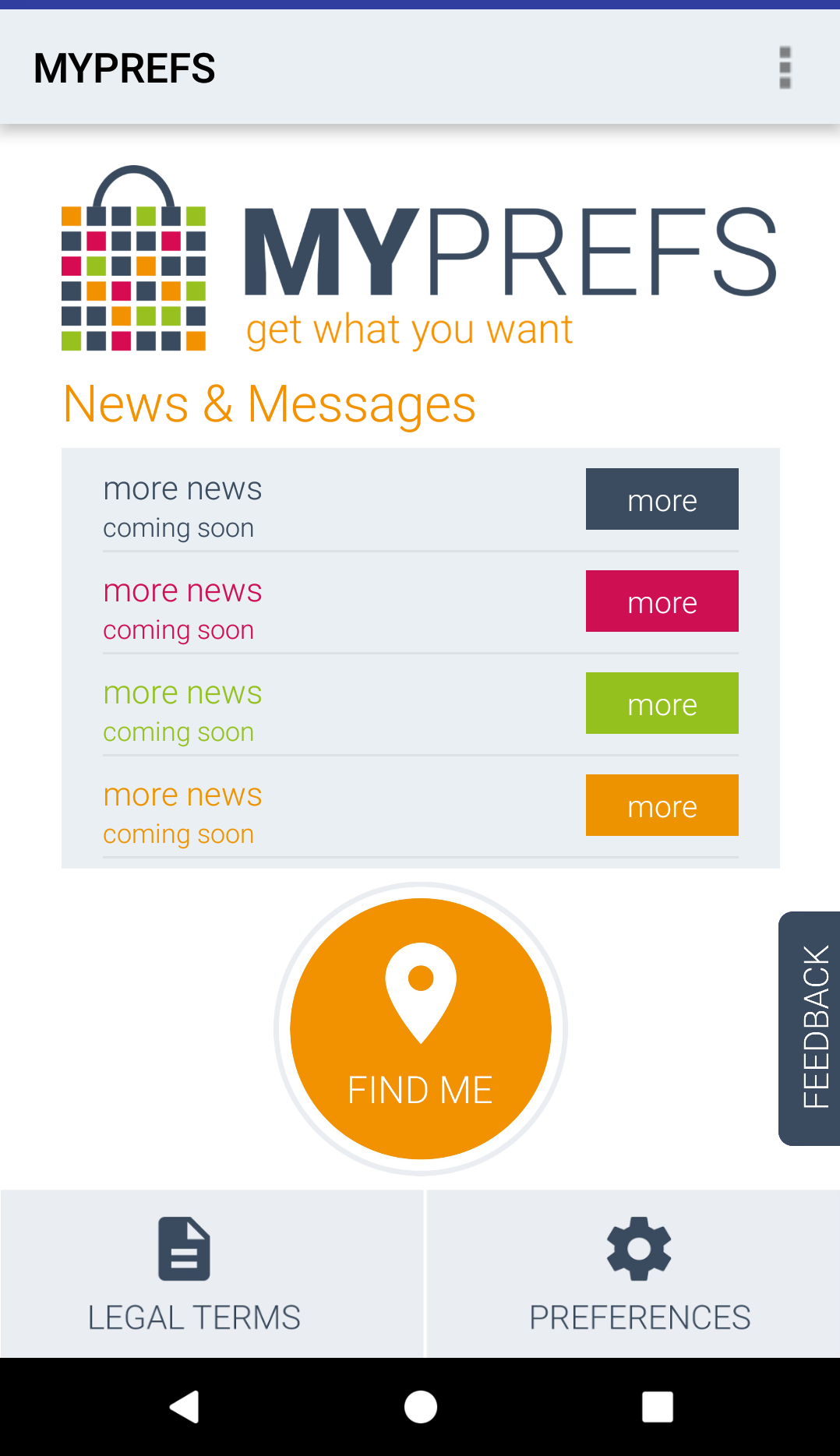}
        \label{fig:facebook:text:clipboard}}\hfill
        \subfloat[Presenting nearby product categories.]{\includegraphics[width=0.31\columnwidth]{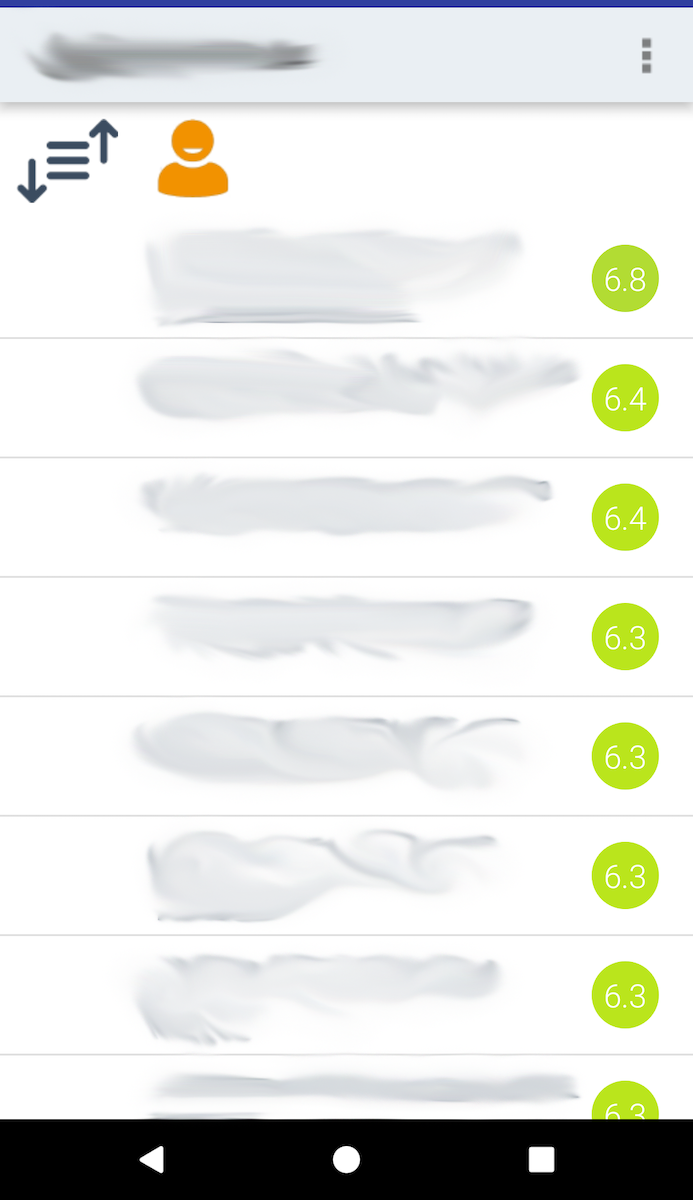}
        \label{fig:facebook:text:group}}\hfill
        \hfill\subfloat[Receiving product ratings.]{\includegraphics[width=0.31\columnwidth]{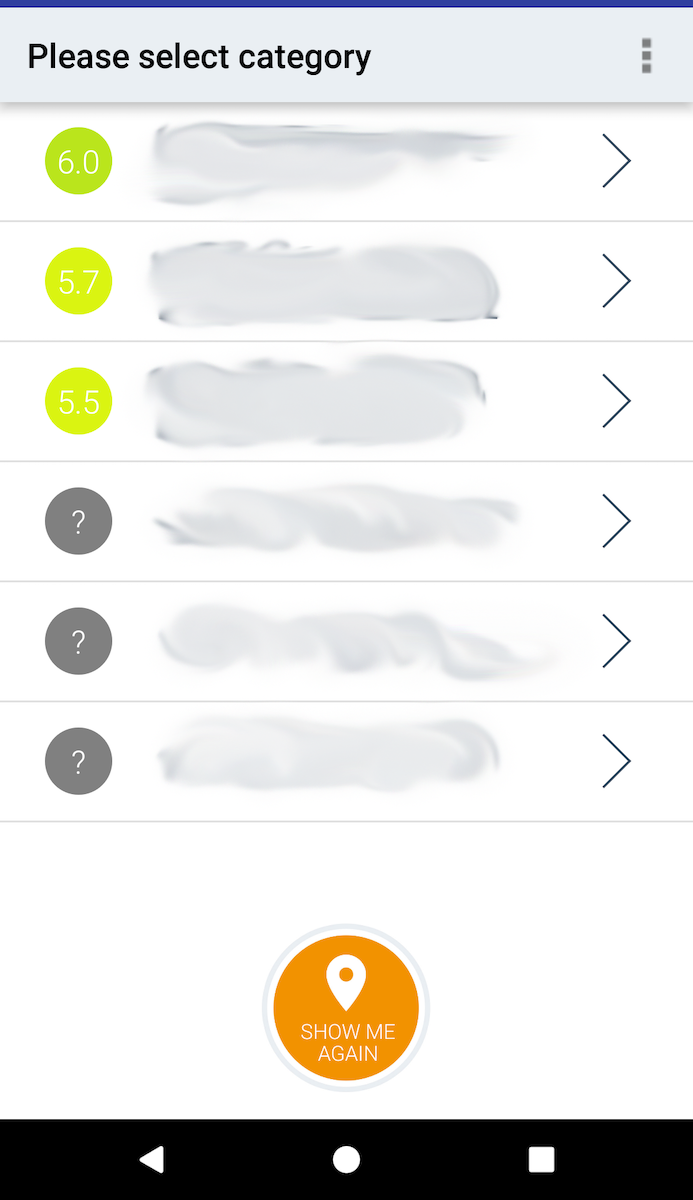}
        \label{fig:facebook:text:post}}\hfill \\
        \caption{Using the application.}
        \label{fig:facebook*}
        \label{fig:app:main}
    \end{figure*}
    \begin{figure*}[ht!]
        \subfloat[The amount a preference contributes to the product rating.]{\includegraphics[width=0.31\columnwidth]{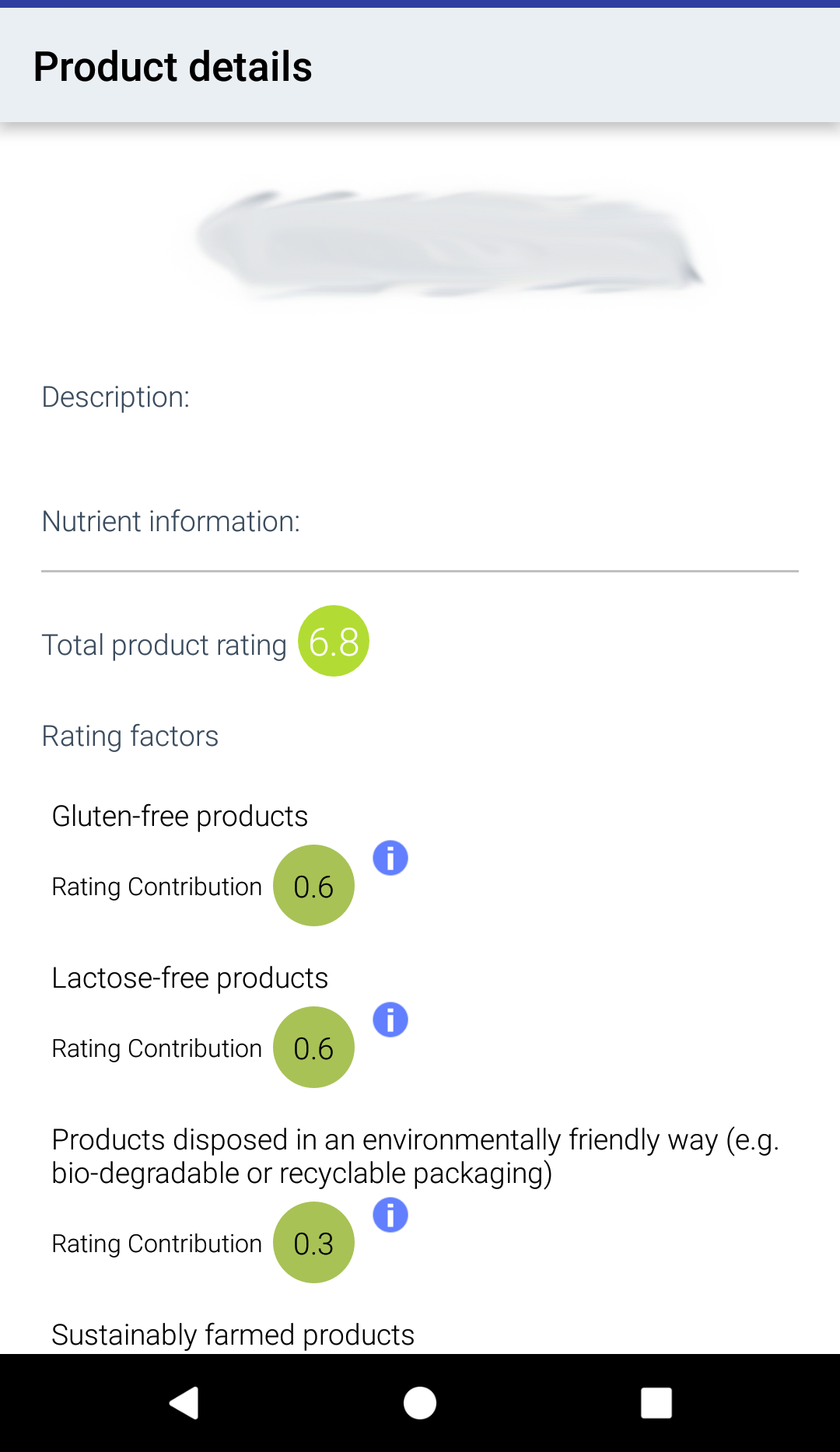}
        \label{fig:contr:pref}}\hfill
        \subfloat[Product tags that contribute a positive amount to the rating.]{\includegraphics[width=0.31\columnwidth]{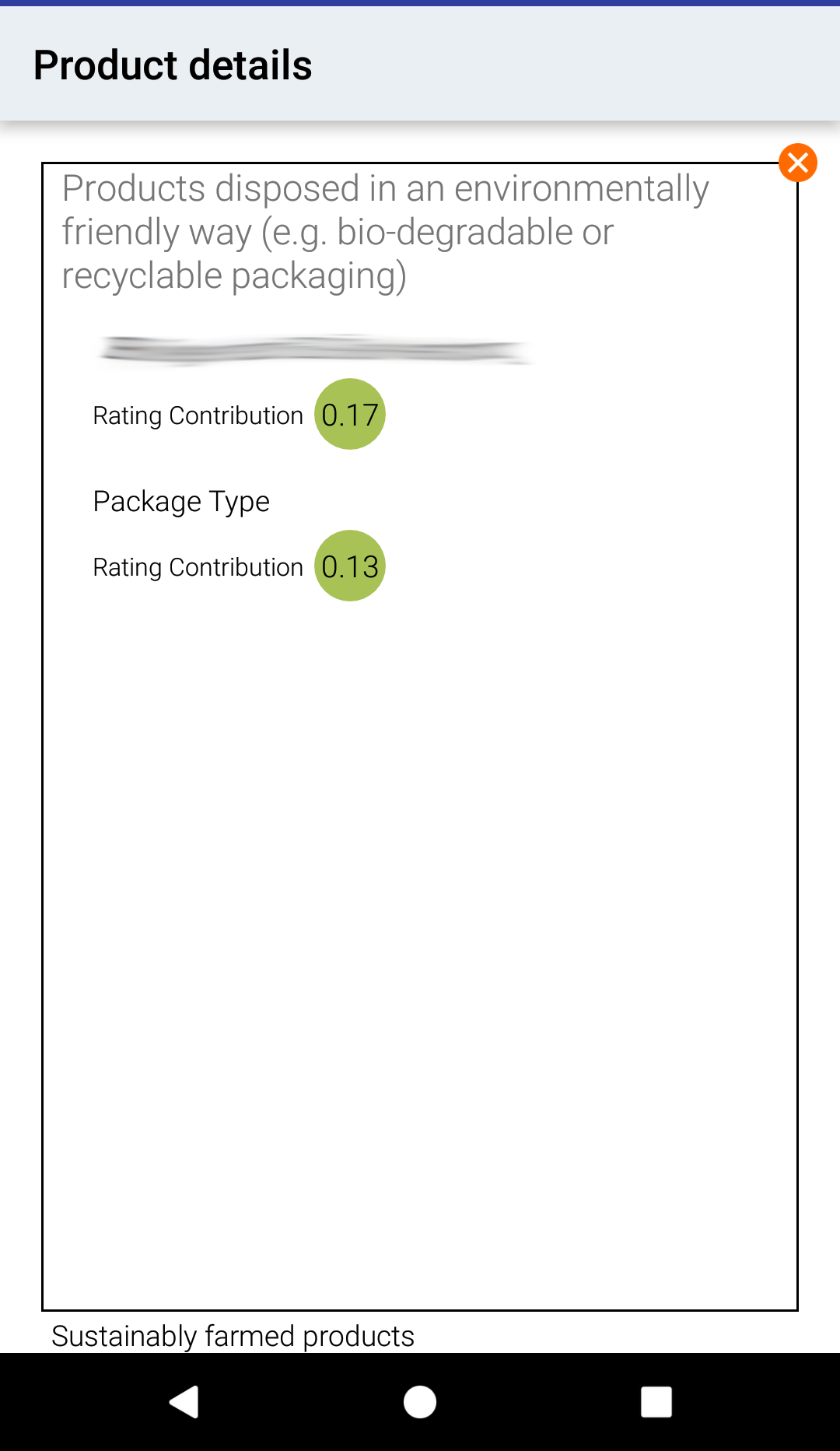}
        \label{fig:contr:pos}}\hfill
        \hfill\subfloat[Product tags that contribute a negative amount to the rating.]{\includegraphics[width=0.31\columnwidth]{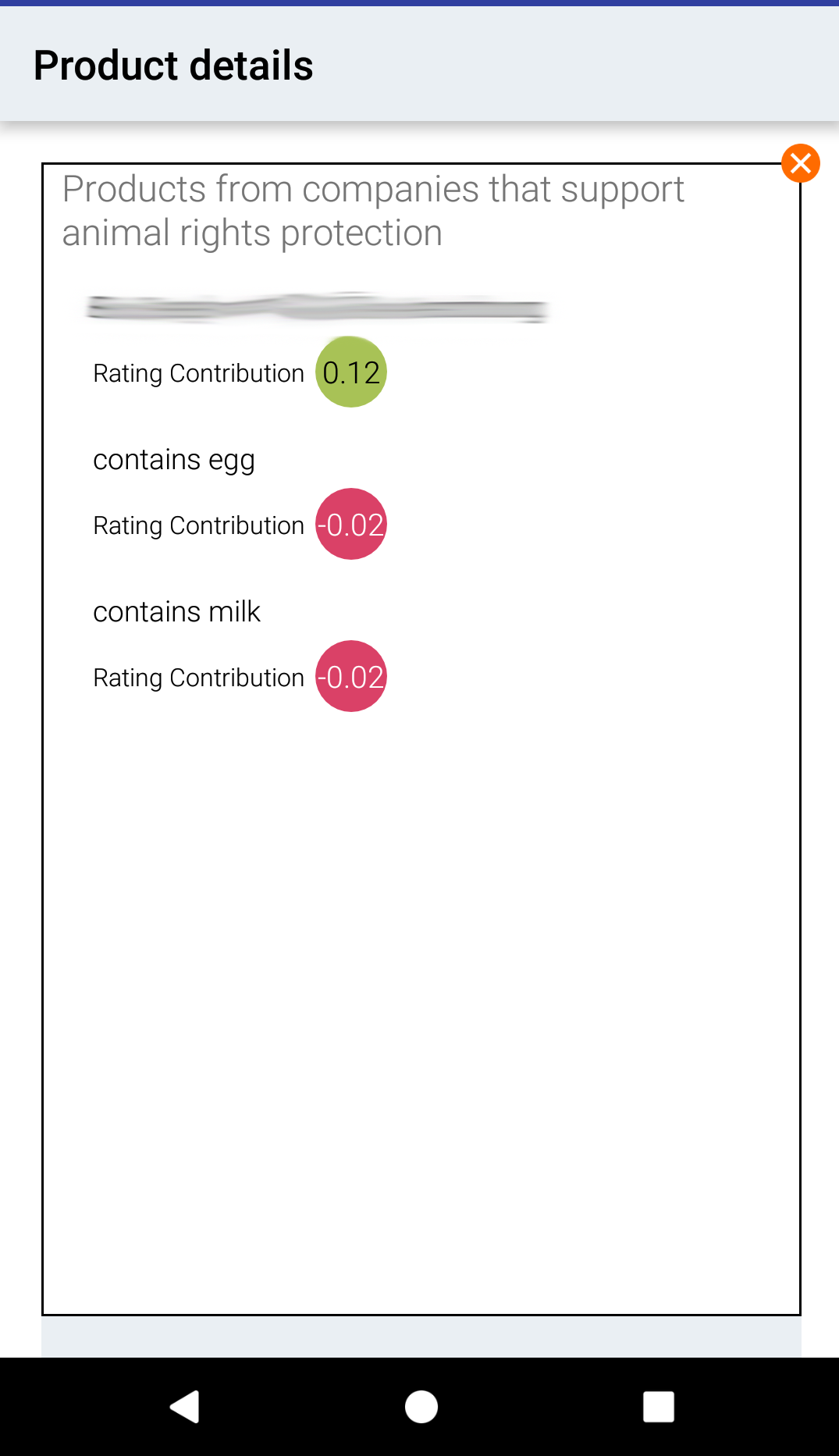}
        \label{fig:contr:neg}}\hfill \\
        \caption{Preference and product tag contribution to the rating.
        This feature improves the explainability of the product rating.}
        \label{fig:contr}
    \end{figure*}

\clearpage

\vspace*{1.5cm} 
\section*{\textbf{Supplementary Tables}}\label{sec:tables}
	 \counterwithin{table}{subsection}
   \renewcommand{\thetable}{S.\arabic{table}}
   \setcounter{table}{0}
   \begin{table}[hb!]
     \caption{{\bf Environmental Preferences}. The preferences setting options provided in value-sensitive smartphone application. }\label{tbl:preferences:env}
     \begin{tabularx}{\linewidth}{eXz}
          \toprule
          \textbf{Preference} &  Description  & \textbf{ID}  \\
          \midrule
            1.
            Products disposed in an environmentally friendly way (e.g. bio-degradable or recyclable packaging)
            &
            Products and product packages that can be recycled efficiently or are biodegradable and do not cause
            environmental hazard when disposed of.
            &
            E.1
          \\
            2.
            Products produced and distributed in an environmentally friendly way
            &
            A product has a sustainable lifecycle if during manufacturing, distribution, consumption and disposal: it
            has low CO2 and water footprint, low transportation costs, it can be efficiently recycled and/or
            biodegradable and produces no waste that is harmful to the environment.
            Some examples include toxic waste, high CO2 emissions during production, usage of palm oil during
            production that leads to deforestation etc.
            &
            E.2
          \\
            3.
            Sustainably farmed products
            &
            Sustainable farming is achieved by a company by applying farming methods such as Integrated Pest
            Management,  No-till agriculture, biodynamic and permaculture.
            For sea products, aquaculture is preferred instead of free fishing.
            The countries where the company produces its products should have clear and fair regulations regarding
            farming and land-management.
            &
            E.3
          \\
          \bottomrule
     \end{tabularx}
\end{table}
\begin{table}[hb!]
     \caption{{\bf Quality Preferences}. The preferences setting options provided in value-sensitive smartphone application.\label{tbl:preferences:qual}}
     \begin{tabularx}{\linewidth}{eXz}
          \toprule
          \textbf{Preference} &  Description  & \textbf{ID}  \\
          \midrule
            1.
            Award winning or high quality certified products.
            &
            There are several officially recognized organizations that offer awards and/or certifications to companies
            that produce high-quality products.
            &   Q.1
          \\
            2.
            Fresh products
            &
            Fresh products are directly brought from production to the shelves of the supermarket.
            Fresh products usually tend to be richer in nutrients.
            Also it is less likely for fresh products to cause health problems to the digestive system.
            & Q.2
          \\
            3.
            Locally originated and domestic products.
            &
            I prefer products that the country of origin is the same as the one I am living in.
            &
            Q.3
          \\
          \bottomrule
     \end{tabularx}
\end{table}
\begin{table}[hb!]
     \caption{{\bf Social Preferences}. The preferences setting options provided in value-sensitive smartphone application.\label{tbl:preferences:social}}
     \begin{tabularx}{\linewidth}{eXz}
          \toprule
          \textbf{Preference} &  Description  & \textbf{ID}  \\
          \midrule
          1.
          Products evaluated with auditing processes that rely on sustainability criteria.
          &
          All the processes that enhance and enable sustainability should be checked and validated from 3rd parties.
          Auditing processes based on sustainability criteria support companies in adopting more sustainable
          processes and also increase their transparency.
          &
          S.1
          \\
            2.
            Products from companies that actively contribute to public and social good
            &
            I prefer supporting companies, that sponsor charities, scholarships, R\&D, social activities
            &
            S.2
          \\
            3.
            Products from companies that support animal rights protection
            &
            Brands that respect and support animal rights should avoid factory farming, animal testing and any kind
            of animal abuse/mistreatment during the production of their products.
            &
            S.3
          \\
            4.
            Products from companies that support fairness and equality in the workplace
            &
            Brands that treat their workers equally and respect their rights.
            Such brands should take action against any discrimination between their employees and promote gender and
            race equality.
            &
            S.4
          \\
            5.
            Products from companies with transparent activities
            &
            I prefer buying products from companies that are open to sharing the impact and nature of their operation.
            &
            S.5
          \\
            6.
            Products from fair trade label companies
            &
            Fair trade focuses on human rights in general.
            Companies should respect human rights and trade with producers in developing and developed countries in a
            fair manner, offering fair prices for  raw materials and not abusing local laws for fair resource
            sharing, like public water sources, fair land management etc.
            &
            S.6
          \\
          \bottomrule
     \end{tabularx}
\end{table}

\begin{table}[hb!]
     \caption{{\bf Health Preferences}. The preferences setting options provided in value-sensitive smartphone application.\label{tbl:preferences:health}}
     \begin{tabularx}{\linewidth}{eXz}
          \toprule
          \textbf{Preference} &  Description  & \textbf{ID}  \\
          \midrule
           1.
           Allergen-free products
           &
           Food allergens are typically naturally-occurring proteins in foods or derivatives of them that cause
           abnormal immune responses.
           Eight ingredients cause about 90\% of food allergy reactions: Milk (mostly regarding
           children), Eggs, Peanuts, Tree nuts, like walnuts, almonds, pine nuts, brazil nuts, and pecans, Soy, Wheat
           and other grains with gluten, including barley, rye, and oats, Fish (mostly in adults), Shellfish (mostly
           in adults)
           &
           H.1
          \\
          2.
          Gluten-free products
          &
          A gluten-free diet is a diet that strictly excludes gluten, a mixture of proteins found in wheat and related grains,
          including barley, rye, oat, and all their species and hybrids (such as spelt, kamut, and triticale).
          The inclusion of oats in a gluten-free diet remains controversial, and may depend on the oat cultivation and
          the frequent cross-contamination with other gluten-containing cereals.
          &
          H.2
          \\
          3.
          High-protein products
          &
          Protein is one of the three macronutrients, along with carbs and fat. It is beneficial for building muscle.
          Protein serves a number of important functions in your body.
          It is made up of individual amino-acids, including many that your body cannot create on its own.
          Protein is the main component of your muscles, bones, skin and hair.
          These tissues are continuously repaired and replaced with new protein.
          &
          H.3
          \\
          4.
          Lactose-free products
          &
          A lactose free diet means eating foods that have no lactose.
          Lactose is a sugar that is a normal part of dairy products.
          Some people do not break down lactose well.
          They may not have enough lactase, the enzyme that breaks lactose down in the body.
          Or, their body may create lactase variants that do not work properly.
          &
          H.4
          \\
          5.
          Low fat products
          &
          Fats are a type of nutrient that you get from your diet.
          It's a major source of energy.
          It helps you absorb some vitamins and minerals.
          Fat is needed to build cell membranes, the vital exterior of each cell, and the sheaths surrounding nerves.
          It is essential for blood clotting, muscle movement, and inflammation.
          It is essential to eat some fats, though it is also harmful to eat too many.
          For long-term health, some fats are better than others.
          Good fats include monounsaturated and polyunsaturated fats.
          Bad ones include industrial-made trans fats.
          Saturated fats fall somewhere in the middle. Saturated fats raise your LDL (bad) cholesterol level.
          High LDL cholesterol puts you at risk for heart attack, stroke, and other major health problems.
          Trans fats  can raise LDL cholesterol levels in your blood.
          They can also lower a person’s HDL (good) cholesterol levels.
          &
          H.5
          \\
            6.
            Low salt products
            &
            There is also some evidence that too much salt can damage the heart, aorta, and kidneys without
            increasing blood pressure, and that it may be harmful for bones, too.
            &
            H.6
          \\
            7.
            Low sugar products
            &
            Added sugar is known to cause heart diseases.
            Sugar delivers “empty calories” — calories unaccompanied by fiber, vitamins, minerals, and other nutrients.
            Too much added sugar can replace healthier foods from a person’s diet.
            &
            H.7
          \\
            8.
            Products rich in antioxidants
            &
            Antioxidants come up frequently in discussions about good health and preventing diseases.
            Their nature is to prohibit (and in some cases even prevent), the oxidation of other molecules in the body.
            Oxidation is a chemical reaction that can produce free radicals, leading to chain reactions that may damage
            cells.
            The term "antioxidant" is mainly used for two different groups of substances: industrial chemicals which
            are added to products to prevent oxidation, and natural chemicals found in foods and body tissue which are
            said to have beneficial health effects.
            It is often debated  whether  they actually prevent diseases, which antioxidant(s) are needed from the diet
            and in what amounts beyond typical dietary intake.
            &
            H.8
          \\
            9.
            Products without artificial colours or flavor enhancers.
            &
            Artificial colors and flavoring enhancers are used to produce coloring effects and improve food taste.
            In general these substances are exhaustively tested in labs before they are used in food production.
            Still, it is not fully determined whether they cause health problems in the long term.
            &
            H.9
          \\
            10.
            Products without preservatives
            &
            A preservative is a substance or a chemical that is added to products such as food, beverages,
            pharmaceutical drugs, paints, biological samples, cosmetics, wood, and many other products to prevent
            decomposition by microbial growth or by undesirable chemical changes.
            Preservatives are used to prolong the shelf-life of the product but may cause health problems in the long
            term.
            &
            H.10
          \\
            11.
            Products without thickeners, stabilizers or emulsifiers
            &
            Emulsifiers allow water and oils to remain mixed together in an emulsion, as in mayonnaise, ice cream,
            and homogenised milk.
            Stabilizers, thickeners and gelling agents, like agar or pectin (used in jam for example) give foods a
            firmer texture.
            While they are not true emulsifiers, they help to stabilize emulsions.
            These additives may cause health problems in the long term.
            &
            H.11
          \\
            12.
            Vegan products
            &
            Vegans choose not to consume dairy, eggs or any other products of animal origin, in addition to not
            eating meat like the vegetarians.
            Veganism was originally defined as "the principle of emancipation of animals from exploitation by man."
            &
            H.12
          \\
            13.
            Vegetarian products
            &
            The Vegetarian Society defines a vegetarian as follows:
            "A vegetarian is someone who lives on a diet of grains, pulses, legumes, nuts, seeds, vegetables, fruits,
            fungi, algae, yeast and/or some other non-animal-based foods (e.g. salt) with, or without, dairy
            products, honey and/or eggs.
            A vegetarian does not eat foods that consist of, or have been produced with the aid of products
            consisting of or created from, any part of the body of a living or dead animal.
            This includes meat, poultry, fish, shellfish*, insects, by-products of slaughter** or any food made with
            processing aids created from these."
            &
            H.13
          \\
          \bottomrule
     \end{tabularx}
\end{table}

\begin{savenotes}
\begin{table}[htb!]
    \centering
    \caption{An example of association score values and their corresponding meaning. RDA refering to ``Recommended Daily Allowance''~\cite{RDA}. LDL refering to ``Low-density lipoprotein''.}
    \label{tbl:correlation}
    \bgroup \def\arraystretch{1.15}
    \begin{tabular}{C{0.22\linewidth}R{0.18\linewidth}R{0.18\linewidth}R{0.27\linewidth}}
        \hline
        association score        & product tag \vmath{\pmb\vproducttag}            &
        preference tag \vmath{\pmb\vpreferencetag} & \textbf{Meaning} \\ \hline
        \vmath{\vcorrelation= 1}                        & vegetable & vegetarian diet & \vproduct fully supports
        \vpreference
        via \vmath{\vproducttag}\\
        \vmath{\vcorrelation \in (0, 1)}                    & 10\% RDA Vitamin C &
        healthy diet &
        \vproduct
        partially supports
        \vmath{\vpreference}
        via \vmath{\vproducttag}\\
        \vmath{\vcorrelation = 0}                    & contains sugar & animal rights & \vproducttag is irrelevant to
        \vmath{\vpreferencetag}\\
        \vmath{\vcorrelation \in (-1,0)}                    & contains LDL & healthy diet
        &\vpreferencetag
partially opposes
        \vmath{\vproducttag}\\
        \vmath{\vcorrelation = -1}                        & contains eggs & vegan diet &\vpreferencetag fully opposes
        \vmath{\vproducttag}\\
        \hline
    \end{tabular}
    \egroup
\end{table}
\end{savenotes}

\clearpage

    \begin{table*}[!htb]
        \caption{\textbf{Alternative approaches.} Overview of online sources that provide
        sustainability ratings. Most of these approaches limit their scope to evaluation of
        brands rather than a broad spectrum of products. They are online web approaches with
        limited integration to shopping processes in retailer shops. Explainability is not explicitly provided and data may only be collected in a centralized fashion. They do not always capture a broad spectrum of sustainability goals. No rigorous evaluation with field studies has been shown how they impact sustainable consumption. }
        \label{tbl:online:ratings}
    \begin{tabularx}{\linewidth}{qqeqqe}
        \toprule
        Organization & Sector Scope & Target & Computation & Main Focus & Rating Type \\
        \midrule
        rankabrand\cite{rankabrand} & General & Brand & Crowdsourced & Environment, Social and Health & Ordinal \\
        GoodGuide\textregistered\cite{goodguide} & Personal Care, Household Supplies, Babies and Kids products &
        Product &
        Ingredient analysis based on official regulations & Environment and Health & Continuous \\
        Shop Ethical!\cite{shopethical} & General & Brand & Screening and reports for praises and/or criticism towards companies
        based
        on expert weighting & Social and Environmental & Ordinal \\
        THE GOOD SHOPPING GUIDE\cite{goodshopping} & General & Companies, Brands & Qualitative and quantitative evaluation
        based on
        expert criteria.
        & Environment, Animals, People, Ethical and Other Factors & Ordinal \\
        The Green Stars Project\cite{greenstars} & General & Companies, Brands, Services, Products etc. & User based
        reviews and rating assignment & Social and Environmental & Ordinal \\
        WikiRate\cite{wikirate} & General & Company & Allows for implementation of any rating based on the data the
        website offers &
        Environmental, Social and Governmental Concerns & Any \\
        Behind the Brands\cite{behindthebrands} & Food & Brand & Expert based & Social, Environmental and Ethical & Continuous,
        ordinal \\
        \bottomrule
    \end{tabularx}
    \end{table*}

	    \begin{table}[hb!]
    \caption{\textbf{Data Sources}. The data sources used for the construction of the used ontology.
             Websites accessed on January 2020.}
    \label{tbl:impl:data-sources}
      \begin{tabularx}{\columnwidth}{fXa}
        \toprule
            \textbf{Source Type} & \textbf{Institute} & \textbf{Website} \\
        \midrule
            \multirow{2}{*}{\shortstack[l]{Retailer\\Database}}  &   Coop Estonia   &  \url{coop.ee}            \\
                                                &   Winkler Markt                   &  \url{www.winklermarkt.at}    \\
            \cline{2-3}
            \multirow{2}{*}{\shortstack[l]{Online\\Datastores}}  &   GS1             &  \url{gs1.org}           \\
                                                &   ecoinvent                       &  \url{ecoinvent.org}      \\
            \cline{2-3}

            \multirow{6}{*}{Experts}            &   AINIA                           &  \url{ainia.es}           \\
                                                &   BIA                             &  \url{bia.ee}             \\
                                                &   ETH                             &  \url{ethz.ch}            \\
                                                &   Ethical Consumer Workshop 2017  &  \url{ethicalconsumer.org}\\
                                                &   LCM                             &  \url{lcm.at}             \\
                                                &   VKI                             &  \url{vki.at}             \\
            \cline{2-3}
            \multirow{2}{*}{Crowdsourcing}      &   Wikipedia                       &    \url{wikipedia.org}      \\
                                                &   Social Impact Data Hack         &    \url{sidh2017.ut.ee}     \\
        \bottomrule
      \end{tabularx}
    \end{table}
		
				    \begin{table}[ht!]
        \centering
        \caption{Approximate confidence intervals and error margins calculated based on
        available sample size.
        The population size for the calculations is estimated as the average number of loyalty card visitors per month .}
        \label{tbl:surveys:part}
        \footnotesize
        \begin{tabularx}{0.5\columnwidth}{Xrr}
            \toprule
            \textbf{Retailer} & \textbf{Survey} & \textbf{\# Users} \\
            \midrule
            \multirow{3}{=}{Retailer A} & Entry & 323 \\
                                        & Exit & 44    \\
                                        & Preferences & 315\\
            \cline{1-3}
            \multirow{3}{=}{Winkler Markt}    & Entry & 69 \\
                                              & Exit & 12 \\
                                              & Preferences & 66  \\
            \bottomrule
        \end{tabularx}
    \end{table}
	
	\begin{table}[!htb]
\caption{\textbf{Causal Impact Analysis}. Best \vmath{k\text{NN}} matches for treatment group selected on the DTW euclidean distance on high mean rating
products weekly expenditure. The control group is selected by using covariates for matching
between users.
Since different values of \vmath{k} result in different control group per combination of criteria the best value of
\vmath{k} is selected via a dynamic time warping of weekly expenditure before treatment.
The warping window size is \vmath{1}, meaning that the algorithm searches 1 week before or after to match the
expenditure of the current weekly expenditure between control and treatment.
 }
\label{tbl:distance:explanations}
\begin{tabularx}{\columnwidth}{XYYYee}
\toprule
Retailer & Monthly Total Expenditure & \raggedright{Distribution of Monthly Expenditure per
Category} &
Monthly
Representations &  Distance &
\vmath{k} \\
\midrule
\multirow{7}{=}{retailer A}     &                         - &                                           - &
\checkmark &  0.030852 &         5 \\
     &                         - &                                  \checkmark &                       - &  0.037871 &         5 \\
     &                         - &                                  \checkmark &              \checkmark &  0.032995 &         2 \\
     &                \checkmark &                                           - &                       - &  0.032736 &         4 \\
     &                \checkmark &                                           - &              \checkmark &  0.033386 &         5 \\
     &                \checkmark &                                  \checkmark &                       - &  0.033722 &         2 \\
     &                \checkmark &                                  \checkmark &              \checkmark &  0.035229 &         4 \\
\cline{1-6}
\multirow{6}{=}{retailer B} &                         - &                                  \checkmark & - &  0.056523 &         1 \\
     &                         - &                                  \checkmark &              \checkmark &  0.061028 &         2 \\
     &                \checkmark &                                           - &                       - &  0.066083 & 2 \\
     &                \checkmark &                                           - &              \checkmark &  0.055792 &         2 \\
     &                \checkmark &                                  \checkmark &                       - &  0.061923 &         1 \\
     &                \checkmark &                                  \checkmark &              \checkmark &  0.059597 &         2 \\
\bottomrule
\end{tabularx}
\end{table}

 \begin{table}[htb!]
        \caption{The summary of the causal impact analysis for Retailer A and Retailer A users.
        The reported decimal values in the table are rounded.}
        \label{table:causal-impact}
        \footnotesize{
         {\setlength{\extrarowheight}{1pt}\begin{tabularx}{\linewidth}{X>{\hsize=1.03\hsize}X>{\hsize=0.5\hsize}y>{\hsize=0.5\hsize}yy>{\hsize=0.5\hsize}y>{\hsize=0.5\hsize}yya}
            \toprule
            \multirow{2}{=}{\textbf{Retailer}} &          & \multicolumn{3}{>{\hsize=3\hsize}X}{\centering{\textbf{Average}}}  & \multicolumn{3}{>{\hsize=3\hsize}X}{\centering{\textbf{Cumulative}}} & \multirow{2}{=}{\vmath{p}}           \\
                                               &          & value	        & s.d.		            & \vmath{95\%} c.i.		       & value 		     & s.d.	        & \vmath{95\%} c.i.               &                                      \\
            \midrule
            \multirow{3}{=}{A}          & Actual		      & \vmath{0.8}	    & -	     			& -					           & \vmath{14.7}    & -            & -                               & \multirow{3}{=}{\vmath{\leq 0.001}} \\
                                           & Prediction		  & \vmath{0.6}     & \vmath{0.0}		& \vmath{[0.5,\  0.6]} 		   & \vmath{10.8}    & \vmath{0.3}  & \vmath{[10.1,\ 11.4]}           &                                      \\
                                           & Absolute Effect  & \vmath{0.2}	    & \vmath{0.0}		& \vmath{[0.2,\ 0.2]}    	   & \vmath{4.0}     & \vmath{0.3}  & \vmath{[3.3,\ 4.6]}             &                                      \\
                                           & Relative Effect  & \vmath{36.7\%}  & \vmath{3.1\%}	    & \vmath{[30.7\%,\ 42.8\%]}	   & -	             & -            & -                               &                                      \\

            \cline{1-9}
            \multirow{3}{=}{B} & Actual		      & \vmath{0.6}	    & -		      		& -					           & \vmath{10.1}    & -            & -                               & \multirow{3}{=}{\vmath{\leq 0.001}}  \\
                                           & Prediction		  & \vmath{0.4}     & \vmath{0.0}		& \vmath{[0.3,\ 0.5]} 		   & \vmath{7.1}     & \vmath{0.8}  & \vmath{[5.4,\ 8.7]}             &                                      \\
                                           & Absolute Effect  & \vmath{0.2}	    & 0.0				& \vmath{[0.1,\ 0.3]}    	   & 2.9			 & 0.8          & \vmath{[1.4,\ 4.7]}             &                                      \\
                                           & Relative Effect  & \vmath{41.0\%}  & \vmath{11.8\%}	& \vmath{[19.2\%,\ 65.4\%]}	   & -	             & -            & -                               &                                      \\
            \bottomrule
            \end{tabularx}
        }}
    \end{table}

\clearpage

	\begin{table}[htbp]
        \centering
        \caption{The price bins used for classifying a product according to its price per unit in
        \reffig{fig:price:radial}.
        Each product price is compared against prices of other products in the same category.
        Package specific quantities are not taken into account.}
        \label{tbl:price:bins}
    \begin{tabularx}{\columnwidth}{eXX}
    \toprule
    Label & Normalized Price Range & Description \\
    \midrule
    B1        &  \vmath{(0, 0.33)}      &        The  \vmath{33\%} of lowest product prices in category \\
    B2        &  \vmath{(0.33, 0.66)}   &        Between  \vmath{33\%} and  \vmath{66\%} product prices\\
    B3        &  \vmath{(0.66, 1)}      &        The  \vmath{66\%} and  \vmath{100\%} of highest product prices in
    category   \\
    \bottomrule
    \end{tabularx}
    \end{table}

	  \begin{table}[!htb]
        \caption{(Max-Min) Normalized price per category and normalized  sustainability index for retailer A.
         Relevant to \reffig{fig:price:radial:norm:coop}.}
        \label{tbl:price:corr:coop:norm}
        \footnotesize

        \begin{tabularx}{\linewidth}{yyy}
        \toprule
              Preference      &   Correlation         &  p-value    \\
        \midrule
                    E.1       &         0.03          &     0.21    \\
                    E.2       &        -0.08          &     0.00    \\
 \textsuperscript{+}E.3       &         0.10          &     0.00    \\
                    H.1       &         0.09          &     0.70    \\
                    H.2       &         1.00          &      -      \\
                    H.3       &        -0.05          &     0.47    \\
                    H.4       &          -            &      -      \\
                    H.5       &        -0.02          &     0.21    \\
                    H.6       &        -0.03          &     0.17    \\
                    H.7       &        -0.04          &     0.12    \\
                    H.8       &          -            &      -      \\
 \textsuperscript{-}H.9       &        -0.10          &     0.01    \\
                    H.10      &         0.14          &     0.23    \\
                    H.11      &         0.03          &     0.74    \\
 \textsuperscript{-}H.12      &        -0.88          &     0.02    \\
                    H.13      &         0.13          &     0.11    \\
                    Q.1       &          -            &      -      \\
                    Q.2       &          -            &      -      \\
 \textsuperscript{+}Q.3       &         0.12          &     0.01    \\
                    S.1       &          -            &      -      \\
                    S.2       &          -            &      -      \\
                    S.3       &         0.00          &     0.81    \\
                    S.4       &        -0.05          &     0.00    \\
                    S.5       &         0.02          &     0.21    \\
                    S.6       &         0.07          &     0.00    \\
        \bottomrule
        \end{tabularx}
\end{table}

		\begin{table}[t]
        \caption{(Max-Min) Normalized price per category and normalized  sustainability index for retailer B.
         Relevant to \reffig{fig:price:radial:norm:wm}.}
        \label{tbl:price:corr:wm:norm}
        \footnotesize
        \begin{tabularx}{\linewidth}{yyy}
        \toprule
         Preference      &   Correlation &  p-value                  \\
        \midrule
                                E.1        &   -0.03   &     0.03       \\
             \textsuperscript{-}E.2        &   -0.17   &     0.00       \\
             \textsuperscript{-}E.3        &   -0.34   &     0.00       \\
             \textsuperscript{+}H.1        &    0.23   &     0.00       \\
             \textsuperscript{+}H.2        &    0.22   &     0.00       \\
                                H.3        &    0.02   &     0.92       \\
                                H.4        &    0.09   &     0.00       \\
                                H.5        &   -0.09   &     0.00       \\
             \textsuperscript{+}H.6        &    0.12   &     0.00       \\
                                H.7        &   -0.01   &     0.59       \\
                                H.8        &     -   &        -         \\
                                H.9        &   -0.03   &     0.15       \\
             \textsuperscript{+}H.10       &    0.40   &     0.02       \\
                                H.11       &    0.01   &     0.78       \\
             \textsuperscript{+}H.12       &    0.17   &     0.00       \\
             \textsuperscript{+}H.13       &    0.20   &     0.00       \\
                                Q.1        &     -   &        -         \\
                                Q.2        &     -   &        -         \\
                                Q.3        &     -   &        -         \\
             \textsuperscript{-}S.1        &   -0.30   &     0.00       \\
                                S.2        &     -   &        -         \\
             \textsuperscript{+}S.3        &    0.29   &     0.00       \\
             \textsuperscript{+}S.4        &    0.22   &     0.00       \\
             \textsuperscript{+}S.5        &    0.35   &     0.00       \\
                                S.6        &   -0.08   &     0.05       \\
        \bottomrule
        \end{tabularx}
\end{table}

	\clearpage

\vspace*{1cm}

\section*{\textbf{Supplementary Methods}}
 \renewcommand\thesubsection{SM.\arabic{subsection}}
\renewcommand\thesubsubsection{SM.\arabic{subsection}.\arabic{subsubsection}}

	\subsection{\textbf{Ontology}}\label{sec:ontology}

In the used product ontology, product characteristics are summarized in the form of words or phrases, the \textbf{product tags}.
Each product tag \vmath{\vproducttag} can be assigned to one or more products
and it summarizes concepts and characteristics of a product \vmath{\vproduct}.
Every product \vmath{\vproduct} can be assigned to multiple product tags.
The set of all product tags denotes the semantic space of product characteristics.
Product tags are generated based on the data available from the data sources.
For example \vmath{\vproduct=\text{"cabbage"}} is associated with the product tag \vmath{\vproducttag=\text{"vegetable"}}.

The consumer preferences (see Table~\ref{tbl:preferences:env}-\ref{tbl:preferences:health}) are
also represented in the consumer preference ontology.
 Several challenges arise when defining an universal golden standard
regarding a sustainable consumption behavior.
Therefore, a personalized view on sustainable consumption is evaluated per consumer.
Each preference is presented to the consumer in form of a statement \vmath{\vpreference}.
Each statement is accompanied by a description that explains the sustainability concepts composing this preference,
e.g. \vmath{\vpreference=\text{"I prefer vegan products.".}}
A consumer \vmath{\vuser} then assigns numerical values to each preference statement \vmath{\vpreference}, to express
support, opposition or neutrality towards the preference statement.
This numerical value is referred to as preference score \vmath{\vpreferencescore}.
The preference score is bound in the range of \vmath{\left[0,
2\cdot\vmean{\vprefscore}\right]}, where \vmean{\vprefscore} is the mean value of the range.
The minimum value of the score implies that a consumer fully opposes a sustainability preference.
The maximum value implies fully supports a sustainability preferences.
The mean value of the scale implies no consumer preference regarding a sustainability criterion.
An assigned preference score is stored in the device of a consumer \vmath{\vuser}, and represents personalization.
The rating system calculates the product rating \vmath{\vrating{\vproduct}{\vuser}} for a given product
\vmath{\vproduct} and a consumer \vmath{\vuser} based on the preference scores.
The introduced method extends the semantic differential methodology for evaluating associations of product
characteristics and
sustainability preferences\cite{helbing2010quantitative, osgood1957measurement}.

Preference statements usually express \vchal{abstract} complex sustainability aspects. These aspects are decomposed into simpler ones creating a hierarchical ontology.
A word or phrase, referred to as preference tag \vmath{\vpreferencetag}, represents an aspect of a preference statement.
For example, \vmath{\vpreference =} "I prefer products that can be
disposed in a sustainable manner." is composed by two sustainability preference tags:
\vmath{\vpreferencetag_1=\text{"biodegradability"}}, which is the ability of the product to dissolve
within an acceptable time and without harming the environment \cite{goswami2016developments}, and
\vmath{\vpreferencetag_2=\text{"recycling capability"}}, which denotes whether a product can be recycled in
an efficient and environmentally friendly way \cite{villalba2002proposal}.
In other words, a preference tag \vmath{\vpreferencetag \in
\toSet{\Omega}_{\vpreference}}, which belongs to a preference tag set
\vmath{\toSet{\Omega}_{\vpreference}} of a preference \vmath{\vpreference},
compresses and represents information regarding the concepts that compose the
preference statement \vmath{\vpreference}.

The main challenge of designing an ontology for a sustainable consumption ontology is to define quantifiable
associations between the semantic spaces of product tags and preference tags.
To enable numerical calculation for product ratings we use the ontological design as sketched in 
\reffig{fig:prod:rating:ontology}.
Several ontological connections are introduced to calculate the aggregate support or opposition
of a product to a consumer's preferences.

The ontology of sustainability concepts has hierarchical structures of
concepts~\cite{Rodriguez2003}, where one concept is composed of several other concepts \vmath{q \in \toSet{Q}}.
If it is not feasible to further decompose a concept, then such a concept is referred to
as a \textit{primitive concept}\footnote{The definition of primitive concepts serves more as a
theoretical tool, with which the scope of the sustainability can be determined.}.
A \textbf{semantic association framework} introduces a logic for quantifiable semantic associations between a
product tag \vmath{\vproducttag} and a preference tag \vmath{\vpreferencetag}.
Examples of such associations can be found on \reftab{tbl:correlation}.
The associations between tags are quantified in a shared semantic space \toSet{Q}, which contains all concepts
relevant to products and sustainability preferences.
The space is defined with the following two assumptions:
\begin{enumerate}
    \item The individual elements of this space are \textit{primitive concepts} that cannot be decomposed into other
    concepts within the defined sustainability scope.
    \item Complex concepts are represented as sets that contain all the \textit{primitive concepts} that compose them.
    For instance, the preference tag \vmath{\vpreferencetag=\text{"vegan"}} may be decomposed to the set
    \vmath{\toSet{Q}_{\vpreferencetag} = \{}"no animals involved in production", "no animal products involved in
    production", \vmath{\ldots\}} where \vmath{\toSet{Q}_{\vpreferencetag} \subseteq \toSet{Q}}.
\end{enumerate}

Suppose a product tag expresses one or more \textit{primitive concepts}.
The set of those concepts is defined as \vmath{\toSet{Q}_{\vproducttag} \subseteq \toSet{Q}}.
Following the same logic, let a preference that its existence is denoted by the
union of \textit{primitive concepts} such as \vmath{\toSet{Q}_{\vpreferencetag}^{+}}.
An association between the product and the preference tag is defined as the overlap between the \textit{primitive
concepts} that each tag represents \vmath{\toSet{Q}_\vproducttag \cap \toSet{Q}_{\vpreferencetag}^{+}} as shown in 
\reffig{fig:association}.
This overlap is maximized when all \textit{primitive concepts} that compose the preference tag
also compose the product tag.
The following positive association score is defined:
\begin{equation}
    \label{eq:pos_association}
    r^{+}(\vproducttag, \vpreferencetag) = \dfrac{\absol{\toSet{Q}_\vproducttag \cap
    \toSet{Q}_{\vpreferencetag}^{+}}}{\absol{\toSet{Q}_{\vpreferencetag}^{+}}}
\end{equation}
The score is bounded, since:
\begin{equation}
    \label{eq:pos:bound}
    \toSet{Q}_\vproducttag \cap
    \toSet{Q}_{\vpreferencetag}^{+} \subseteq \toSet{Q}_{\vpreferencetag}^{+} \Leftrightarrow r^{+}\left
    (\vproducttag,
    \vpreferencetag\right) \in \left[0,1\right]
\end{equation}
For example, let a product tag and the relevant \textit{primitive concepts}: \vmath{\vproducttag=}"vegetable" with
\vmath{\toSet{Q}_\vproducttag=\{}"minimal CO\textsubscript{2} footprint", "plant part", "not
animal product"\vmath{\}}.
Suppose the following two preference tags,
which are decomposed to \textit{primitive concepts}, \vmath{\vpreferencetag_1=}"vegetarian
diet" with \vmath{\toSet{Q}_{\vpreferencetag1}=\{}"plant part", "not animal product"\vmath{\}}
and \vmath{\vpreferencetag_2=}"sustainable production" \vmath{\toSet{Q}_{\vpreferencetag2}=\{}"minimal
CO\textsubscript{2} footprint", "minimal water footprint", "no toxic waste"\vmath{\}}.
The "vegetable" product tag is composed of concepts which are enough to guarantee a "vegetarian diet".
Yet, the "vegetable" product tag is composed of some but not all of the concepts that guarantee a sustainable
production.

A product tag may oppose the existence of a preference tag concept.
For example the concept of "animal product" in food, fully opposes a "vegan diet" preference.
When a product contains animal products, then it is definitely not vegan.
Partial opposition is also possible, as for example the concept "contains lactose" indicates that a product
probably contains animal products, as lactose is a protein mainly found in milk\cite{Kretchmer1972}.
Therefore, the need to define opposite associations arises.
The negative \textit{primitive concepts} of a preference tag are defined as a set
\vmath{\toSet{Q}_\vpreferencetag^{\scriptstyle{-}}}.
For that, an association is defined when the concept in a
product tag is mutually exclusive with concepts defining a preference tag.
In this case a negative association score evaluates the overlap of concepts belonging to the
set of simple concepts of the product tag \vmath{\toSet{Q}_{\vproducttag} \cap
\toSet{Q}_\vpreferencetag^{\scriptstyle{-}}}.
The negative association score is defined as:
\begin{equation}
    \label{eq:neg_association}
    r^{-}(\vproducttag, \vpreferencetag) = \dfrac{\absol{\toSet{Q}_{\vproducttag} \cap
    \toSet{Q}_\vpreferencetag^{\scriptstyle{-}}}}{\absol{\toSet{Q}_\vpreferencetag^{\scriptstyle{-}}}}
\end{equation}
The negative score is also bounded, since:
\begin{equation}
    \label{eq:neg:bound}
    \toSet{Q}_\vproducttag \cap
    \toSet{Q}_{\vpreferencetag}^{-} \subseteq \toSet{Q}_{\vpreferencetag}^{-} \Leftrightarrow r^{-}\left
    (\vproducttag,
    \vpreferencetag\right) \in \left[0,1\right]
\end{equation}
The preference tag is defined by the union of its negative and positive concepts
\vmath{\toSet{Q}_{\vpreferencetag} = \toSet{Q}_\vpreferencetag^+ \cup \toSet{Q}_\vpreferencetag^-} as shown in 
\reffig{fig:preference:concepts}.
E.g. the preference tag \vmath{\vpreferencetag=}"vegan" can be decomposed to the supporting \textit{primitive concepts}
\vmath{\toSet{Q}_\vpreferencetag^{+}=}"plant part".

Product tags that contain \textit{primitive concepts} that both oppose and support a preference tag are also
possible.
In such case, the sum of positive and negative scores is calculated.
For this calculation, it is assumed that positive simple concepts cancel out negative concepts and vice versa.
The association score between a product tag and a preference tag is defined as:
\begin{equation}
    \label{eq:total_association}
    \vcorrelation  = r^{+}(\vproducttag, \vpreferencetag) - r^{-}(\vproducttag, \vpreferencetag)
    \stackrel{\ref{eq:pos:bound}, \ref{eq:neg:bound}, \ref{eq:total_association}}{\Leftrightarrow}
    \vcorrelation \in \left[-1, 1\right]
\end{equation}

This is denoted as the \textbf{individual bounding property} of the score and applies to all association scores
between any product and preference tags.

A product is often related to a set of product tags \vmath{\toSet{\vproducttag}_{\vproduct}}, therefore it can be
represented by the union of all \textit{primitive concepts} of these tags:
\begin{equation}
    \label{eq:sets:union}
    \toSet{Q}_{\vproduct} = \bigcup_{z \in \toSet{\vproducttag}_{\vproduct}}\toSet{Q}_{\vproducttag}
\end{equation}
An example of how a product \vmath{\vproduct} is decomposed to the \textit{primitive concepts} of the product tags
it consists of is found in \reffig{fig:product:concept}.

In practice, the calculation of the association score is performed by knowledge systems that
rely on (i) expert knowledge, (ii) crowdsourcing and (iii) machine learning.
As illustrated on line 4 in \reftab{tbl:correlation}, the products that contain high quantities of
LDL, oppose the preference tag "healthy diet", e.g.\
the health of the cardiovascular system\cite{Miller1975}.
This is quantified via a negative association score value in the range \vmath{\left(-1, 0\right)}.
The negative threshold value is assigned to the "contains eggs" product tag, which fully opposes the
"vegan" preference tag, as shown
on row 5 of \reftab{tbl:correlation}.

Associations between products \vmath{\vproduct} and preference tags are calculated by an aggregated
association score \vmath{\vrelevance}.
According to \refeq{eq:sets:union}, the intersection between all \textit{primitive concepts} related to a product and a
preference tag is used for such calculation:
\begin{equation}\label{eq:relevance:agg}
    \vrelevance = \dfrac{\absol{{\toSet{Q}_{\vproduct}}\cap\toSet{Q}_{\vpreferencetag}^{+}}}
            {\absol{\toSet{Q}_{\vpreferencetag}^{+}}}
            -
            \dfrac{\absol{{\toSet{Q}_{\vproduct}}\cap\toSet{Q}_{\vpreferencetag}^{-}}}
            {\absol{\toSet{Q}_{\vpreferencetag}^{-}}}
            \stackrel{\refeq{eq:sets:union}}{\Leftrightarrow{}}
\end{equation}
Note that, it is challenging to assign \textit{primitive concepts} directly to products.
A fixed sustainability scope supports the identification of overlaps and the existence of
primitive concepts.
The definition of primitive concepts is challenging when the ontology is under construction and
the sustainability scope is not fixed.
Conceptual overlaps are common in real world scenarios where product tags are usually derived from
labels or certifications, which are related to several \textit{primitive concepts}, e.g.the fair
trade label.
Intersection between a product tag and a preference tag \vmath{\vcorrelation} quantifies their shared
semantic space, consisting of at least one primitive concept.
The individual intersections between product tags and preference tags \vmath{\vcorrelation} can be used for the
approximation of the aggregated association \vmath{\vrelevance}.
The number of available product tags is considerably lower than the number of products worldwide.
When the product tags share \textit{primitive concepts}, it may prove challenging to calculate this intersection,
especially when the \textit{primitive concepts} are not identified.
Shared product tag \textit{primitive concepts} introduce overlaps between intersections of different product tags and a
preference tag.
Such overlaps introduce an error \vmath{\epsilon} in the approximation of the aggregate associations
\vmath{\vrelevance}.
Reducing the overlaps between product tags of the same product, minimizes this error, as shown in \textit{Lemma}
\ref{lemm:errors}.

\begin{lemma}\label{lemm:errors}
    The aggregated association \vmath{\vrelevance} between a product \vmath{\vproduct} and a preference tag
    \vmath{\vpreferencetag} is approximated with error \vmath{\epsilon} by the sum of tag associations
    \vmath{\vcorrelation} of each related  product tag \vmath{\vproducttag \in \toSet{\vproducttag}_{\vproduct}}
    with the preference tag \vmath{\vpreferencetag}, assuming that the \textit{primitive concept} overlaps
    between product tags are minimized \vmath{\bigcap_{z\in\toSet{\vproducttag}_\vproduct}\toSet{Q}_{\vproducttag
    } \to 0}, such that \vmath{\epsilon \to 0}.
\end{lemma}

\begin{proof}
    A product is defined as the union of associated \textit{primitive concepts} that product tags represent.
    Therefore the aggregated association for \vmath{\vproducttag \in \toSet{\vproducttag}_{\vproduct}} is:
    \begin{equation}
    \label{eq:association:1}
        \begin{gathered}
            \vrelevance = \dfrac{\absol{{\toSet{Q}_{\vproduct}}\cap\toSet{Q}_{\vpreferencetag}^{+}}}
            {\absol{\toSet{Q}_{\vpreferencetag}^{+}}}
            -
            \dfrac{\absol{{\toSet{Q}_{\vproduct}}\cap\toSet{Q}_{\vpreferencetag}^{-}}}
            {\absol{\toSet{Q}_{\vpreferencetag}^{-}}}
            \stackrel{\refeq{eq:sets:union}}{\Leftrightarrow{}} \\
            \vrelevance = \dfrac{\absol{\left(\bigcup_{z}
            \toSet{Q}_{\vproducttag}\right)\cap\toSet{Q}_{\vpreferencetag}^{+}}}
            {\absol{\toSet{Q}_{\vpreferencetag}^{+}}}
            -
            \dfrac{\absol{\left(\bigcup_{z}\toSet{Q}_{\vproducttag}\right)\cap\toSet{Q}_{\vpreferencetag}^{-}}}
            {\absol{\toSet{Q}_{\vpreferencetag}^{-}}}
            \Leftrightarrow{} \\
            \vrelevance = \dfrac{\absol{\bigcup_{\vproducttag}\left
            (\toSet{Q}_{\vproducttag}\cap\toSet{Q}_{\vpreferencetag}^{+}\right)}}
            {\absol{\toSet{Q}_{\vpreferencetag}^{+}}}
            -
            \dfrac{\absol{\bigcup_{\vproducttag}\left(\toSet{Q}_{\vproducttag}\cap\toSet{Q}_{\vpreferencetag}^{-}\right)}}
            {\absol{\toSet{Q}_{\vpreferencetag}^{-}}}
            \\
        \end{gathered}
    \end{equation}
    Each nominator can be further analyzed using the general form of the Inclusion-Exclusion principle\cite{rota1964foundations}.
    The first fraction nominator is expanded as, for any subset of product tags
    \vmath{\emptyset\neq\toSet{\vproducttag}_{\vproduct}'\subseteq\toSet{\vproducttag}_{\vproduct}}, and then all
    intersection between a preference tag and more than one product are isolated to determine overlaps:
    \begin{equation}
        \label{eq:association:2}
        \begin{aligned}
        \absol{\bigcup_{\vproducttag}\left(\toSet{Q}_{\vproducttag}\cap\toSet{Q}_{\vpreferencetag}^{+}\right)}
            &=  \sum_{\emptyset\neq\toSet{\vproducttag}_{\vproduct}'\subseteq\toSet{\vproducttag}_{\vproduct
        }}
        \left[(-1)^{\absol{\toSet{\vproducttag}_{\vproduct}'}-1}
                \biggl|\bigcap_{\vproducttag\in\toSet{\vproducttag}_\vproduct'}
                        \toSet{Q}_{\vproducttag}\cap\toSet{Q}_\vpreferencetag^{+}
                \biggr|
        \right] \\
            &= \sum_{\vproducttag\in\toSet{\vproducttag}_{\vproduct}}\absol{\toSet{Q}_{\vproducttag}
            \cap
            \toSet{Q}_\vpreferencetag^{+}
            } \\
        &+
        \underbrace{\sum_{\substack{\absol{\toSet{\vproducttag}_{\vproduct}'}>1\\ \toSet{\vproducttag}_{\vproduct
        }'\subseteq
        \toSet{\vproducttag}_{\vproduct}}}
        \left[
                (-1)^{\absol{\toSet{\vproducttag}_{\vproduct}'}-1}
                \biggl|\bigcap_{\vproducttag\in\toSet{\vproducttag}_\vproduct'}
                    \toSet{Q}_{\vproducttag}\cap\toSet{Q}_\vpreferencetag^{+}
                    \biggr|
        \right]}_{\epsilon^{+}} \\
        \end{aligned}
    \end{equation}
    Thus:
    \begin{equation}
        \label{eq:overlap:error:pos}
        \absol{\bigcup_{\vproducttag}\left(\toSet{Q}_{\vproducttag}\cap\toSet{Q}_{\vpreferencetag}^{+}\right)}  =
        \sum_{\vproducttag\in\toSet{\vproducttag}_{\vproduct}}\absol{\toSet{Q}_{\vproducttag}
            \cap
            \toSet{Q}_\vpreferencetag^{+}
            } + \epsilon^{+}
    \end{equation}
    Respectively it is shown that:
    \begin{equation}
        \label{eq:overlap:error:neg}
        \absol{\bigcup_{\vproducttag}\left(\toSet{Q}_{\vproducttag}\cap\toSet{Q}_{\vpreferencetag}^{-}\right)}  =
        \sum_{\vproducttag\in\toSet{\vproducttag}_{\vproduct}}\absol{\toSet{Q}_{\vproducttag}
            \cap
            \toSet{Q}_\vpreferencetag^{-}
            } + \epsilon^{-}
    \end{equation}

    The terms \vmath{\epsilon^+} and \vmath{\epsilon^-} are the overlap correction terms introduced by the
    Inclusion-Exclusion principle.
    These terms express the semantic overlap between different product tags of a product and a single preference tag.
    Therefore, it is possible now to expand \textit{Relation} \ref{eq:association:1}:
     \begin{equation}
        \label{eq:association}
        \begin{gathered}
            \vrelevance = \dfrac{\sum_{z}{\absol{\toSet{Q}_z\cap\toSet{Q}_{\vpreferencetag}^{+}}} +
            \epsilon^{+}
            }
            {\absol{\toSet{Q}_{\vpreferencetag}^{+}}}
            -
            \dfrac{\sum_{z}{\absol{\toSet{Q}_z\cap\toSet{Q}_{\vpreferencetag}^{-}}} +
            \epsilon^{-}}
            {\absol{\toSet{Q}_{\vpreferencetag}^{-}}}
            \stackrel{\textit{Eq.}~\ref{eq:pos_association}~\&~\ref{eq:neg_association}}{\Leftrightarrow{}} \\
            \vrelevance =  \sum_{z}{r^{+}(\vproducttag, \vpreferencetag)} - \sum_{z}{r^{-}(\vproducttag,
            \vpreferencetag)} + \epsilon
            \stackrel{\textit{Eq.}~\ref{eq:total_association}}{\Leftrightarrow{}} \\
            \vrelevance =  \sum_{z}{\vcorrelation} + \epsilon
        \end{gathered}
    \end{equation}
    where:
    \begin{equation}
        \label{eq:overlap:error}
        \begin{gathered}
            \epsilon = \dfrac{\epsilon^{+}}{\absol{\toSet{Q}_{\vpreferencetag}^{+}}}
            + \dfrac{\epsilon^{-}}{\absol{\toSet{Q}_{\vpreferencetag}^{-}}}
        \end{gathered}
    \end{equation}
    Since the intersection between sets is a commutative operation, it can be derived from \refeq{eq:association:2}
    that:
    \begin{equation}
    \biggl|\bigcap_{\vproducttag\in\toSet{\vproducttag}_\vproduct'}
                    \toSet{Q}_{\vproducttag}\cap\toSet{Q}_\vpreferencetag^{+}
                    \biggr| =
                    \biggl|\toSet{Q}_\vpreferencetag^{+}  \cap
    \left(\bigcap_{\vproducttag\in\toSet{\vproducttag}_\vproduct'}
                    \toSet{Q}_{\vproducttag}
    \right)
                    \biggr|
     \end{equation}
    \textit{Equations} \ref{eq:association:2}, \ref{eq:overlap:error:pos} and \ref{eq:overlap:error:neg} show that if
    the overlaps \vmath{\bigcap_{z\in\toSet{\vproducttag}_\vproduct}\toSet{Q}_{\vproducttag} \to 0} between
    product tags are minimized then \vmath{\epsilon^{+}, \epsilon^{-} \to 0 \Rightarrow
    \epsilon \to 0}.
    Thus the lemma is proved.
\end{proof}

For example assume the product \vmath{\vproduct=}"orange-lettuce-rice salad" in \reffig{fig:product:concept}, which is
associated with the product tags \vmath{\toSet{\vproducttag}_{\vproduct} = \{}"vegetable", "fruit", "cereal"\vmath{\}}.
The product tags \vmath{\vproducttag_{1}} = "cereal", \vmath{\vproducttag_{2}} = "fruit" and
\vmath{\vproducttag_{3}}="vegetable" share several \textit{primitive concepts}, such as "plant part".
Each product tag has a positive association score with the preference tag \vmath{\vpreferencetag=} "vegetarian".
As it is showcased in \reffig{fig:overlap} there are several overlaps between associations, due to the shared primitive
concepts of the product tags.
Summing all association scores with the preference tag "vegetarian" introduces errors because of the shared "primitive
concepts".
The calculation of error correction terms \vmath{\epsilon^+ , \epsilon^- } requires all possible combinations of
intersections of product tags with shared \textit{primitive concepts}.
Such calculation in the worst case requires an exponential time complexity of \vmath{O(2^{n})} for every
aggregate association score.
Therefore, the aggregation of association scores may become infeasible for an ontology with a high number of
shared \textit{primitive tags} amongst associations per product.
This challenge can be addressed in the construction of the ontology, by indentifying and isolating overlapping
\textit{primitive concepts}.
In the previous example, this can be achieved by creating a new product tag \vmath{\vproducttag_{4}} = "plant part" and
assigning it to all products that have the "vegetable" or "fruit" tag.
All the association scores between the "vegetable" and "fruit" product tags are now reduced by an
amount \vmath{\delta}, such that \vmath{\vcorrelation = \vcorrelation - \delta} with \vmath{\vproducttag \in
\{\vproducttag_1,
\vproducttag_2\}}.
All associated preference tags with the product tags "fruit" and "vegetable" can now be associated to the product
tag "plant part" with association score \vmath{\vcorrelationindexed{\vproducttag_{4}}{\vpreferencetag} = \delta}, as illustrated in
\reffig{fig:reduction:design}.
In such case, all terms of \refeq{eq:overlap:error} are equal to \vmath{0}, since all possible intersection are
equal to the empty set.

Based on the above example to avoid overlaps, a generic \textbf{reduction design principle} is proposed
during the ontology design:
\begin{enumerate}
    \item If a product tag contains all the \textit{primitive concepts} of another product tag, then only one is
    chosen and assigned to a product.
     \item If there are overlaps of \textit{primitive concepts} between two product tags of the same product, but
    neither can be omitted because their non-shared \textit{primitive concepts} are important, then the
    intersection of their \textit{primitive concepts} should be treated as a separate product tag and assigned to all
    products these tags are associated with.
    The shared \textit{primitive concepts} are omitted from the original product tags.
\end{enumerate}
The introduction of the reduction design principle minimizes overlap error \vmath{\epsilon\to0} and therefore the
aggregate association can now be calculated as:
\begin{equation}
    \label{eq:additive}
    \vrelevance =  \sum_{z}{\vcorrelation}
\end{equation}

This calculation has linear complexity \vmath{O(n)} to the number of product tag concepts in each aggregate
association calculation.
Yet, the reduction design principle methodology introduces a quadratic \vmath{O(n^{2})} complexity in regards to
product tags, once during the creation of the ontology, as each tag needs to be compared against each other to
determine overlaps.
The reduction design principle is illustrated in \reffig{fig:reduction:design}.

If the aggregated association is \vmath{0}, then it is not possible to determine whether a product supports or
opposes a preference tag, since it contains equally enough positive and negative concepts.
The uncertainty is treated as a product having no information.
Analyzing and comparing the aggregate associations between different preferences can be used for the
identification of possible \textit{trade-offs} and \textit{rebound effects}. 
If no overlapping occurs (\vmath{\epsilon=0}) between the associations, then the individual bounding property is
also extended to the aggregated association.
The \textbf{aggregated association bounding property} is the following:
\begin{equation}
    \begin{gathered}
        \dfrac{\absol{{\toSet{Q}_{\vproduct}}\cap\toSet{Q}_{\vpreferencetag}^{+}}}
        {\absol{\toSet{Q}_{\vpreferencetag}^{+}}}
        -
        \dfrac{\absol{{\toSet{Q}_{\vproduct}}\cap\toSet{Q}_{\vpreferencetag}^{-}}}
        {\absol{\toSet{Q}_{\vpreferencetag}^{-}}}
        \in [-1,1] \stackrel{\refeq{eq:association}, \epsilon = 0}{\Leftrightarrow}\\
        \vrelevance \in [-1,1]
    \end{gathered}
\end{equation}

Application of the reduction design principle reduces overlaps, yet it may increase the amount of tags and
associations in the ontology, when tags are decomposed to disjoint tags with less primitive concepts.
Furthermore, successful application relies on the ability of the expert or system to identify and break down
associations between preference and product tags.
Therefore a trade-off is introduced between efficiency, performance and maintenance of the ontology.

\vspace*{1cm}
	\subsection{\textbf{Product Rating Mechanism}}\label{sec:rating}
		The ontology is used to calculate a distributed and privacy-preserving product rating value \vmath{\varrho(\vproduct,
\vuser)\in\mathbb{R}} between a
product \vmath{\vproduct} and a user \vmath{\vuser}.
This is achieved by implementing a product rating system with rating design principles of the content based
recommender systems\cite{Jan10, Isinkaye2015}.
The product rating is designed to use the aggregated association scores \vmath{\vrelevance}.

\subsubsection{\textbf{Comparable aggregated associations}}\label{subsec:comparable-aggregated-associations}
Different products may satisfy the same preference tag via product tags that are related to different \textit{primitive
concepts}.
For that reason, comparison of aggregated associations for the same preference tag and different products are
not easy to interpret, i.e. each product may have completely unique characteristics that satisfy each preference.
To be able to compare how different products satisfy a preference, a reference product can be used
to
normalize
all aggregate association scores.
The reference products are defined as follows:
(i) a reference product \vmath{\vproduct+} that maximally satisfies the user's preferences and
(ii) a reference product \vmath{\vproduct-} that maximally opposes them.
For example, a reference product can be a
theoretical or existing product that contain all ontology product tags related to a preference tag.
The maximum possible positive aggregated association score product is defined as
the aggregated association score of a product that contains all product tags positively
associated with a preference tag:
\begin{equation}
    \begin{gathered}
        \vrelpos = \sum_{\vproducttag}{\vcorrelation},\\
        \left\{\vproducttag\,|\,\vproducttag \in
        \toSet{\vproducttag}_{\vproduct+} \wedge \vcorrelation > 0 \right\}
    \end{gathered}
    \label{eq:relevance:max}
\end{equation}
Such product is referred to as \emph{positive reference product}, and its association score is the \emph{positive
reference association}.

Following the same principle, a product with the minimum possible negative aggregated association
score is also introduced.
Such product is referred to as \emph{negative reference product}, and its association score is the \emph{negative
reference association}.
The calculation for such score is:
\begin{equation}
    \begin{gathered}
        \vrelneg = \sum_{\vproducttag}{\vcorrelation},\\
        \left\{\vproducttag\,|\,\vproducttag \in
        \toSet{\vproducttag}_{\vproduct-} \wedge \vcorrelation < 0 \right\}
    \end{gathered}
    \label{eq:relevance:min}
\end{equation}

Once the aggregated association scores are calculated, a normalization is applied by dividing with the preference
scores related to a preference tag.
Therefore, comparison between different products is possible:
\begin{equation}
    \begin{gathered}
        \vrelnorm =  \left\{ \,
        \begin{IEEEeqnarraybox}[][c]{l?s}
            \dfrac{\vrelevance}{\absol{\vrelneg}}\ , \quad \vrelevance < 0 \\
            \quad0\qquad, \quad  \vrelevance = 0 \\
            \dfrac{\vrelevance}{\vrelpos}\ , \quad \vrelevance > 0
        \end{IEEEeqnarraybox}
        \right.
    \end{gathered}
    \label{eq:relevance:norm}
\end{equation}
Existence of overlaps between product tags when calculating the aggregated association of theoretical reference
scores based causes the normalized aggregated association to approach \vmath{0}, as the
denominator in \refeq{eq:relevance:norm} increases.
Still, it is guaranteed that support or opposition towards a preference tag are not switched due to
the overlaps in normalization, since the denominator is positive in both \textit{Relations}
\ref{eq:relevance:max} and \ref{eq:relevance:min}.
The normalized association score is bound in the range \vmath{\left[-1, 1\right]}, since an actual product may have at
most the maximum positive or minimum negative aggregated association score towards a preference tag.
Therefore, the term  \vmath{\vrelnorm} is also bound in range \vmath{\left[-1, 1\right]}.
The choice of the reference products is left to the ontology designer.
Possible choices for a reference product are:
\begin{enumerate}
    \item an existing product that achieves the highest/negative aggregate association score.
    \item a theoretical or existing product that shares all positive/negative associations with a
    preference tag, while respecting the reduction design principle.
    \item a theoretical or existing product that shares all positive/negative associations with a
    preference tag, without respecting the reduction design principle.
    This option calculates the highest possible denominator value in \refeq{eq:relevance:norm}.
    This clipping mechanism is illustrated in \reffig{fig:clip:aggregate}.
\end{enumerate}
The above choices were evaluated with the app testers, and the normalized aggregated
scores produced with the third choice were used during the ASSET field test.

\subsubsection{\textbf{Sustainability Index: Non-personalized Product Representations}}\label{subsec:product-preference-embeddings}
The next step to calculate the product ratings is to establish a relationship between preferences and products.
Aggregation over the normalized association score of a product and preference tags of a preference state estimates
this relationship.
This aggregation is performed by using a measure of central tendency as the estimator.
More specifically, here the expected value is chosen.
The expected value of the normalized aggregated association scores of all the preference tags that are
related to the preference is an estimate that quantifies the support or opposition of a product by the
preference:
\begin{equation}
    \label{eq:connection}
    \begin{gathered}
        \vconnection = \dfrac{\sum_{\vpreferencetag \in \Omega_{\vpreference}}
        {\vrelnorm}}{\absol{\Omega_{\vpreference}}}
    \end{gathered}
\end{equation}
Each preference tag represent a sustainability goal considered in a preference statement.
When a product has positive normalized aggregated association score with a preference tag, then it
supports in achieving the sustainability goal represented by that tag.
In the opposite case, the product may cause the failure of achieving a sustainability goal.
Therefore, the above calculation is referred to as the sustainability index of a product for a
preference, as it
indicates whether purchasing a product supports in achieving or failing a sustainability goal.
Calculating the expected preference-product associations for all preferences and a product yields a numerical
vector representation for that product in the preference semantic space.
Such representations are used to compare products and determine whether a product is expected to be supported or
opposed by a set of preferences.
Measures of central tendency over a product representation, such as the mean \vmath{\overline{\vconnection}} can be
used to calculate the non-personalized ontology estimate of the product sustainability.

\subsubsection{\textbf{Self-determined User Personalization}}\label{subsec:self-determined-product-ratings}
High preference scores indicate that a preference is important for a user when calculating the product rating.
Therefore, a user's preference scores are used as weights of importance for each preference.
A user can express both opposition and support towards any preference.
Thus, any inaccuracy or bias introduced by the ontology design may be mitigated by
the user by adjusting preference scores.
This is also considered as an implicit user-determined extra correction on overlapping \textit{primitive concepts}.
Opposition and support of a user towards preferences are modeled via the offset of a preference score from the
preference score median \vmean{s} as shown in \refeq{eq:offset}:
\begin{equation}
    \label{eq:offset}
    \begin{gathered}
        \vscoreoffset{\vpreferencescore} = \vpreferencescore - \vmean{s}
    \end{gathered}
\end{equation}
The higher the support or opposition of a user towards a preference, the higher the absolute value of the
offset.

A weighted average between product-preference association scores and the user preference offsets are used to
calculate a personalized association between a user and a product, given the user's self-determined preference
scores.
The sum of absolute offsets is used as denominator to preserve the sign of the rating while normalizing.
Preserving the signs allows to extend the association logic to user-product level.
Positive product ratings indicate that the product mostly supports the user
supported preferences and opposes user opposed preferences.
For negative product ratings, the product opposes user supported preferences and supports user opposed
preferences.
Although self-determined personalization allows for a user's \vchal{subjectivity} to influence the ratings, the user
is made aware about which preferences produce such ratings.
This results in a learning effect, that increases user awareness towards their sustainability preferences.
For example when a person tunes the preferences to allow their favorite products achieve high scores, they are
aware that to do so, they have to go against the preference they originally support.
Thus the unscaled product rating is calculated as follows:
\begin{equation}
    \label{eq:rating:raw}
    \begin{gathered}
        \varrho^{*}(\vproduct,\vuser) =
        \dfrac{\sum_{\vpreference}\vconnection\cdot\vscoreoffset{\vpreferencescore}}
        {\sum_{\vpreference}\absol{\vscoreoffset{\vpreferencescore}}}
    \end{gathered}
\end{equation}

The scale of \vmath{[-1, 1]} can be transformed to any range of real numbers by applying a linear scaling with
parameters \vmath{\alpha} and \vmath{\beta}.
Transforming the rating scale has several applicability scenarios.
The rating can be scaled to different ranges to match the most preferred grading system of the
country, where the algorithm is deployed\cite{blanton1999better}.
Another possible usage of the scaling coefficients is to attribute rating for asymmetric perception of negative
and positive rating values\cite{kwon2016third, parguel2011sustainability}.
For example, based on the work of Parguel et al.\cite{parguel2011sustainability}, negative values have
a higher impact on user perception.
In such case, a different lower scaling coefficients \vmath{\alpha, \beta} can be used to reduce
the impact.
\begin{equation}
    \label{eq:rating}
    \begin{gathered}
        \vrating{\vproduct}{\vuser} =
        \alpha \cdot \rho^{*}(\vproduct,\vuser)
        + \beta
    \end{gathered}
\end{equation}
The product rating scale is designed to utilize user preference scores and create an association between a user
and a product based on the product-preference associations.
The rating value is expected to be bounded in a range \vmath{[\beta-\alpha, \beta+\alpha]}.
Products that neither support nor oppose the user's preferences are assigned the mean value of the range
\vmath{\beta}.
If the product supports a user supported preference or opposes a user opposed preference, then the product rating
increases.
On alternative scenarios the product rating decreases.

The rating value compresses \textit{overwhelming information} and shows to the user an estimate of the personalized
product sustainability.
Depending on the UI design, it is possible to allow the user to further explore the ontology
dynamics that result in this rating value.
In the \reffig{fig:prod:rating:ontology} all related ontology entities and rating calculations are presented.

\subsubsection{\textbf{Algorithm complexity}}\label{subsec:algorithm-complexity}
All calculations to compute the product-preference representations rely on information that is not related to the
user.
Therefore, \textit{Equations~\ref{eq:additive}-\ref{eq:connection}} can be computed without privacy intrusion risks.
The computational cost is significantly reduced for the user's device, since it is possible to calculate, store and
distribute the product representations \vmath{\vconnection} by using a central database system.
The calculation and storage complexity for the worst case scenario, where all products are
connected to every product tag, all product tags are associated to all
preference tags and all preference tags are connected to all preferences is:
\begin{equation}
    \label{eq:complexity:equi}
    O(n) = P \cdot T \cdot \Omega \cdot C
\end{equation}
where:
\begin{conditions*}
    \vmath{P} & The number of all products. \\
    \vmath{T} & The number of all product tags. \\
    \vmath{\Omega} & The number of all preference tags.\\
    \vmath{C} & The number of all preferences.
\end{conditions*}
Assuming that all the above sizes are equal, the worst case computational and
storage complexity for the algorithm is polynomial to the power of \vmath{4}:
\begin{equation}
    \label{eq:complexity}
    \begin{gathered}
        (\text{Relation }~\ref{eq:complexity:equi}) \land  (P = T = \Omega = C) =
        O(n^4)
    \end{gathered}
\end{equation}
Since the total products and product tags are often significantly more than the
preference statements and preference tags, the expected time and space
complexity is reduced to quadratic polynomial time \vmath{O(n^{2})}.
Therefore, modern CPUs on mobile phones and database servers can handle up to
hundreds of thousands of product rating calculations per minute and store several
thousands of products and product tags\cite{papadimitriou2019adaptive}.

\subsubsection{\textbf{Overlaps and possible
errors}}\label{subsec:overlaps-and-possible-error-overflows}
As discussed in \refsec{sec:ontology}, the application of the reduction design principle introduces several
trade-offs between the efficiency of the rating calculations and maintenance of the ontology.
Time constraints and limited resources regarding the construction and testing of the ontology may allow overlaps
in the ontology, which introduce errors and biases that may break the bounding properties of \textit{Equation}
\ref{eq:total_association}.
Thus, in such cases a clipping normalization is introduced to avoid numerical instabilities due to overflows outside
the theoretical aggregated association bounds:
\begin{equation}
    \label{eq:normalization}
    \begin{gathered}
        \vnorm{a} =  \left\{ \,
        \begin{IEEEeqnarraybox}[][c]{l?s}
            -\tau\ , \text{if}\quad a \leq -\tau \\
            a\quad, \text{if}\quad -\tau < a < \tau \\
            \tau\quad, \text{if}\quad a \geq \tau
        \end{IEEEeqnarraybox}
        \right.
    \end{gathered}
\end{equation}
The introduction of the clipping changes \textit{Relations}~\ref{eq:additive},~\ref{eq:relevance:max}
and~\ref{eq:relevance:min} to the following:
\begin{equation}
    \label{eq:relevance:norm:asset}
    \begin{gathered}
        \vrelevance = \vnorm{\sum_{\vproducttag \in \toSet{\vproducttag}_{\vproduct} }\vcorrelation}
    \end{gathered}
\end{equation}
\begin{equation}
    \begin{gathered}
        \vrelpos = \vnorm{\sum_{\vproducttag}{\vcorrelation}},\\
        \left\{\vproducttag\,|\,\vproducttag \in
        \toSet{\vproducttag} \wedge \vcorrelation > 0 \right\}
    \end{gathered}
    \label{eq:relevance:max:norm}
\end{equation}
\begin{equation}
    \begin{gathered}
        \vrelneg = \vnorm{\sum_{\vproducttag}{\vcorrelation}},\\
        \left\{\vproducttag\,|\,\vproducttag \in
        \toSet{\vproducttag} \wedge \vcorrelation < 0 \right\}
    \end{gathered}
    \label{eq:relevance:min:norm}
\end{equation}
Even if overlaps affect the rating process, the users are able to mitigate the error by readjusting their
preference scores, introducing an extra correction mechanism from their side.
An illustration of the clipping mechanism on the normalized aggregate score is found in
\reffig{fig:clip:aggregate} in step 1.

		\subsection{\textbf{Tractable and Explainable Ratings}}\label{subsec:tractable-and-explainable-ratings}

The proposed ontology design and rating calculations rely on a fully tractable analytical framework.
It is possible to calculate the exact amount that a preference, preference or product tag contributed to the rating.
More specifically following \cref{eq:rating:raw,eq:rating}, one could solve to calculate the exact contribution of a
specific preference \vmath{\vpreference^*} to the product rating as follows:
\begin{equation}
    \label{eq:rating:raw:contr}
    \begin{aligned}
        \varrho(\vproduct,\vuser) &= \alpha\left(
        \dfrac{\sum_{\vpreference \in \toSet{\vpreference} -
        \{\vpreference^*\}}\vconnection\cdot\vscoreoffset{\vpreferencescore}}
        {\sum_{\vpreference}\absol{\vscoreoffset{\vpreferencescore}}} \right.\\
        &+  \left. \dfrac{\vconn{\vproduct}{\vpreference^*}\cdot\vscoreoffset{\vpreferencescore^*}}
        {\sum_{\vpreference}\absol{\vscoreoffset{\vpreferencescore}}}
         \right) + \beta
    \end{aligned}
\end{equation}
Therefore the contribution of preference \vmath{\vpreference^*} to the final rating is calculated as:
\begin{equation}
    \label{eq:contr:preference}
    \begin{gathered}
        \alpha\dfrac{\vconn{\vproduct}{\vpreference^*}\cdot\vscoreoffset{\vpreferencescore^*}}
        {\sum_{\vpreference}\absol{\vscoreoffset{\vpreferencescore}}}
    \end{gathered}
\end{equation}
The contribution of a specific preference tag \vmath{\vpreferencetag^*} is calculated following the same
decomposition logic on \cref{eq:rating:raw,eq:rating,eq:connection}, the contribution of a specific preference tag
to the rating can be calculated as:
\begin{equation}
    \label{eq:contr:prftag}
    \begin{gathered}
        \alpha
        \dfrac{\sum_{\vpreference \in \toSet{\vpreference}}
         \dfrac{\vrelsupp{*}{\vproduct, \vpreferencetag^*}}{\absol{\Omega_{\vpreference}}}
        \cdot\vscoreoffset{\vpreferencescore}}
        {\sum_{\vpreference}\absol{\vscoreoffset{\vpreferencescore}}} \quad
    \end{gathered}
\end{equation}
And the contribution of a specific product tag \vmath{\vproducttag^*} to the product rating can be calculated by
decomposing
\cref{eq:rating:raw,eq:rating,eq:connection,eq:relevance:norm}:
\begin{equation}
    \label{eq:contr:prftag}
    \begin{gathered}
        \alpha
        \dfrac{\sum_{\vpreference \in \toSet{\vpreference}}
         \dfrac{ \sum_{\vpreferencetag \in \Omega_{\vpreference}}
         \dfrac{
          \vcorr{\vproducttag^*,\vpreferencetag}
         }{
          \eta_{norm}
         }
         }{\absol{\Omega_{\vpreference}}}
        \cdot\vscoreoffset{\vpreferencescore}}
        {\sum_{\vpreference}\absol{\vscoreoffset{\vpreferencescore}}} \quad
    \end{gathered}
\end{equation}
where \vmath{\eta_{norm}} has the corresponding denominator value of either \vmath{\{\vrelpos, \vrelneg\}}  that
normalizes the aggregated
association in
\cref{eq:relevance:norm}.
In case that clipping is used, as shown in \refeq{eq:relevance:norm:asset}, the contribution is
calculated
proportionally:
\begin{equation}
    \label{eq:contr:prftag}
    \begin{gathered}
        \alpha
        \dfrac{\sum_{\vpreference \in \toSet{\vpreference}}
         \dfrac{ \sum_{\vpreferencetag \in \Omega_{\vpreference}}
         \dfrac{
            \dfrac{
            \vcorr{\vproducttag^*,\vpreferencetag}
            }{
             \sum_{z}{\vcorrelation}
            }
         }{
          \eta_{norm}
         }
         }
         {\absol{\Omega_{\vpreference}}}
        \cdot\vscoreoffset{\vpreferencescore}}
        {\sum_{\vpreference}\absol{\vscoreoffset{\vpreferencescore}}} \quad
    \end{gathered}
\end{equation}
Product tag contributions can also be calculated per preference or preference tag.
Such calculations are possible by ommitting the calculation related to the product rating and
calculate the amount a product tag contributes to other scores, such as the sustainability index discussed in \Cref{eq:connection}.
Finally, contributions can either be stored when calculating the product rating, increasing memory complexity or
recalculated after the rating calculation, thus increasing computation complexity.

	\subsection{\textbf{Crowd Sourced Product Data}}\label{sec:crowd_sourced_product_data}
			To facilitate wisdom of the crowds, a datathon took place at the University of Tartu,
			Estonia on November 2017~\cite{datathon}.
			In total three teams of 3-4 people participated in the datathon, and produced data files and code within 72 hours.
			The datathon teams processed text data from Wikipedia pages related to product characteristics and sustainability goals.
			Each team used statistical methods and machine learning techniques based on NLP and sentiment analysis to extract an association score between product tags and sustainability topics.
			Each score should cite the wikipedia page id that was used to extract such data. 
			This resulted in several thousand association scores between product and preference tags.
			The ASSET consortium used the majority of these results to evaluate existing association score and discover new ones.

			Several new product tags and associations were introduced in the ontology, mainly in the environmental and social categories.
			Further collaboration with members of the winning datathon team took place in order to enrich and evaluate the created ontology.
				
			As new product tags and association scores were discovered, the datathon provides a real world example on how machine learning can be used to populate a sustainability ontology.
			Furthermore, the datathon showcases a working case, where crowd-sourced data can be used for populating a sustainability ontology.
				
\subsection{\textbf{Software Libraries}}
		Data were collected, stored and processed for categorization and classification by AINIA according to GDPR.  The smart phone app integrates a library version~\cite{nervousnet} of the Nervousnet system~\cite{Pournaras2015}. While no Nervousnet functionality has been used during the lifetime of this project and work, its integration has been part of a deliverable for the ASSET EU-project. Its modular and flexible sensor data management system has potential future applicability in shopping scenarios from multiple retailer shops.
		
		For the development and analysis of the presented results several libraries were used. Some
notable references are: (i) pycausalimpact\footnote{\url{https://github.com/dafiti/causalimpact}},
(ii) scikit-learn~\cite{scikit-learn}, (iii) scipy~\cite{scipy}, (iv) the discrete information
theory package~\cite{dit}, (v)pyclustering~\cite{andrei_novikov_2018_1491324}, (vi)
plotly~\cite{plotly}.
Several other libraries that were used, are also mentioned in the paper code repository.

\section*{\textbf{Supplementary Notes}}	
\setcounter{subsection}{0}
\renewcommand\thesubsection{SN.\arabic{subsection}}
		\subsection{\textbf{Entry and Exit Survey}}\label{subsec:behavioral-change:qualitative}
		Entry and exit surveys are presented to the user during the field test.
The aim of the surveys is to collect data regarding user opinions on sustainable consumption,
sustainability preferences and the usage of the application.
The number of participants for each survey is summarized in \reftab{tbl:surveys:part}
Answers on the exit survey provide useful insights regarding the behavioral change and sustainability awareness
of users during the field test.
Users confirmed that the rating methodology offers product
ratings that justify their preference settings.
Access to product information and the product tags is visible in the application for users that click on a product.
Such information assists the decision or the product purchase.
User declared that it is relatively easy to access the product information and that the provided information
justifies the rating (\textit{Supplementary Figures}~\ref{fig:feedback:pos:extrainfo} and~\ref{fig:feedback:pos:justify}).
The highly rated products tend to match the preferences of the majority of the users as shown in
\textit{Supplementary Figures}~\ref{fig:feedback:pos:prefmatch},~\ref{fig:feedback:pos:preferences:account} and
~\ref{fig:feedback:pos:capture:preferences}.
Finally, an increase in discovery of novel products is stated by the users in \reffig{fig:feedback:pos:novelty}.

Shopping behavior is affected for some users by using the app.
Mainly in retailer B users support that they would not buy products that achieve low ratings and buy high
rated products before using the app, which emphasizes their sustainability focus
(\textit{Supplementary Figures}~\ref{fig:feedback:awareness:high:before} and~\ref{fig:feedback:awareness:high:before}).
Regarding future purchases, users support that they will buy products that achieve high ratings and also avoid
buying products that achieve low rating, as shown in \textit{Supplementary Figures}~\ref{fig:feedback:awareness:high:rating}
and~\ref{fig:feedback:awareness:low:rating}.
Furthermore, users in retailer A estimated that they buy over \vmath{5\%} of highly rated products in retailer A, whereas
most retailer B users estimate a value over \vmath{10\%}, as illustrated in
\reffig{fig:feedback:percentage:bought}.
In both retailers, the majority of users agree to an extent that the usage of the application raised their
awareness towards sustainability \reffig{fig:feedback:awareness:awareness}

Product information and price seem to be the main decision factors when
users decide regarding a product purchase, as illustrated in \reffig{fig:feedback:final}.

Concluding, weakly qualitative evaluation indicates that the application provides enough information to
the users to purchase highly rated products, thus increasing the sustainability of their consumption.
Screenshots from the app UI of the survey and preferences is illustrated in \reffig{fig:app:usage}.
Users perceive that the product rating takes into account their preferences and is well justified by the available
data. Other factors may affect the participation of users such as technical problems with the app, low usability of UI, seasonality (traveling), limited access to retailers shops for some consumers, etc.

	\subsection{\textbf{Evaluation of Product Prices}}\label{subsec:prices:eval}
		
\subsubsection*{Sustainability index vs product price}
This evaluation studies whether products that highly support a specific preference have a higher price
compared to products from the same category.
Evaluating sustainability index and price over all products provides insights about the general expenditure.
Consumers might be interested in replacing products they buy with more sustainable products from the same
category.
To evaluate whether more expensive products achieve higher sustainability scores compared to
competitive products from the same category, both price per unit and sustainability index are
rescaled using the min-max normalization for each product per category.
I.e. Each value is normalized by substracting the minimum value and then dividing by the
difference between maximum and minimum value\cite{dodge2006oxford}.
For the normalization to work, each category needs to contain at least two products with different sustainability
index and price values.
Therefore, only such categories are used for the analysis.
As illustrated in \reffig{fig:price:radial:norm:coop}, purchasing more expensive products per category in retailer A seems to
have almost no impact on the sustainability index for most categories.
The vegan diet preference "H.12" is negative correlated with price, possibly indicating that the most
expensive products per category often contain an animal product.
For retailer B, the sustainability index for most preferences is either positively correlated or uncorrelated with
prices.
Negative correlated preferences with prices are "E.2", "E.3" and "S.1", indicating that one can
purchase cheaper products per category, which are farmed in a sustainable manner, distributed in an environmentaly
friendly way and passed sustainability auditing process.
This result may indicate that in most categories the most expensive product originates from a place that the
transportation has high environmental impact, sustainable farming techniques are not followed or sustainable
auditing processes do not hold.
Furthermore, since retailer B is a sustainability focused retailer, it is expected that the suppliers may mitigate
sustainable production costs in their product prices.

Note that the lower expenditure level of Retailer A during the summer is a seasonality effect that is prominent due to the students' holidays and constructions in the neighborhood.


\printnomenclature
	
\bibliography{026_document_no_url}
\bibliographystyle{naturemag}